\documentclass[12pt]{article}

\usepackage{bm}
\usepackage{float}
\usepackage{relsize}
\usepackage{array}
\usepackage{fullpage}
\usepackage{cancel}
\usepackage{amsfonts}
\usepackage{amsmath}
\usepackage{setspace}
\usepackage{easybmat}
\usepackage{slashed}
\usepackage{amsmath}
\usepackage{amssymb}
\usepackage{graphicx}
\usepackage{verbatim}
\usepackage{caption}
\usepackage{subcaption}
\usepackage{makecell}
\usepackage{hyperref}

\usepackage{hyperref}

\usepackage{environ}
\usepackage{tikz}

\usepackage{environ}
\usepackage{tikz}
\tikzset{Witten diagram/.style={execute at begin picture={%
\draw[black,fill=white!20] circle[radius=\pgfkeysvalueof{/tikz/Witten/radius}];
}},vertex/.style={circle,fill,inner sep=1.5pt,node
contents={}},
Witten/.cd,radius/.initial=2.0cm}
\NewEnviron{wittendiagram}[1][]{\vcenter{\hbox{\begin{tikzpicture}[Witten diagram,#1]%
\BODY
\end{tikzpicture}}}}

\usetikzlibrary{decorations.pathmorphing}
\usetikzlibrary{positioning}
\usetikzlibrary{intersections}
\tikzset{snake it/.style={decorate, decoration=snake}}

\usetikzlibrary{decorations.markings}
\usetikzlibrary{arrows.meta}
\usetikzlibrary{decorations.pathreplacing}

\allowdisplaybreaks
\def\timesbox{\hbox{$\scriptscriptstyle\times$}}
\def\ant{ {{\lower 1ex  \timesbox} \atop {\raise 1.5ex  \timesbox}}}
\def\f{\frac}
\def\pa{\partial}

\def\a{\alpha}
\def\b{\beta}
\def\c{\chi}
\def\d{\delta}
\def\e{\epsilon}
\def\g{\gamma}
\def\k{\kappa}

\def\n{\nu}

\def\t{\tau}

\def\D{\Delta}

\def\G{\Gamma}

\newcommand{\mathsym}[1]{{}}
\newcommand{\unicode}[1]{{}}

\newcommand{\be}{\begin{equation}}
\newcommand{\ee}{\end{equation}}
\newcommand{\beqa}{\begin{eqnarray}}
\newcommand{\eeqa}{\end{eqnarray}}
\newcommand{\bsp}{\begin{split}}
\newcommand{\esp}{\end{split}}
\newcommand{\bgth}{\begin{gather}}
\newcommand{\egth}{\end{gather}}

\newcommand{\del}{\partial}

\newcommand{\cO}{{\cal O}}

\title{Thermal AdS two-point functions and loop induced effective one-point vertices}
\author{Astha Kakkar $^{a}$\footnote{asthakakkar8@gmail.com}   \\
$^a$\small{{\em Indian Institute of Technology Kanpur,
 Kalyanpur, Kanpur 208016, India}}
}

\date{}

\begin{document}

\maketitle
\abstract{We present a study of thermal two-point functions from Witten diagrams in thermal AdS$_{d+1}$, defined with the conformal boundary $S_\beta^1 \times \mathbb{R}^{d-1}$, with an aim to establish a connection between thermal CFT data and perturbative bulk dynamics. We analyze one-loop tadpole and bench diagrams arising from the cubic $\phi\phi\chi$ interaction in the bulk, segregating the different relative thermal winding sectors for the bench, and also extend the analysis to include quartic scalar contact diagrams with and without derivative interactions. We also analyze the spinning exchange thermal two-point function. Utilizing the split representation of bulk propagators and the method of images we reorganize the thermal Witten diagrams into the standard AdS integrals and obtain spectral and image-sum representations that make KMS periodicity of the correlators manifest. Additionally, we show possible position dependent effective one-point vertices of scalar and spinning fields in the bulk arising from the reduction of the coincident- point thermal bulk to bulk propagator loop, allowing tadpole-type diagrams to be reconstructed from effective local bulk interactions. The bench diagram however retains an intrinsically bi-local structure. We further investigate the thermal one-point coefficients and perturbative OPE data encoded in these thermal correlators, including contributions associated with anomalous dimensions. Finally, we examine their relation to thermal dispersion relations and the role of contributions that are not determined by discontinuity data alone. Our results provide a set of explicit bulk constructions for exploring the analytic structure of thermal CFT correlators and their connection to the thermal conformal bootstrap. }

\newpage

\tableofcontents

\section{Introduction}

Correlation functions in thermal Anti-de Sitter (AdS) space contain a wealth of dynamical data.
Ranging from shedding light on aspects of thermal phase transitions, especially near critical points, to providing the building blocks for the recent advances in thermal bootstrap they form a crucial set of objects to be studied. In the holographic dictionary black holes in asymptotically AdS space are understood as holographic duals of thermal states in the boundary CFT \cite{Witten:1998zw}. At high temperatures across the Hawking-Page transition \cite{Hawking:1982dh}
the black hole is the stable background while at lower temperatures thermal AdS is the dominant saddle and thus provides a useful playground for exploration of QFTs on a fixed background without the complexities introduced by the presence of black hole horizons. Thus from the holographic perspective, a thermal state on the boundary CFT defined on \(S^1_\beta \times \mathbb{R}^{d-1}\) at low temperatures maps to a fixed background in the bulk endowed with a compact thermal cycle. Consequently thermal correlation functions in the finite temperature boundary CFT can be computed by evaluating thermal AdS bulk Witten diagrams,  thereby explicitly relating boundary thermal data to the spectrum and interactions of fields propagating in the bulk geometry. The analytic properties of thermal correlators, the Kubo--Martin--Schwinger (KMS) periodicity, and the OPE data acquire thus a direct bulk interpretation. Consequently, thermal AdS offers a natural laboratory for understanding the dual bulk origin of thermal bootstrap constraints and for exploring how boundary thermal correlators encode information about the bulk dynamics. We thus embark on a journey to study and formulate the thermal bootstrap directly as a consequence of bulk analysis in thermal AdS, and evaluate bulk Witten diagrams in this regard to derive the allowed analytic structure, study the appearance of the KMS invariant two-point functions, derive thermal OPE data, and compare to the dispersion relations of boundary two-point functions. \\

\noindent
These thermal correlators on $S^1_{\beta}\times \mathbb{R}^{d-1}$ contain information that is absent from vacuum correlation
functions such as non vanishing thermal one-point coefficients. This richness arises  as on compactifying the Euclidean time the conformal symmetry gets correspondingly reduced. However, thermal correlators in this case are constrained by the powerful Kubo--Martin--Schwinger (KMS) condition that relates correlation functions at different points on the thermal circle and plays, at finite temperature, an analogous role to crossing symmetry in the vacuum bootstrap. There have been major developments in the recent times strengthening this domain of the thermal bootstrap programme. The modern thermal bootstrap was launched in \cite{Iliesiu:2018fao} and was applied to the 3d Ising model in \cite{Iliesiu:2018zlz}. Holographic approaches were also explored in \cite{Alday:2020eua}. Recent developments include \cite{Barrat:2025nvu} \cite{Barrat:2025twb} where analytic approaches have included the formulation of thermal two-point functions in terms of dispersion relations in the complexified Euclidean time plane and also extended to non zero spatial separations. These dispersion relations have been used to reconstruct the correlators from their discontinuities,
up to an additive constant and theory-dependent information,
and further the generalized method of images construction has been deployed to extend the results away from the OPE convergence
regime. There also have been holographic studies on the black brane background \cite{Barrat:2025twb}\cite{Rodriguez-Gomez:2021pfh} 
\cite{Karlsson:2022osn}\cite{Dodelson:2022yvn}.  Related past studies have also explored anomalous dimensions from thermal AdS partition functions \cite{Kraus:2020nga}, computation of thermal AdS Witten diagrams \cite{Alday:2020eua} \cite{Buric:2026pes} \cite{Buric:2024kxo} and conformal blocks for thermal one-point-functions \cite{Kraus:2017ezw} \cite{Gobeil:2018fzy}. More recently there have been attempts to relate OPE and QNM data \cite{Arnaudo:2026axe} and proposals for the thermal Polyakov blocks \cite{Barrat:2026jfg}, \cite{Guo:2026xyl}.\\

\noindent
Evaluating bulk thermal Witten diagrams thus provides a natural setting to compliment these recent developments.  We therefore attempt to understand these developments in thermal bootstrap directly from the bulk dynamics. An important initial step in this direction appeared as a study of weakly coupled QFTs in AdS at finite temperature \cite{Alday:2020eua}, where a contact thermal two-point Witten diagram was computed from bulk interactions in Mellin space. For local quartic interactions, the resulting correlators were shown to satisfy analyticity, OPE and KMS constraints. Towards the goals described above, in this paper, we employ the split representation of the AdS propagators to evaluate our Witten diagrams in the position space. In the vacuum, Witten-diagram technology in position space is highly developed \cite{Costa:2014kfa} - \cite{Carmi:2024tzj}. The main aim of this paper hence is to bolster the development of the thermal bulk analysis in a systematic manner. We thus extend the computation of Witten diagrams to the thermal AdS by a careful rearrangement of the bulk thermal indices which allows us to utilize the usual vacuum AdS integrals. We thermally identify both ends of every bulk-to-bulk
and bulk-to-boundary propagator, the most general setup, and then shift the thermal arrangements of the bulk points such that the bulk integrals take their familiar vacuum form. The method of images implements thermal periodicity by summing Euclidean-time translations.\\

\noindent
 Another second motivation comes the dispersion relation approaches where a thermal two-point function can be reconstructed from its singularity structure and appropriate asymptotic information. From the bulk point of view, it is therefore natural to ask whether the
thermal dispersion relation itself admits a direct bulk counterpart. This gives a bulk interpretation of the quantities appearing in the boundary bootstrap and the constraints satisfied by them. Our bulk Witten diagrams thus would compliment the dispersive relations approach to reconstruct the two point functions while also capturing the contributions not determined by the discontinuities. Our analysis also provides a controlled laboratory for the characterization of bulk effective interaction vertices that arise from the reduction of coincident point thermal bulk loops allowing us to rebuild the tadpole like diagrams. A complete understanding of how these bootstrap constraints systematically translate to the holographic bulk remains an active frontier in the AdS/CFT correspondence.\\

\noindent
The paper is organized as follows. In section \ref{section2}
we lay down the basic notations and concepts that we utilize in and build our analysis on. Section \ref{section3} then gives the computation of the one-loop tadpole and bench diagrams for scalars. In section \ref{section4} we revisit the contact diagrams without and with derivative interactions. In section \ref{section5} we extend our analysis to include spinning exchanges while section \ref{section6} explores the effective vertices from the reduction of the bulk loops. In section \ref{drarc} we examine how our thermal Witten diagrams relate to the reconstruction of correlators from the thermal CFT dispersion relations. Finally we conclude and provide some future directions of study in section \ref{section7}.

\section{Generalities of (thermal) AdS and thermal CFT review }\label{section2}
In this section we lay down the building blocks for the computations of thermal two point functions from Witten diagrams and how they relate to the recent thermal bootstrap studies.

\subsection{Euclidean and thermal AdS geometry}

Euclidean AdS$_{d+1}$ is realized as the hyperboloid embedded in $\mathbb{R}^{1,d+1}$
\begin{equation}
X^2
\equiv
-X_0^2+X_1^2+\cdots+X_d^2+X_{d+1}^2
=
-R^2.
\label{eq:hyperboloid}
\end{equation}
We may introduce light-cone embedding coordinates with $~X^\pm=X_0\pm X_{d+1}$ as
\begin{equation}
X^A=(X^-,X^+,X^i)
\label{eq:lc}
\end{equation}
and further choose Poincar\'e patch coordinates
\begin{equation}
X^A
=
\left(
\frac{R^2}{z},
\frac{z^2+x^2}{z},
\frac{R\,x^\mu}{z}
\right)
\label{eq:poincare}
\end{equation}
where $x^\mu=(\tau,\mathbf{x})$. In the Poincar\'e coordinates, the Euclidean AdS metric takes the form
\begin{equation}
ds^2
=
\frac{R^2}{z^2}
\left(
dz^2+d\tau^2+d\mathbf{x}^2
\right).
\label{eq:EAdS}
\end{equation}
where $R$ is the AdS radius and the conformal boundary is located at \(z=0\). Thermal AdS$_{d+1}$ is then defined as the quotient space $\mathbb{H}^{d+1}/\mathbb{Z}$ by compactifying the Euclidean time direction as $\t\sim\t+\beta$,  where $\b = T^{-1}$, in the Euclidean Poincar\'e coordinates and depicted with the following action of $\g^{n}\in \mathbb{Z}$ on the coordinates $X=(z,\t,\mathbf{x})$
\beqa
\g^n(z,\t,\mathbf{x})=(z,\t+n\b,\mathbf{x}).
\eeqa
The conformal boundary in this case is $S^1_\b\times\mathbb{R}^{d-1}$.
We define another set of coordinates, which will be helpful in our analysis, below
\beqa
r=\frac{\sqrt{\t^2+|\mathbf x|^2}}{\b} ,~~~~ \eta=\frac{x\cdot e}{|x|} =\f{\t}{\sqrt{\t^2+|\mathbf x|^2}}
\eeqa
where $e^\mu$ is a unit vector along the thermal circle. To make connection with the CFT results more concrete we also define the following set of coordinates
\beqa
z=\t+ix~~,~~\bar{z}=\t-ix
\eeqa
and thus can write
\beqa
r^2 = z \bar{z} \,, \qquad 2\eta r=z+\bar{z} \,, \qquad w^2 = \frac{z}{\bar{z}} \,.
\eeqa
We further follow the convention under which $X,Y,Z$ represent the bulk embedding coordinates and $P_i$ the coordinates on the boundary. Boundary points are again null coordinates $P^2 = 0$ and $P \simeq \lambda P$. Additionally while writing the coordinate dependence in our thermal Witten diagram computations we follow the notation $x+n\g=(\t+n,\mathbf{x})$ to indicate thermally identified boundary points and $X^{a\g}\equiv(z,\t+a,\mathbf{x})$ as the thermally identified bulk points with $\b$ being set to 1.

\subsection{Propagators and normalizations}

The bulk to boundary propagator for a scalar in AdS$_{d+1}$ in embedding coordinates is
\beqa
K_{\D}(P,X)=\frac{\sqrt{\mathcal{C}_{\D}}}{(-2P.X)^{\D}}
\eeqa
while in Poincar\'e coordinates this is written as
\beqa
K_{\D}(z,x')= \sqrt{\mathcal{C}_\D}\left(\f{z}{z^2+|x-x'|^2}\right)^{\Delta}
\eeqa
with the normalization defined as
\beqa
\mathcal{C}_\D=\left(\f{\Gamma(\D)}{2\pi^{d/2}\Gamma(\D-d/2+1)}\right)^{\f{1}{2}}.
\eeqa
Aditionally, the scalar bulk to bulk propagator is written as
\beqa
G_\D(X_1,X_2)=\f{\mathcal{C}_\D}{\xi(x_1,x_2)^\D}{}_2 F_1\left(\D,\D-\f{d}{2}+\f{1}{2},2\D-d+1,-\f{4}{\xi(X_1,X_2)}\right) \nonumber
\eeqa
where $\xi$ is the squared chordal distance between the two bulk points, given in Poincare coordinates as
\beqa
\xi(X_1,X_2)=R^2\f{(z_1-z_2)^2+(x^\mu_1-x^\mu_2)^2}{z_1 z_2}.
\eeqa
The scalar bulk to bulk propagator also admits a very useful spectral representation allowing us to write
\begin{equation}
G_{\Delta}(X_1, X_2) =  \int \frac{d\nu}{2\pi} \frac{2\nu^2 \sqrt{\mathcal{C}_{\frac{d}{2}+i\nu}\mathcal{C}_{\frac{d}{2}-i\nu}}}{\nu^2 + \left(\Delta - \frac{d}{2}\right)^2} \int dP \, K_{\frac{d}{2}+i\nu}(P, X_1) K_{\frac{d}{2}-i\nu}(P, X_2),
\end{equation}
There also exists the related harmonic function $\Omega_{\nu}$ which can be expressed as
\begin{equation}
\Omega_{\nu}(X_1, X_2) = \frac{\nu^2 \sqrt{\mathcal{C}_{\frac{d}{2}+i\nu}\mathcal{C}_{\frac{d}{2}-i\nu}}}{\pi} \int dP \, K_{\frac{d}{2}+i\nu}(P, X_1) K_{\frac{d}{2}-i\nu}(P, X_2)
\end{equation}
The thermal AdS bulk to boundary and bulk to bulk propagators can now be constructed from those presented above by using the method of images as
\beqa
K_\D^\b(P,X)=\sum_{n\in \mathbb{Z}}K_\D(\g^n P,X)
\eeqa
and 
\beqa
G_\D^\b(X_1,X_2)=\sum_{n\in \mathbb{Z}}G_\D(X_1,\g^n X_2)
\eeqa
where the $n=0$ for the coincident point propagator gives the vacuum AdS divergent term. In our diagrams with coincident point bulk to bulk propagators we consider $\sum_{n\neq 0} G_\D(X_1,X_2^{n\g})$ where the $n=0$ divergent term will be assumed to be regularized by some appropriate scheme. This shall not change the finite temperature analysis. As a particularly useful case we have the following coincident point thermal bulk to bulk propagator
\beqa
G^\b_{\D}(X,\g^n X)=\sum_{n \neq 0}\mathcal{C}_\D \xi_n^{-\D} ~{}_2F_1\left(\D,\D-\f{d}{2}+\f{1}{2},2\D-d+1,-\f{4}{\xi_n}\right)
\eeqa
where for the two bulk points \(X^{n_1\g}\) and \(X^{n_2\g}\), separated by a translation
$\t \rightarrow \t+n ~,~ n=n_1-n_2$, we can define the chordal invariant as
\beqa
\xi_n=\xi(X,\g^n X)=(X_{}-\g^n X)^2=\left(\f{n}{z} \right)^2
\eeqa
We will use these expressions together with $R=1$ and $\b=1$ in our computations that follow.

\subsection{Spinning propagators and cubic couplings}
\label{sec::sbd}

Next we review the basics for spinning fields in AdS following the analysis in \cite{Costa:2014kfa} \cite{Giombi:2017hpr} and \cite{Sleight:2016hyl}.
A totally symmetric field of mass $m^2 = \Delta(\Delta - d) - J$ and with spin $J$, subject to the constraints of tangentiality and homogeneity, can be written in the ambient space as
\begin{equation}
\phi_J(X;U) = \frac{1}{J!} \phi_{A_1 \dots A_J}(X) U^{A_1} \dots U^{A_J}
\end{equation}
where $U^A$ is the constant ambient auxiliary vector. The Levi-Civita connection in the ambient space representation and the projection operator respectively are
\beqa
\nabla_A = \mathcal{P}_A^{B} \f{\partial}{\partial X^B}~~\mathcal{P}_A^{B} = \delta_A^B - X_A X^B / X^2
\eeqa
The ambient representative $f_{A_1,\cdots A_J}$ of a spin $J$ symmetric boundary field is traceless with respect to the ambient metric and also tangent to the light-cone. In addition to $P^2=0$ we also have $Z^2 = 0$ and $Z \cdot P = 0$. Here $Z^2 = 0$ enforces tracelessness and $Z \cdot P = 0$ enforces transversality and we write
 \beqa
 f_J(P;Z)=\f{1}{J!}f_{A_{1},\cdots A_J}(P)Z^{A_1}\cdots Z^{A^J}. 
 \eeqa
We also come across the following Thomas derivative, which replaces the partial derivative,  as a result of the modification of the operator calculus
\beqa
\hat{\partial}_{ZA} = \partial_{ZA} - \frac{1}{d-2+2Z \cdot \partial_Z} Z_A \partial_Z^2.
\eeqa
The bulk-to-boundary propagator for such a totally symmetric field can be expressed in the ambient space formalism \cite{Mikhailov:2002bp} \cite{Costa:2014kfa}:
\begin{equation}
    K_{\Delta,s}\left(X,U;P,Z\right)=
    \left(U \cdot Z-\frac{U \cdot P Z \cdot X}{P \cdot X}\right)^J\frac{\sqrt{\mathcal{C}}_{\Delta,J}}{\left(-2P \cdot X\right)^{\Delta}},
\end{equation}
with the normalisation
\begin{equation}
    \mathcal{C}_{\Delta,J}=\frac{\left(\Delta+J-1\right)\Gamma\left(\Delta\right)}{2\pi^{d/2}\left(\Delta-1\right)\Gamma\left(\Delta+1-\frac{d}{2}\right)}.
\end{equation}
We can also express the bulk-to-boundary propagator in the form \cite{Sleight:2016hyl}
\begin{equation} \label{bubos}
     K_{\Delta,J}\left(X, U;P, Z \right) = \frac{1}{\left(\Delta-1\right)_J}\left({\cal D}_P\left(Z;U\right)\right)^JK_{\Delta,0}\left(X;P\right),
\end{equation}
where the differential operator reads
\begin{equation}
 {\cal D}_P\left(Z;U\right) = \left(Z \cdot U\right)\left(Z \cdot \frac{\partial}{\partial Z} - P \cdot \frac{\partial}{\partial P}\right) + \left(P \cdot U \right)\left(Z \cdot \frac{\partial}{\partial P}\right),
\end{equation}
and acts on a scalar bulk-to-boundary propagator with the same dimension. This in particular leads to the following useful relation
\begin{equation}
  \left(U_i \cdot \partial_X\right)^n  K_{\Delta,J}\left(X, U;P, Z \right) = \frac{2^n\left(\Delta+1-\frac{d}{2}\right)_n}{\left(\Delta-1\right)_J}\left({\cal D}_P\left(Z;U\right)\right)^J\left(U_i \cdot P\right)^n K_{\Delta+n,0}\left(X;P\right),
\end{equation}
which we shall use to evaluate our Witten diagrams.
The spectral representation for the traceless transverse (TT) bulk to bulk propagator can now be written as
\begin{equation}
G_{\Delta,s}^{\text{TT}} = \int_{-\infty}^{\infty} d\nu \, \frac{\Omega_{\nu,s}(X_1, X_2)}{\nu^2 + \left(\Delta - \frac{d}{2}\right)^2}. 
\end{equation}
where now
\begin{equation}
\Omega_{\nu,J}(X_1, X_2) = \frac{\nu^2}{\pi} \int dP \, K_{\frac{d}{2}+i\nu, J}(X_1; P, \partial_Z) \cdot K_{\frac{d}{2}-i\nu, J}(X_2; P, Z). 
\end{equation}
For the cubic couplings of interest to us we follow the conventions laid down in \cite{Giombi:2017hpr} where the basis of on-shell cubic vertices, in the ambient space formalism, for totally symmetric fields was given by
\begin{multline}\label{Ical}
{\cal I}_{J_1,J_2,J_3}^{n_1,n_2,n_3}=\sum_{m_i} C_{J_1,J_2,J_3;m_1,m_2,m_3}^{n_1,n_2,n_3}{\cal Y}^{J_1-m_2-m_3}_1{\cal Y}^{J_2-m_3-m_1}_2{\cal Y}^{J_3-m_1-m_2}_3  {\cal H}^{m_1}_1{\cal H}^{m_2}_2{\cal H}^{m_3}_3\, \\
\times \phi_{J_1}\left(X_1,U_1\right)\phi_{J_2}\left(X_2,U_2\right)\phi_{J_3}\left(X_3,U_3\right)\Big|_{X_i=X},
\end{multline}
and the cumulative sum was defined as
\beqa
\sum\limits_{m_i} = \sum\limits^{\text{min}\left\{J_1,J_2,n_3\right\}}_{m_3=0}  \sum\limits^{\text{min}\left\{s_1-n_3,J_3,n_2\right\}}_{m_2=0}\sum\limits^{\text{min}\left\{J_2-n_3,J_3-n_2,n_1\right\}}_{m_1=0}
\eeqa
with the coefficients reading
\begin{multline}
   C_{J_1,J_2,J_3;m_1,m_2,m_3}^{n_1,n_2,n_3}= \left(\tfrac{d-2 (J_1+J_2+J_3-1)-\left(\tau_1+\tau_2+\tau_3\right)}{2} \right)_{m_1+m_2+m_3}\\\times\, \prod_{i=1}^3 \left[2^{m_i}\,\binom{n_i}{m_i} (n_i+\delta_{(i+1)(i-1)}-1)_{m_i}\right]\,,
\end{multline}
and $\delta_{(i-1)(i+1)}=\frac12(\tau_{i-1}+\tau_{i+1}-\tau_i)$, $i \cong i+3$. These are known to be built from the following six basic $SO\left(d+1,1\right)$-covariant contractions 
\begin{subequations}\label{6cont}
\begin{align}
\mathcal{Y}_1&=\partial_{U_1}\cdot\pa_{X_2}\,,&\mathcal{Y}_2&=\pa_{U_2}\cdot\pa_{X_3}\,,&\mathcal{Y}_3&=\pa_{U_3}\cdot\pa_{X_1}\,,\nonumber\\
    \mathcal{H}_1&=\pa_{U_2}\cdot\pa_{U_3}\,,&\mathcal{H}_2&=\pa_{U_3}\cdot\pa_{U_1}\,,&\mathcal{H}_3&=\pa_{U_1}\cdot\pa_{U_2}\,.\nonumber
\end{align}
\end{subequations}
The basis \eqref{Ical} is convenient for Witten diagram computations and for our case the vertex simplifies further as we consider $m_i=0$. Thus for the simple coupling of two scalars to a single spin $J$ field, we have the cubic vertex $\mathcal{V}^{(3)} = g \mathcal{Y}_1^s \chi_s \phi \phi$.

\subsection{Thermal CFT recap}
We now provide a quick recap of the thermal CFT programme. With the emergence of temperature as a result of the the Euclidean time being identified as $\t \sim \t+\b$ in $x^\mu\equiv  (\tau, \mathbf{x}) \in S^1_\beta \times \mathbb{R}^{d-1}$, we get an additional scale in our theory thus partially breaking the conformal symmetry. One consequence of this is the widening of the set of observables to include non vanishing thermal one-point functions of operators which are written as
\beqa
\langle\mathcal{O}^{\mu_1\dots \mu_J}(x)\rangle_{\b} &= \frac{b_\cO}{\b^\D}(e^{\mu_1}\cdots e^{\mu_J} -\text{traces}),
\eeqa
and where due to the residual symmetries, only primary even-spin symmetric traceless tensors $\cO^{\mu_1\dots \mu_J}(x)$ can have the above nonzero one-point functions. Here $\D$ denotes the dimension of operator $\cO$ and $b_\cO$ is the thermal one-point function coefficient. The thermal bootstrap program proceeds by placing certain constraints on the thermal two point functions, namely the the KMS condition, which acts as an analogue for the crossing equation. Acting on the two-point function $g(\t,x)$ of identical real scalars $\phi$ at inverse temperature $\b$ defined below
\beqa
g(\t,x)=\langle\phi(x_1)\phi(x_2)\rangle_{\b}
\eeqa
the KMS condition is expressed as the periodicity along the thermal circle thus imposing
\beqa
\label{eq:KMS condition}
\langle\phi(\tau,{\bf x})\phi(0)\rangle_\b = \langle\phi(\tau+\b,{\bf x})\phi(0)\rangle_\b \, .
\eeqa
 where $x_2=0$ has been set by translations, and the coordinates $x^\mu = (\tau,\bf{x})$ are such that $\tau\in S_\b^1$ and ${\bf x} \in \mathbb{R}^{d-1}$ as already described. 
The OPE in this case has a radius of convergence $|x| < \b$ and for the thermal two-point function, it takes the well known form \cite{Iliesiu:2018fao}
\beqa\label{eq:OPE}
\langle\phi(x)\phi(0)\rangle_\b &=& \sum_{\cO\in \phi\times \phi} \frac{a_\cO }{\b^\D} C_J^{(\nu)}\left(\frac{x\cdot e}{|x|}\right) {|x|}^{\Delta-2\Delta_\phi}, \qquad \text{for } \quad |x|< \b,\nonumber \\ 
a_{\mathcal{O}} &=& f_{\phi\phi\mathcal{O}}b_\mathcal{O} \frac{J!}{2^J (\nu)_J}\, .
\eeqa
The sum above runs over operators $\cO$ in the $\phi\times\phi$ OPE, with $\D$ and $J$ the dimension and spin of $\cO$, and $f_{\phi\phi\cO}$ are OPE coefficients. The $C_J^{(\nu)}$ are Gegenbauer polynomials, while we have defined $\nu =\tfrac{d-2}{2}$.
Further, we can write the thermal two-point function as a function of the previously described $z$ and $\bar z$ coordinates as 
\beqa
g(z,\bar z)\equiv \langle \phi(z,\bar z) \phi(0)\rangle_\b \,.
\eeqa
the KMS condition, combined with the symmetry $x \rightarrow -x$ in these coordinates imposes the following constraint
\beqa
\label{eq:KMS in z zbar}
g(z,\bar z) = g(1-z,1-\bar z).
\eeqa
where $\b$ has been set to one for simplicity.
A simple yet important solution that is often  encountered is given by the mean field theory (MFT) two-point function and is given by
\beqa
\langle \phi(x)\phi(0) \rangle_\beta^{\text{MFT}} = \sum_{m=-\infty}^{\infty} \frac{1}{((\tau + m)^2 + \mathbf{x}^2)^\Delta} =  \sum_{m=-\infty}^\infty \frac{1}{(z+m)^\D(\bar z+m)^\D}\, .
\eeqa
 The OPE of the MFT two-point function is known to be particularly simple, containing only the identity ${\bf 1}$ and the double-trace operators $[\phi\phi]_{n,\ell}$ with dimension $\D = 2\D_\phi + 2n+\ell$ and restricted to only even spin $\ell$. The OPE thus takes the form
\beqa
\label{eq:MFT OPE}
g(z,\bar{z})^{\text{MFT}}= \frac{1}{(z \bar z)^{\D_\phi}} + \sum_{n=0}^\infty \sum_{\ell=0,2,\dots} a_{[\phi\phi]_{n,\ell}} C_{\ell}^{(\nu)}(\eta) (z \bar z)^{n+\ell/2} \, ,
\eeqa
where the MFT thermal coefficients are given by~\cite{Iliesiu:2018fao}
\beqa
\label{eq:MFT thermal coefficients}
a_{\bf 1}&=&1 \nonumber\\
 a_{[\phi\phi]_{n,\ell}}&=& 2\zeta(2\D_\phi+2n+\ell)\frac{(\ell+\nu)(\D_\phi)_{\ell+n}(\D_\phi-\nu)_{n}}{n!(\nu)_{\ell+n+1}} \, .
\eeqa
and where $(\alpha)_k = \frac{\Gamma(\alpha+k)}{\Gamma(\alpha)}$ denotes the Pochhammer symbol. 

An alternate expression for the zero relative spatial separation correlator has also been obtained by re-summing the discontinuities \cite{Barrat:2025nvu} across branch cut
\begin{equation}
g(\tau) = \sum_{\Delta} a_{\Delta} \left[ \zeta_H(2\Delta_\phi - \Delta, \tau) + \zeta_H(2\Delta_\phi - \Delta, 1 - \tau) \right] + \kappa
\end{equation}
where the Hurwitz $\zeta$-functions make apparant the KMS invariance and are multiplied by the OPE coefficients $a_{\Delta}$ that encode the dynamical information and $\kappa$ is some constant. This form can also be interpreted as an infinite sum over the fundamental scalars' two-point function in generalized free field (GFF) theory \cite{Marchetto:2023xap}, with coefficients that are given by the thermal OPE data and where
\begin{equation}
g_{\text{GFF}}(\tau, \Delta) = \zeta_H(2\Delta, \tau) + \zeta_H(2\Delta, 1 - \tau) \,. \tag{2.28}
\end{equation}
Recent bootstrap studies have also attempted to reconstruct the correlator from the dispersion-relation and the generalized method of images for non zero spatial separation as \cite{Barrat:2025nvu}
\begin{equation}
g(z, \bar{z}) = \frac{1}{2} \sum_{m \in \mathbb{Z}} g_{\text{dr}}(z - m, \bar{z} - m) + g_{\text{arcs}}(z, \bar{z}).
\end{equation}
where $g_{\text{dr}}$ is computed from the discontinuity of the thermal OPE data and $g_{arcs}$ is the contribution not captured by the discontinuity and is theory dependent.
In this paper our aim was to compute these KMS invariant two-point correlation functions from bulk Witten diagrams. The KMS invariance in our results at non zero spatial separations is made manifest by the more general Epstein Hurwitz zeta functions.

\section{One-loops diagrams}\label{section3}
In this section we compute two classes of two-point one-loop Witten diagrams involving only scalar fields: the tadpole and bench diagrams as depicted in figures (\ref{loopY}) and (\ref{bench}) respectively.  For this we consider a massive real scalar field $\phi$ with a cubic $\phi \phi\chi$ interaction term given by the Euclidean action
\beqa
S_E=\int_{TAdS_{d+1}}d^{d+1}X~\sqrt{g}~\f{1}{2}\left[\phi\mathcal{L}_\phi\phi+\chi\mathcal{L}_\chi\chi+g\phi\phi\chi\right]
\eeqa
where $\mathcal{L}_\a=-\square_E^{2}+m_\a^2$ and $m_\a^2=\Delta_\a(\D_\a-d)$ with $\a\in(\phi,\chi)$.\\

\subsection{The tadpole} \label{opetadpole}

We first illustrate the evaluation of the tadpole diagram depicted in figure (\ref{loopY}). We here lay down the procedure for evaluating the diagram in thermal AdS explaining in detail the procedure involved. Initially we begin with thermally identifying both the ends of each bulk to bulk and bulk to boundary propagators, which is the most general set-up. We can then shift the thermal arrangements of the bulk points to arrange the bulk thermal AdS integrals in the familiar vacuum forms. This procedure is explained in detail below.
\vspace{1cm}
\[\begin{wittendiagram}

\draw[color=black, fill=white,](-0.1,0.5) circle (0.5);
 \draw (-1.15,-1.6) node[vertex] -- (-0.1,-1) node[vertex];
 \draw (1.15,-1.6) node[vertex] -- (-0.1,-1) node[vertex];
 \draw (-0.1,-1) node[vertex] -- (-0.1,0) node[vertex]node[vertex];
  \draw (-0.1,0) node[vertex] -- (-0.1,0) node[vertex]node[vertex];
  \node at (-1.9, -1.7) {$x_1+n_1$};
  \node at (1.9, -1.7) {$x_2+n_2$};
  \node at (0.35, -0.2) {$Y$};
  \node at (0.35, -0.8) {$X$};
    \node at (-1, -1.2) {$\D_{\phi_1}$};
  \node at (1, -1.2) {$\D_{\phi_2}$};
    \node at (-0.5, -0.5) {$\D'$};
  \node at (0, 1.3) {$\tilde{\D}$};
\end{wittendiagram}\hspace{0.3cm}\quad
\bm{\xrightarrow{\hspace{1cm}} }
\quad
\hspace{0.8cm}\begin{wittendiagram}
 \draw (-1.15,-1.6) node[vertex] -- (-0.1,-1) node[vertex]node[vertex];
 \draw (1.15,-1.6) node[vertex] -- (-0.1,-1) node[vertex]node[vertex];
   \draw[dashed] (-0.1,0) node[vertex] -- (-1.9,-0.5) node[vertex];
    \draw[dashed] (-0.1,-1) node[vertex] -- (-1.9,-0.5) node[vertex, color=blue];
      \draw[dashed] (-0.1,0) node[vertex] -- (-1.7,1) node[vertex, color=blue];
       \draw[dashed] (-0.1,0) node[vertex] -- (1.7,1) node[vertex];
  \node at (-1.9, -1.7) {$x_1+n_1$};
  \node at (1.9, -1.7) {$x_2+n_2$};
  \node at (0.4, -0.2) {$Y$};
  \node at (0.4, -0.8) {$X$};
  \node at (-2.4,-0.4) {$x'$};
   \node at (-2.4,1.1) {$\tilde{x}$};
    \node at (2.9,1.1) {$\tilde{x}-n_3\gamma$};
        \node at (-1.1, -1) {{\tiny $\f{d}{2}-i\nu'$}};
    \node at (-1.4, 0) {{\tiny $\f{d}{2}+i\nu'$}};
        \node at (0.9, 1.1) {{\tiny $\f{d}{2}-i\tilde{\nu}$}};
    \node at (-0.9, 1.1) {{\tiny $\f{d}{2}+i\tilde{\nu}$}};
\end{wittendiagram}\] 

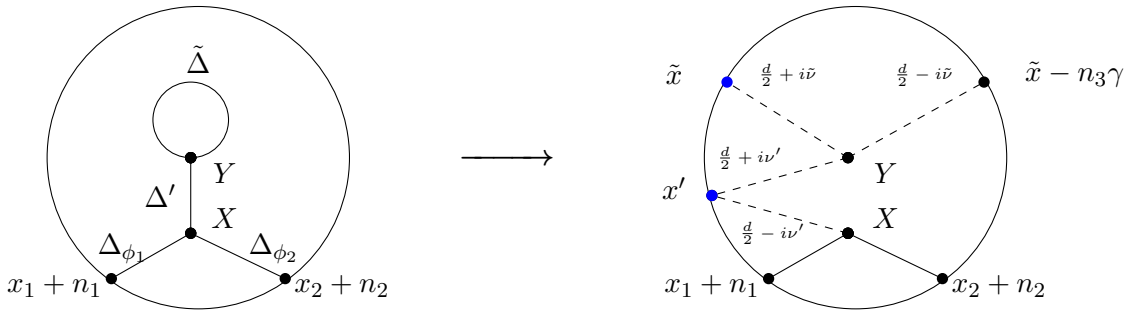
\captionof{figure}{The tadpole diagram in thermal AdS with details of the split representation and the resultant thermal rearrangement of the thermal points. Thermally identified points are shown in black while non identified points are shown in blue.}\label{loopY}

\vspace{0.5cm}

The diagram becomes
\beqa
\mathcal{W}&=&\f{1}{\mathcal{N}^4}\sum_{\substack{a,b,c,d,e,f \neq 0\\n_1,n_2}}\f{g^2}{2}\int_{TAdS}dX dY~K_{\Delta_{\phi_1}}(X^{a\gamma},x_1+n_1\gamma)K_{\Delta_{\phi_2}}(X^{b\gamma},x_2+n_2\gamma)G_{\Delta'}(X^{c\gamma},Y^{d\gamma})G_{\tilde{\Delta}}(Y^{e\gamma},Y^{f\gamma})\nonumber\\
&=&\f{1}{\mathcal{N}^4}\sum_{\substack{a,b,c,d,e,f \neq 0\\n_1,n_2}}\f{g^2}{2}\int_{TAdS}dX dY~K_{\Delta_{\phi_1}}(X^{a\gamma},x_1+n_1\gamma)K_{\Delta_{\phi_2}}(X^{a\gamma},x_2+\underbrace{(n_2+b-a)}_{n_2'=n_2}\gamma)\nonumber\\
&\times&G_{\Delta'}(X^{a\gamma},\underbrace{Y^{(d-c+a)\gamma}}_{Y^{d'\gamma=d\gamma}})G_{\tilde{\Delta}}(Y^{d\gamma},\underbrace{Y^{(f-d+e)\gamma}}_{Y^{f'\gamma=f\gamma}})\nonumber\\
&=&\f{1}{\mathcal{N}}\sum_{\substack{a,d \neq 0\\n_1,n_2,n_3}}\f{g^2}{2}\int_{-\infty}^{\infty}\f{2\nu'^2\sqrt{\mathcal{C}_{\f{d}{2}+i\nu}C_{\f{d}{2}-i\nu'}}~d\nu'}{2\pi\left[ \nu'^2+(\Delta'-d/2)^2\right]}\int_{-\infty}^{\infty}\f{2\tilde{\nu}^2\sqrt{\mathcal{C}_{\f{d}{2}+i\tilde{\nu}}C_{\f{d}{2}-i\tilde{\nu}'}}~d\tilde{\nu}}{2\pi\left[ \tilde{\nu}^2+(\tilde{\Delta}-d/2)^2\right]}\nonumber\\
&\times& \int_{TAdS}dX dY\int d x' \ d \tilde{x}K_{\Delta_{\phi_1}}(X^{a\gamma},x_1+n_1\gamma)K_{\Delta_{\phi_2}}(X^{a\gamma},x_2+n_2\gamma)K_{\f{d}{2}+i\nu'}(X^{a\gamma},x')\nonumber\\
&\times&K_{\f{d}{2}-i\nu'}(Y^{d\gamma},x')K_{\f{d}{2}+i\tilde{\nu}}(Y^{d\gamma},\tilde{x})K_{\f{d}{2}-i\tilde{\nu}}(Y^{d\gamma},\underbrace{\tilde{x}+(-d+f)\gamma}_{\tilde{x}-n_3\gamma})
\eeqa
\\

Below we describe the details on the rearrangement that allows us to perform the bulk integrals:
\begin{itemize}
    \item In the first line we divide by $\mathcal{N}^4$, where $\mathcal{N}=|\mathbb{Z}|$, to compensate for the overcounting introduced by mirror-imaging both the initial point and the end points of the propagators.
    \item We next adjust the bulk points in order to be able to do the bulk integrals in thermal AdS using the usual vacuum integral \ref{bulkint}. For this, in the second line we have redefined $n_2$,$d$ and $f$ as $n_2+b-a=n_2'$ (relabeled as $n_2$), $d-c+a=c'$ (relabeled as $c$) and $f-d+e=f'$ (relabeled as $f$). This kills three factors of $\mathcal{N}$ from the denominator. Essentially, each value of $b$, $c$ and $e$ gives the same series in $n_2$, $d$ and $f$ and the sums over $b$, $c$ and $e$ being divergent cancel a factor of $\mathcal{N}$ each in the denominator.
    \item We have also introduced the split representation in the last line and redefined $f-d=n_3$.
\end{itemize}

Now the integral over the bulk points in thermal AdS can be performed as

\beqa
\int_{TAdS}d\ X=\int_0^{\infty}\f{d z}{(z)^{d+1}}\int d^{d-1}\mathbf{x}\int_0^\beta d \tau
\eeqa
realizing that
\beqa
\int_0^\infty d \tau=\sum_{m=-\infty}^{\infty}\int_0^\beta d \tau=\mathcal{N}\beta
\eeqa
we can write the integral over the periodically identified coordinate as $\f{1}{\mathcal{N}}\int_0^{\infty}d \tau$. 
This allows us to write
\beqa
\int_{TAdS}d\ X=\f{1}{\mathcal{N}}\int_{AdS}d\ X.
\eeqa

We shall see that the fictitious factors of $\mathcal{N}$ will cancel in our final answers. We also note that the thermally identified points are just translated points on the boundary  $\t \sim \t+n$.
We can now perform the integrals over $X$ and $Y$ as in the usual way in eq (\ref{bulkint}). These give after defining: $P_{ij}=(x_i-x_j)^2=x_{ij}^2$ (where from now on we drop the $\mu$ superscript from the boundary points $x^\mu$)

\beqa
&&\f{1}{\mathcal{N}}\int d X ~K_{\Delta_{\phi_1}}(X,x_1+n_1\g) K_{\Delta_{\phi_2}}(X,x_2+n_2\g)K_{\f{d}{2}+i\nu'}(X,x')\\
&=&\f{1}{\mathcal{N}}\f{B_{\Delta_{\phi_1},\Delta_{\phi_2},\f{d}{2}+i\nu}}{((x_1-x_2+n_1\g-n_2\g)^2)^{\f{\Delta_{\phi_1}+\Delta_{\phi_2}-\f{d}{2}-i\nu'}{2}}((x_2-x'+n_2\g)^2)^{\f{-\Delta_{\phi_1}+\Delta_{\phi_2}+\f{d}{2}+i\nu'}{2}}((x'-x_1-n_1\g)^2)^{\f{\Delta_{\phi_1}-\Delta_{\phi_2}+\f{d}{2}+i\nu'}{2}}}\nonumber
\eeqa

Similarly the integral over $Y$ can be performed as

\beqa
&&\f{1}{\mathcal{N}}\int d Y K_{\f{d}{2}-i\nu'}(Y,x') K_{\f{d}{2}+i\tilde{\nu}}(Y,\tilde{x})K_{\f{d}{2}-i\tilde{\nu}}(Y,\tilde{x}-n_3\g)\\
&=&\f{1}{\mathcal{N}}\f{B_{\f{d}{2}-i\nu',\f{d}{2}+i\tilde{\nu},\f{d}{2}-i\tilde{\nu}}}{((x'-\tilde{x})^2)^{\f{2i\tilde{\nu}+\f{d}{2}-i\nu'}{2}}((n_3\g)^2)^{\f{\f{d}{2}+i\nu'}{2}}((\tilde{x}-x'-n_3\g)^2)^{\f{-2i\tilde{\nu}+\f{d}{2}-i\nu'}{2}}}\nonumber
\eeqa

Next we need to do the integral over the boundary points which go as

\beqa
&&\int  \f{dx'}{((x_2-x'+n_2)^2)^{\f{-\Delta_{\phi_1}+\Delta_{\phi_2}+\f{d}{2}+i\nu'}{2}}((x'-x_1-n_1)^2)^{\f{\Delta_{\phi_1}-\Delta_{\phi_2}+\f{d}{2}+i\nu'}{2}}}\nonumber\\
&\times&\f{d\tilde{x}}{((x'-\tilde{x})^2)^{\f{\f{d}{2}-i\nu'+2i\tilde{x}}{2}}((\tilde{x}-x'-n_3)^2)^{\f{\f{d}{2}-i\nu'-2i\tilde{x}}{2}}}
\eeqa

We first do the $\tilde{x}$ integral and then the $x'$ integral. These are two point boundary integrals that can be performed using eq (\ref{2ptbint}). The $\tilde{x}$ integral first gives

\beqa
 \int \f{d\tilde{x}}{((x'-\tilde{\nu})^2)^{\f{\f{d}{2}-i\nu'+2i\tilde{x}}{2}}((\tilde{\nu}-x'-n_3\g)^2)^{\f{\f{d}{2}-i\nu'-2i\tilde{x}}{2}}}=\f{\pi^{d/2}\G\left( \f{\f{d}{2}+i\nu'-2i\tilde{\nu}}{2} \right)\G\left( \f{\f{d}{2}+i\nu'+2i\tilde{\nu}}{2} \right)\G(-i\nu')}{\G\left(\f{\f{d}{2}-i\nu'\pm 2i\tilde{\nu}}{2} \right)\G\left(\f{d+2i\nu'}{2}\right)((n_3\g)^2)^{-i\nu'}}
\eeqa
Then the $x'$ integral gives
\beqa
&&\int d x' \f{1}{((x_+n_1\g-x')^2)^{\f{\D_{\phi_1}-\D_{\phi_2}+\f{d}{2}-i\nu'}{2}}((x_2+n_2\g-x')^2)^{\f{\D_{\phi_2}-\D_{\phi_1}+\f{d}{2}+i\nu'}{2}}}\nonumber\\
&=&\f{\pi^{d/2}\G\left( \f{\D_{\phi_2}-\D_{\phi_1}+\f{d}{2}-i\nu'}{2} \right)\G\left( \f{\D_{\phi_1}-\D_{\phi_2}+\f{d}{2}-i\nu'}{2} \right)\G(i\nu')}{\G\left(\f{\f{d}{2}+i\nu'+\D_{\phi_2}-\D_{\phi_1}}{2} \right)\G\left(\f{\f{d}{2}+i\nu'-\D_{\phi_2}+\D_{\phi_1}}{2} \right)\G\left(\f{d-2i\nu'}{2}\right)}\f{1}{((x_2-x_1+n_2\g-n_1\g)^2)^{i\nu'}}
\eeqa

At this stage we observe that the sums over $a,d$ cancel the factors of $\mathcal{N}^2$ in the denominator. Clubbing all remaining terms together we have

\beqa
\mathcal{W}&=&\f{g^2}{64\pi^{\f{d+2}{2}}}\f{1}{\mathcal{N}}\int_{-\infty}^{\infty}\f{d\nu'}{\left[ \nu'^2+(\Delta'-\f{d}{2})^2\right]}\int_{-\infty}^{\infty}\f{d\tilde{\nu}}{\left[ \tilde{\nu}^2+(\tilde{\Delta}-\f{d}{2})^2\right]}\Gamma\left(\f{\f{d}{2}\pm i\nu'}{2}\right)\Gamma\left(\f{\f{d}{2}\pm 2i\tilde{\nu}+i\nu'}{2}\right)\nonumber\\
&\times&\f{ \Gamma\left(\f{\Delta_{\phi_1}+\Delta_{\phi_2}-\f{d}{2}+i\nu'}{2}\right)\Gamma\left(\f{\Delta_{\phi_1}+\Delta_{\phi_2}-\f{d}{2}-i\nu'}{2}\right)\Gamma\left(\f{-\Delta_{\phi_1}+\Delta_{\phi_2}+\f{d}{2}-i\nu'}{2}\right)\Gamma\left(\f{\Delta_{\phi_1}-\Delta_{\phi_2}+\f{d}{2}-i\nu'}{2}\right)}{\G(\pm i\tilde{\nu})\G(\f{d\pm2i\nu'}{2})\G(\D_{\phi_1})\G(\D_{\phi_2})\G(\D_{\phi_1}-\f{d}{2}+1)\G(\D_{\phi_2}-\f{d}{2}+1)}\nonumber\\
&\times& \sum_{n_1,n_2,n_3}\f{1}{((x_2-x_1+n_2\g-n_1\g)^2)^{\f{\D_{\phi_1}+\D_{\phi_2}-d/2+i\nu'}{2}}((n_3\g)^2)^{\f{d/2-i\nu'}{2}}}
\eeqa
Further, the sums over $n_1,n_2$ can be simplified as follows. Each term in $n_2$ gives the same series in $n_1$ and therefore the $n_2$ sum when performed cancels the factor of $\mathcal{N}$. For $\mathbf{x_1}=\mathbf{x_2}$ or the zero spatially separated points we can write the image sum in terms of Hurwitz zeta function as
\beqa
\sum_{n_1}\f{1}{(x_2-x_1-n_1\g)^{\D_{\phi_1}+\D_{\phi_2}-d/2+i\nu'}}&=&\zeta_{H}\left(\D_{\phi_1}+\D_{\phi_2}-d/2+i\nu',\t_2-\t_1\right)\nonumber\\
&+&\zeta_{H}\left(\D_{\phi_1}+\D_{\phi_2}-d/2+i\nu',1-(\t_2-\t_1)\right)
\eeqa
with a structure which respects the KMS condition and as also expected from the thermal bootstrap studies. While for a non zero spatial separation the image sum is written in terms of the Epstein - Hurwitz zeta function 
\beqa
\sum_{n_1}\f{1}{(x_2-x_1-n_1\g)^{\D_{\phi_1}+\D_{\phi_1}-d/2+i\nu'}}&=&\zeta_{\text{EH}}\left(\D_{\phi_1}+\D_{\phi_1}-d/2+i\nu',\t_{12},\mathbf{x}_{12}\right)
\eeqa
with its analytically continued form beyond Re$(s)>1/2$ with $s=\D_{\phi_1}+\D_{\phi_2}-d/2+i\nu'$
\beqa
   \xi_{\text{EH}}(s,\tau,\mathbf{x}) = \frac{\sqrt{\pi} \, \Gamma\!\left(\frac{s-1}{2}\right)}{\Gamma\!\left(\frac{s}{2}\right)} \, x^{1-s} + \frac{4\sqrt{\pi}}{\Gamma\!\left(\frac{s}{2}\right)} \sum_{k \ge 1} \left( \frac{\pi k}{x} \right)^{\frac{s-1}{2}} K_{\frac{s-1}{2}}(2\pi k x) \cos (2\pi k \tau)
\eeqa
where $K_{\frac{s-1}{2}}(2\pi k x)$ is the modified Bessel function of the second kind. Writing 
\beqa
\left((\tau_{12} + n_1)^2 + \mathbf{x}_{12}^2\right)^{-s/2} = |\tau_{12} + n_1|^{-s} \left(1 + \f{\mathbf{x}_{12}^2}{(\tau_{12} + n)^2}\right)^{-s/2}
\eeqa
the above we can also re-express the result in a KMS invariant form as
\beqa
    \xi_{EH}(s,\tau_{12},\mathbf{x}_{12}) = \sum_{j \ge 0} \binom{-s/2}{j} (\mathbf{x}_{12}^2)^{j} \Big[ \zeta_H(s+2j, \tau_{12}) + \zeta_H(s+2j, 1-\tau_{12}) \Big].
\eeqa
Finally the sum over $n_3$ can be written as

\beqa
\sum_{m=-\infty}^{\infty}\f{1}{((n_3)^2)^{\f{d/2-i\nu'}{2}}}=\left[1+(-1)^{-\f{d}{2}-i\nu'} \right]\zeta\left(\f{d}{2}-i\nu'\right).
\eeqa

The final expression thus is

\beqa
\mathcal{W}&=&\f{g^2}{64\pi^{\f{d+2}{2}}}\f{1}{\mathcal{N}}\int_{-\infty}^{\infty}\f{d\nu'}{\left[ \nu'^2+(\Delta'-\f{d}{2})^2\right]}\int_{-\infty}^{\infty}\f{d\tilde{\nu}}{\left[ \tilde{\nu}^2+(\tilde{\Delta}-\f{d}{2})^2\right]}\Gamma\left(\f{\Delta_{\phi_1}+\Delta_{\phi_2}-\f{d}{2}+i\nu'}{2}\right)\Gamma\left(\f{\Delta_{\phi_1}+\Delta_{\phi_2}-\f{d}{2}-i\nu'}{2}\right)\nonumber\\
&\times& \f{\Gamma\left(\f{\f{d}{2}+i\nu'}{2}\right)\Gamma\left(\f{\f{d}{2}-i\nu'}{2}\right) \Gamma\left(\f{\f{d}{2}-2i\tilde{\nu}+i\nu'}{2}\right)\Gamma\left(\f{\f{d}{2}+2i\tilde{\nu}+i\nu'}{2}\right)\Gamma\left(\f{-\Delta_{\phi_1}+\Delta_{\phi_2}+\f{d}{2}-i\nu'}{2}\right)\Gamma\left(\f{\Delta_{\phi_1}-\Delta_{\phi_2}+\f{d}{2}-i\nu'}{2}\right)}{\G(\D_{\phi_1})\G(\D_{\phi_2})\G(\D_{\phi_1}-\f{d}{2}+1)\G(\D_{\phi_2}-\f{d}{2}+1)\G(\pm i\tilde{\nu})\G(\f{d\pm2i\nu'}{2})}\nonumber\\
&\times&   \zeta_{\text{EH}}\left(\D_{\phi_1}+\D_{\phi_2}-\f{d}{2}+i\nu',\tau_{12},\mathbf{x}_{12}\right)  \zeta\left(\f{d}{2}-i\nu'\right)
\eeqa
We thus obtain a manifestly KMS invariant two point correlator.\\

We now need to evaluate the spectral integrals to get the full correlator
\beqa
\int_{-\infty}^{\infty}d\nu' \int_{-\infty}^{\infty} d\tilde{\nu} ~ F(\nu',\tilde{\nu})~ \zeta_{\text{EH}}\left(\D_{\phi_1}+\D_{\phi_1}-\f{d}{2}+i\nu',\tau_{12},\mathbf{x}_{12}\right)\zeta\left(\f{d}{2}-i\nu'\right)\nonumber\\
\eeqa
where
\beqa
F(\nu',\tilde{\nu})&=&\f{1}{\left[ \nu'^2+(\Delta'-d/2)^2\right]\left[ \tilde{\nu}^2+(\tilde{\Delta}-d/2)^2\right]}\Gamma\left(\f{\Delta_{\phi_1}+\Delta_{\phi_2}-\f{d}{2}+i\nu'}{2}\right)\Gamma\left(\f{\Delta_{\phi_1}+\Delta_{\phi_2}-\f{d}{2}-i\nu'}{2}\right)\nonumber\\
&\times&\Gamma\left(\f{\f{d}{2}-2i\tilde{\nu}+i\nu'}{2}\right)\Gamma\left(\f{\f{d}{2}+2i\tilde{\nu}+i\nu'}{2}\right)\Gamma\left(\f{-\Delta_{\phi_1}+\Delta_{\phi_2}+\f{d}{2}-i\nu'}{2}\right)\Gamma\left(\f{\Delta_{\phi_1}-\Delta_{\phi_2}+\f{d}{2}-i\nu'}{2}\right) \nonumber\\
&\times& \f{\Gamma\left(\f{\f{d}{2}+i\nu'}{2}\right)\Gamma\left(\f{\f{d}{2}-i\nu'}{2}\right)}{\G(\pm i\tilde{\nu})\G(\f{d\pm2i\nu'}{2})}
\eeqa
The spectral integrals here need to be performed. We notice the following pole structure for this in the UHP:\\

\begin{enumerate}
    \item  \textbf{Double trace poles} from $\Gamma\left(\frac{\D_{\phi_1}+\D_{\phi_2}-\f{d}{2}+i\nu'}{2}\right)$

       These poles occur when $\frac{\D_{\phi_1}+\D_{\phi_2}-\f{d}{2}+i\nu'}{2}=-2n \rightarrow \nu'=\D_{\phi_1}+\D_{\phi_2}+2n$\\

\item \textbf{Physical poles} from $\nu'=\D'-d/2$ of the physical exchange\\
       \item \textbf{Physical poles} from $\tilde{\nu}=\tilde{\D}-d/2$ of the loop, these further affect the next set of poles in point (4)\\
    
    \item Poles from $\Gamma\left(\f{i\nu+2\tilde{\D}}{2}\right)$
    
   These poles occur when $\f{i\nu+2\tilde{\D}}{2}=-2m \rightarrow \nu'=2\tilde{\D}+2m$\\

    \item There is a cancellation of poles between $\Gamma\left(\f{\f{d}{2}\pm i\nu'}{2}\right)$ and $\G(\f{d}{2}\pm i\nu')$.

\end{enumerate} 

Double poles occur from the collision of poles in (1) and (4) when $\D_{\phi_1}+\D_{\phi_2}+2n=2\tilde{\D}+2m$. In order to evaluate the anomalous dimensions we instead evaluate the Witten diagram with an effective bulk vertex coming from the bulk loop described in section \ref{section6} and as evaluated for the exchange diagram in section \ref{effY}. We further show in appendix \ref{effverapp} that the computation for the tadpole diagram in this section, with the integrals performed in a slightly different order, indeed reduces to the one with effective vertices described in section (\ref{section6}). Below in figure (\ref{figure2}) we schematically show how the boundary integrals simplify and give rise to the factor of $\sum_m (n_3^2)^{\f{d/2-i\nu'}{2}}$.

\vspace{1cm}

        \begin{tikzpicture}[ arrowstyle/.style={->, >=stealth, line width=1pt},scale=0.9]

  \draw[dashed] (-1.1,-1) node[vertex] -- (0.55,0.7) node[vertex]node[vertex];
     \draw[dashed] (-1.1,-1) node[vertex] -- (-2.85,0.7) node[vertex]node[vertex];
    \draw[dashed] (0.55,0.7) node[vertex] -- (-2.85,0.7) node[vertex]node[vertex];
   \draw[dashed] (-1.1,-1) node[vertex] -- (0.55,-2.7) node[vertex]node[vertex];
     \draw[dashed] (-1.1,-1) node[vertex] -- (-2.85,-2.7) node[vertex]node[vertex];
       \draw[dashed] (0.55,-2.7) node[vertex] -- (-2.85,-2.7) node[vertex]node[vertex];
         
  \node at (-2.4, -2.95) {$x_1+n_1$};
  \node at (0.4, -2.95) {$x_2+n_2$};
  \node at (-0.4,-1) {$x'$};
   \node at (-2.7,0.95) {$\tilde{x}$};
    \node at (0.7,0.95) {$\tilde{x}+n_3$};

  \hspace{5cm} \draw[arrowstyle] (-4, -1) -- (-3, -1);
   \draw[dashed] (-1.1,-1) node[vertex] -- (-0.05,0.7) node[vertex]node[vertex];
     \draw[dashed] (1.1,-1) node[vertex] -- (-0.05,0.7) node[vertex]node[vertex];
  \draw[dashed] (1.1,-1) node[vertex] -- (-1.1,-1) node[vertex]node[vertex];
   \draw[dashed] (-1.1,-1) node[vertex] -- (2.2,-2.7) node[vertex]node[vertex];
    \draw[dashed] (1.1,-1) node[vertex] -- (2.2,-2.7) node[vertex]node[vertex];
    \draw[dashed] (1.1,-1) node[vertex] -- (-2.05,-2.7) node[vertex]node[vertex];
    \draw[dashed] (-1.1,-1) node[vertex] -- (-2.05,-2.7) node[vertex]node[vertex];
      \draw[dashed] (2.2,-2.7) node[vertex] -- (-2.05,-2.7) node[vertex]node[vertex];
         
  \node at (-2.4, -2.95) {$x_1+n_1$};
  \node at (2.6, -2.95) {$x_2+n_2$};
  \node at (-1.5,-1) {$x'$};
  \node at (2.0,-1) {$x'+n_3$};
    \node at (0.7,0.9) {$\tilde{x}$};

  \hspace{3cm}  \draw[arrowstyle] (1, -1) -- (2, -1);

   \draw[dashed] (4.1,-1) node[vertex] -- (6.2,-2.7) node[vertex]node[vertex];
    \draw[dashed] (4.1,-1) node[vertex] -- (2.2,-2.7) node[vertex]node[vertex];
      \draw[dashed] (2.2,-2.7) node[vertex] -- (6.2,-2.7) node[vertex]node[vertex];
         
  \node at (2, -2.95) {$x_1+n_1$};
  \node at (6.5, -2.95) {$x_2+n_2$};
  \node at (3.2,-1) {$x'$};
     
    \end{tikzpicture}


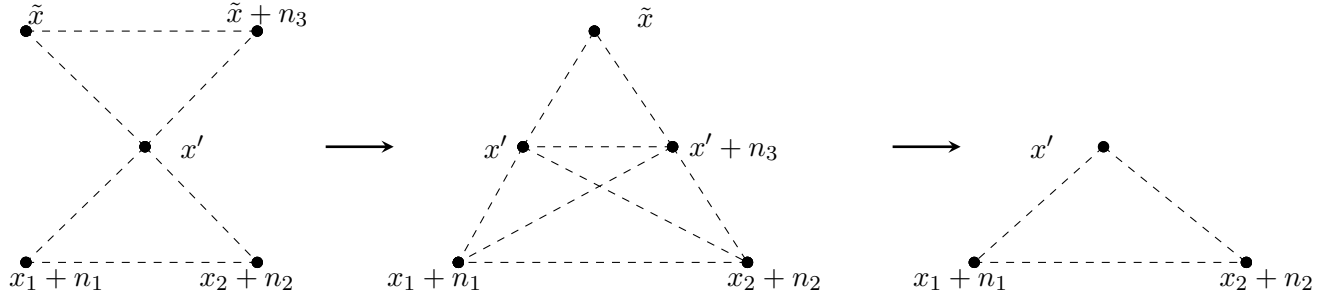
\captionof{figure}{Schematic explanation for the evaluation of the boundary integrals in the tadpole diagram.}\label{figure2}

\subsection{The Bench}
We now study the bench two-point function in thermal AdS$_{d+1}$. We do this for the zero and non-zero relative thermal separations between the two bulk points and further split the later analysis for the $d=1$ and $d>1$ cases.

\vspace{0.5cm}

\[\begin{wittendiagram}
\begin{scope}
    \clip (-0.9,-0.3) rectangle (0.9,1);
    \draw (0,-0.2) circle(0.9);
    \draw (-0.9,-0.3) -- (0.9,-0.3);
\end{scope}
 
 \draw (-1.4,-1.4) node[vertex] -- (-0.85,-0.3) node[vertex]node[vertex];
 \draw (1.4,-1.4) node[vertex] -- (0.85,-0.3) node[vertex]node[vertex];

  \node at (-2.3, -1.5) {$x_1+n_1$};
  \node at (2.3, -1.5) {$x_2+n_2$};
  \node at (1.3, -0.1) {$Y$};
  \node at (-1.3, -0.1) {$X$};
  \node at (-1.5, -0.7) {$\D_{\phi_1}$};
  \node at (1.5, -0.7) {$\D_{\phi_2}$};
  \node at (0, -0.6) {$\D'$};
  \node at (0, 1.1) {$\tilde{\D}$};
\end{wittendiagram}\hspace{0.3cm}\quad
\bm{\xrightarrow{\hspace{1cm}} }
\quad
\hspace{0.5cm}
\begin{wittendiagram}

 \draw (-1.65,-1.1) node[vertex] -- (-0.85,-0.3) node[vertex]node[vertex];
 \draw (1.65,-1.1) node[vertex] -- (0.85,-0.3) node[vertex]node[vertex];

  \draw[dashed] (-0.75,-0.4)  -- (0,-1.97) node[vertex]node[vertex];
   
  \draw[dashed] (0.75,-0.5) -- (0,-1.97) node[vertex, color=blue];
   
 \draw[dashed] (0.85,-0.3)  -- (1.4,1.4) node[vertex];   
 
  \draw[dashed] (-0.85,-0.3)  -- (-1.4,1.4) node[vertex, color=blue];

  \node at (-2.3, -1.5) {$x_1+n_1$};
  \node at (2.3, -1.5) {$x_2+n_2$};
  \node at (1.3, -0.2) {$Y$};
  \node at (-1.3, -0.2) {$X$};
  \node at (0,-2.4) {$x'$};
   \node at (-2,1.4) {$\tilde{x}$};
    \node at (2.6,1.4) {$\tilde{x}+n_3\g$};
       \node at (-1, -1.2) {{\tiny $\f{d}{2}-i\nu'$}};
    \node at (1, -1.2) {{\tiny $\f{d}{2}+i\nu'$}};
        \node at (0.8, 1) {{\tiny $\f{d}{2}-i\tilde{\nu}$}};
    \node at (-0.8, 1) {{\tiny $\f{d}{2}+i\tilde{\nu}$}};
\end{wittendiagram}\]
\captionof{figure}{The thermal bench diagram showing the thermal arrangement of the points and the details of the split representation. The thermally identified points are given in black, while non identified points are in shown in blue.}\label{bench}

\vspace{1cm}
The two point bench diagram is evaluated as
\beqa
\mathcal{W}&=&\f{1}{\mathcal{N}^4}\sum_{\substack{a,b,c,d,e,f \neq 0\\n_1,n_2}}g^2\int_{TAdS}~g^2~dX dY~K_{\Delta_{\phi_1}}(X^{a\gamma},x_1+n_1\gamma)G_{\Delta'}(X^{c\gamma}Y^{d\gamma})G_{\tilde{\Delta}}(X^{e\gamma},Y^{f\gamma})\nonumber\\
&\times &K_{\Delta_{\phi_2}}(Y^{b\gamma},x_2+n_2\gamma)\nonumber\\
&=&\f{1}{\mathcal{N}^2}\sum_{\substack{a,b,d,f \neq 0\\n_1,n_2}}g^2\int_{TAdS}~g^2~dX dY~ K_{\Delta_{\phi_1}}(X^{a\gamma},x_1+n_1\gamma)K_{\Delta_{\phi_2}}(Y^{b\gamma},x_2+n_2\gamma)G_{\Delta'}(X^{a\gamma},\underbrace{Y^{(d+a-c)\gamma}}_{Y^{(d'=d)\gamma}})\nonumber\\
&\times&G_{\tilde{\Delta}}(X^{a\gamma},\underbrace{Y^{(f+a+e)\gamma}}_{Y^{(f'=f)\gamma}})\nonumber\\
&=&\f{1}{\mathcal{N}}\sum_{\substack{a,d \neq 0\\n_1,n_2,n_3}}\int_{-\infty}^{\infty}~g^2~\f{2\nu'^2\sqrt{\mathcal{C}_{\f{d}{2}+i\nu'}\mathcal{C}_{\f{d}{2}-i\nu'}}~d\nu'}{2\pi\left[ \nu'^2+\left(\Delta'-\f{d}{2}\right)^2\right]}\int_{-\infty}^{\infty}\f{2\tilde{\nu}^2\sqrt{\mathcal{C}_{\f{d}{2}+i\tilde{\nu}}\mathcal{C}_{\f{d}{2}-i\tilde{\nu}'}}~d\tilde{\nu}}{2\pi\left[ \tilde{\nu}^2+\left(\tilde{\Delta}-\f{d}{2}\right)^2\right]}\nonumber\\
&\times &\int_{TAdS}dX dY \int d x' d \tilde{x}~K_{\Delta_{\phi_1}}(X^{a\gamma},x_1+n_1\gamma)K_{\Delta_{\phi_2}}(Y^{d\gamma},x_2+n_2\gamma)K_{\f{d}{2}+i\nu'}(X^{a\gamma},x')\nonumber\\
&\times&K_{\f{d}{2}-i\nu'}(Y^{d\gamma},x')K_{\f{d}{2}-i\tilde{\nu}}(X^{a\gamma},\tilde{x})K_{\f{d}{2}+i\tilde{\nu}}(Y^{d\gamma},\underbrace{\tilde{x}+(f-d)\g}_{\tilde{x}+n_3\gamma})
\eeqa
Where to write the above we follow a similar set of steps as for the tadpole diagram to arrange the bulk integrals and thus the diagram. We can again perform the two bulk integrals as

\beqa
&&\f{1}{\mathcal{N}}\int dX ~K_{\Delta_{\phi_1}}(X,x_1+n_1\gamma)
K_{\f{d}{2}+i\nu'}(X,x')K_{\f{d}{2}-i\tilde{\nu}}(X,\tilde{x})\nonumber\\
&=&\f{B_{\D_{\phi_1},\f{d}{2}+i\nu',\f{d}{2}-i\tilde{\nu}}}{((x_1+n_1\g-x')^2)^{\f{\D_{\phi_1}+i\nu'+i\tilde{\nu}}{2}}((x'-\tilde{x})^2)^{\f{d-\D_{\phi_1}+i\nu'-i\tilde{\nu}}{2}}((x_1+n_1\g-\tilde{x})^2)^{\f{\D_{\phi_1}-i\nu'-i\tilde{\nu}}{2}}}
\eeqa

and

\beqa
&&\f{1}{\mathcal{N}}\int dY K_{\Delta_{\phi_2}}(Y,x_2+n_2\gamma)K_{\f{d}{2}-i\nu'}(Y,x')K_{\f{d}{2}+i\tilde{\nu}}(Y,\tilde{x}_3+n_3\g)
\\
&=&\f{B_{\D_{\phi_2},\f{d}{2}-i\nu',\f{d}{2}+i\tilde{\nu}}}{((x_2+n_2\g-x')^2)^{\f{\D_{\phi_2}-i\nu'-i\tilde{\nu}}{2}}((x'-\tilde{x}-n_3\g)^2)^{\f{d-\D_{\phi_2}-i\nu'+i\tilde{\nu}}{2}}((x_2+n_2\g-\tilde{x}-n_3\g)^2)^{\f{\D_{\phi_2}+i\nu'+i\tilde{\nu}}{2}}}\nonumber
\eeqa

We now need to do the boundary integrals.

\beqa
&&\int d x' ~ d\tilde{x} ~ \f{1}{((x_1+n_1\g-x')^2)^{\f{\D_{\phi_1}+i\nu'+i\tilde{\nu}}{2}}((x'-\tilde{x})^2)^{\f{d-\D_{\phi_1}+i\nu'-i\tilde{\nu}}{2}}((x_1+n_1\g-\tilde{x})^2)^{\f{\D_{\phi_1}-i\nu'-i\tilde{\nu}}{2}}}\\
&\times& \f{1}{((x_2+n_2\g-x')^2)^{\f{\D_{\phi_2}-i\nu'-i\tilde{\nu}}{2}}((x'-\tilde{x}-n_3\g)^2)^{\f{d-\D_{\phi_2}-i\nu'+i\tilde{\nu}}{2}}((x_2+n_2\g-\tilde{x}-n_3\g)^2)^{\f{\D_{\phi_2}+i\nu'+i\tilde{\nu}}{2}}}\nonumber
\eeqa\\

We note that each of these is a four point boundary integral. Here $n_3$ represents the relative thermal winding between the two bulk points. We now proceed to evaluate the above boundary integral in two different cases as explained below.

\noindent
\subsubsection{Case I : zero relative thermal shift between the two bulk points}
This happens when $c-d=e-f$ or $n_3=0$. The boundary integrals can then simplify and can be evaluated as follows.
\beqa
&&\int d x' ~ d\tilde{x} ~ \f{1}{((x_1+n_1\g-x')^2)^{\f{\D_{\phi_1}+i\nu'+i\tilde{\nu}}{2}}((x'-\tilde{x})^2)^{\f{d-\D_{\phi_1}+i\nu'-i\tilde{\nu}}{2}}((x_1+n_1\g-\tilde{x})^2)^{\f{\D_{\phi_1}-i\nu'-i\tilde{\nu}}{2}}}\\
&\times& \f{1}{((x_2+n_2\g-x')^2)^{\f{\D_{\phi_2}-i\nu'-i\tilde{\nu}}{2}}((x'-\tilde{x})^2)^{\f{d-\D_{\phi_2}-i\nu'+i\tilde{\nu}}{2}}((x_2+n_2\g-\tilde{x})^2)^{\f{\D_{\phi_2}+i\nu'+i\tilde{\nu}}{2}}}\nonumber
\eeqa\\
We first do the $x'$ integral
\beqa
&&\int ~ \f{d x'}{((x_1+n_1\g-x')^2)^{\f{\D_{\phi_1}+i\nu'+i\tilde{\nu}}{2}}((x'-\tilde{x})^2)^{\f{2d-\D_{\phi_1}-\D_{\phi_2}}{2}}((x_2+n_2\g-x')^2)^{\f{\D_{\phi_2}-i\nu'-i\tilde{\nu}}{2}}}\nonumber\\
&=&\pi^{\f{d}{2}}\f{\G\left(\f{\D_{\phi_1}+\D_{\phi_2}-d}{2}\right)\G\left(\f{d-\D_{\phi_1}-i\tilde{\nu}-i\nu'}{2}\right)\G\left(\f{d-\D_{\phi_2}+i\nu'+i\tilde{\nu}}{2}\right)}{\G\left(\f{2d-\D_{\phi_1}-\D_{\phi_2}}{2}\right)\G\left(\f{\D_{\phi_1}+i\nu'+i\tilde{\nu}}{2}\right)\G\left(\f{\D_{\phi_2}-i\nu'-i\tilde{\nu}}{2}\right)}\\
&\times&\f{1}{((x_1+n_1\g-\tilde{x})^2)^{\f{d-\D_{\phi_2}+i\nu'+i\tilde{\nu}}{2}}((x_2+n_2\g-\tilde{x})^2)^{\f{d-\D_{\phi_1}-i\nu'-i\tilde{\nu}}{2}}((x_1+n_1\g-x_2-n_2\g)^2)^{\f{-d+\D_{\phi_1}+\D_{\phi_2}}{2}}}\nonumber
\eeqa\\
Now the $\tilde{x}$ integral is
\beqa
I&=&\int ~ \f{d \tilde{x}}{((x_1+n_1\g-\tilde{x})^2)^{\f{\D_{\phi_1}-\D_{\phi_2}+d}{2}}((x_2+n_2\g-\tilde{x})^2)^{\f{d-\D_{\phi_1}+\D_{\phi_2}}{2}}}\nonumber\\
&=&\pi^{\f{d}{2}}\f{\G(\f{\D_{\phi_1}-\D_{\phi_2}}{2})\G\left(\f{\D_{\phi_
1}-\D_{\phi_2}}{2}\right)\G\left(\f{d}{2}\right)}{\G\left(\f{d+\D_{\phi_1}-\D_{\phi_2}}{2}\right)\G\left(\f{d-\D_{\phi_1}+\D_{\phi_2}}{2}\right)}\f{1}{((x_1+n_1\g-x_2-n_2\g)^2)^{\f{d}{2}}}
\eeqa\\

Thus the diagram finally is
\beqa \label{bench0}
\mathcal{W}&=&\f{1}{\mathcal{N}}\sum_{n_1,n_2=-\infty}^\infty\f{~g^2~}{32 \pi^{\f{d+8}{2}}}\f{\mathcal{C}(\D_{\phi_1})\mathcal{C}(\D_{\phi_2})}{\G(\D_{\phi_1})\G(\D_{\phi_2})}\f{\G\left(\f{\D_{\phi_1}+\D_{\phi_2}-d}{2}\right)}{\G(d-\f{\D_{\phi_1}+\D_{\phi_2}}{2})}\f{\G\left(\f{\D_{\phi_1}-\D_{\phi_2}}{2}\right)\G\left(\f{\D_{\phi_1}-\D_{\phi_2}}{2}\right)\G\left(\f{d}{2}\right)}{\G\left(\f{d+\D_{\phi_1}-\D_{\phi_2}}{2}\right)\G\left(\f{d-\D_{\phi_1}+\D_{\phi_2}}{2}\right)}
\nonumber\\
&\times& \f{1}{((x_1+n_1\g-x_2-n_2\g)^2)^{\f{\D_{\phi_1}+\D_{\phi_2}}{2}}} \int d\nu' ~ d\tilde{\nu}~\mathcal{F}(\nu',\tilde{\nu})
\eeqa
Thus again getting a form respecting the KMS constraint with
\beqa
\sum_{n_1}\f{1}{((x_1+n_1\g-x_2)^2)^{\f{\D_{\phi_1}+\D_{\phi_2}}{2}}} =\xi_{EH}\left(\D_1+\D_2-\f{d}{2}+i\nu',\tau_{12},x_{12}\right)
\eeqa
where again we have clubbed the two sums into one thus canceling the factor of $\mathcal{N}$. This diagram at zero temperature as studied in \cite{Giombi:2017hpr} provides a self correction diagram giving correction to the $\D_\phi$.

\noindent
\subsubsection{Case II : non-zero relative thermal shift between the two bulk points}
This is the more general case when $c-d\neq e-f$ or $n_3 \neq 0$. Here $\tilde{x}$ and $\tilde{x}+n_3$ are genuinely different boundary points. In this case we do the boundary integrals in the following manner by splitting the $d=1$ and $d>1$ bulk analysis for the given integral
\beqa
&&\int d x' ~ d\tilde{x} ~ \f{1}{((x_1+n_1\g-x')^2)^{\f{\D_{\phi_1}+i\nu'+i\tilde{\nu}}{2}}((x'-\tilde{x})^2)^{\f{d-\D_{\phi_1}+i\nu'-i\tilde{\nu}}{2}}((x_1+n_1\g-\tilde{x})^2)^{\f{\D_{\phi_1}-i\nu'-i\tilde{\nu}}{2}}}\\
&\times& \f{1}{((x_2+n_2\g-x')^2)^{\f{\D_{\phi_2}-i\nu'-i\tilde{\nu}}{2}}((x'-\tilde{x}-n_3\g)^2)^{\f{d-\D_{\phi_2}-i\nu'+i\tilde{\nu}}{2}}((x_2+n_2\g-\tilde{x}-n_3\g)^2)^{\f{\D_{\phi_2}+i\nu'+i\tilde{\nu}}{2}}}\nonumber
\eeqa

\vspace{0.5cm}
\textbf{\underline{AdS$_{2}$ (with $d=1$):}}\\

The first $x'$ is a four point integral which arranges itself into a partial wave. In this case we have the following definition of the conformal block and partial wave \cite{Alkalaev:2023evp}

\begin{equation}
G_{\Delta}^{\D_1,\D_2,\D_3,\D_4}(x_1, x_2, x_3, x_4) = \mathcal{L}^{\D_1,\D_2,\D_3,\D_4}(x_1, x_2, x_3, x_4) g_{\Delta}^{\D_1,\D_2,\D_3,\D_4}(\chi_1) 
\end{equation}

with the cross ratio and leg factor 
\begin{equation}
\chi_1=\f{x_{12}^2x_{34}^2}{x_{13}^2x_{24}^2}~~,~~\mathcal{L}^{\D_1,\D_2,\D_3,\D_4}(x_1, x_2, x_3, x_4) = \frac{1}{|x_{12}|^{\D_1+\D_2} |x_{34}|^{\D_3+\D_4}} \left| \frac{x_{23}}{x_{13}} \right|^{\D_{12}} \left| \frac{x_{24}}{x_{23}} \right|^{\D_{34}} 
\end{equation}
where $\D_{ij}=\D_i-\D_j$ and
\begin{equation}
g_{\Delta}^{\D_1,\D_2,\D_3,\D_4}(\chi_1) = \chi_1^{\Delta} \, {}_2F_1 \left[ \left. \begin{matrix} \Delta + \D_{12}, \, \Delta - \D_{34} \\ 2\Delta \end{matrix} \right| \chi_1 \right] \,. \label{eq:2.23}
\end{equation}
thus we get a conformal partial wave for the $x'$ integral and can write, after leaving out the shadow part,
\beqa
&&\int \f{dx'}{((x_1+n_1\g-x')^2)^{\f{\D_{\phi_1}+i\nu'+i\tilde{\nu}}{2}}((x'-\tilde{x})^2)^{\f{d-\D_{\phi_1}+i\nu'-i\tilde{\nu}}{2}}((x_2+n_2\g-x')^2)^{\f{\D_{\phi_2}-i\nu'-i\tilde{\nu}}{2}}}\nonumber\\
&\times &\f{1}{((x'-\tilde{x}-n_3\g)^2)^{\f{d-\D_{\phi_2}-i\nu'+i\tilde{\nu}}{2}}}\nonumber\\
&=& G_{\Delta'}^{\f{1}{2}+i\tilde{\nu}, \,\Delta_1, \,\Delta_2, \f{1}{2}-i\tilde{\nu}}(\tilde{x}, x_1+n_1 \g, x_2+n_2 \g, \tilde{x}+n_3 \g)
\eeqa

which can be expanded as
\beqa
&& G_{\Delta'}^{\f{1}{2}+i\tilde{\nu}, \,\Delta_{\phi_1}, \,\Delta_{\phi_2}, \f{1}{2}-i\tilde{\nu}}(\tilde{x}, x_1+n_1 \g, x_2+n_2 \g, \tilde{x}+n_3 \g)= \left( \frac{(\tilde{x}-x_1-n_1 \g)^{-\f{1}{2}-i\tilde{\nu}-\Delta_{\phi_1}}}{(x_2+n_2 \g - \tilde{x} - n_3 \g)^{\D_{\phi_2}+\f{1}{2}-i\tilde{\nu}}} \right)\nonumber\\
&\times& \left( \frac{x_1+n_1 \g - x_2 - n_2 \g}{\tilde{x}-x_2-n_2\g } \right)^{\f{1}{2}+i\tilde{\nu}-\Delta_{\phi_1}} \left( \frac{x_1+n_1 \g - \tilde{x}-n_3 \g}{x_1+n_1 \g - x_2 - n_2 \g} \right)^{\Delta_{\phi_2}-\f{1}{2}+i\tilde{\nu}}\nonumber \\
 & \times& \sum_{m=0}^{\infty} \frac{(\f{1}{2}+i\tilde{\nu}+\Delta'-\Delta_{\phi_1})_m (\f{1}{2}-i\tilde{\nu}-\Delta_{\phi_2}+\Delta'))_m}{(2\Delta')_m m!} \left( \frac{(\tilde{x}-x_1-n_1 \g)(x_2+n_2 \g - \tilde{x} - n_3 \g)}{(\tilde{x}-x_2-n_2 \g)(x_1+n_1 \g - \tilde{x} - n_3 \g)} \right)^{m+\Delta'}
\eeqa

Thus we again have for the $\tilde{x}$ integral the form
\beqa\label{partialw2}
& & \f{1}{((x_1+n_1 \g - x_2 - n_2 \g)^2)^{\f{-\Delta_{\phi_1}-\Delta_{\phi_2}+1}{2}} }\sum_{m=0}^{\infty} \frac{(\f{1}{2}+i\tilde{\nu}+\Delta'-\Delta_{\phi_1})_m (\f{1}{2}-i\tilde{\nu}-\Delta_{\phi_2}+\Delta')_m}{(2\Delta')_m m!} \nonumber \\
& \times & \int  \frac{d\tilde{x}}{((\tilde{x}-x_1-n_1 \g)^2)^{\f{\Delta_{\phi_1}+\f{1}{2}+i\tilde{\nu}-m-\Delta'}{2}} ((\tilde{x}-x_2-n_2 \g)^2)^{\f{-\Delta_{\phi_1}+\f{1}{2}+i\tilde{\nu}+m+\Delta'}{2}}} \nonumber \\
& \times & \frac{1}{\left((x_1+n_1 \g - \tilde{x} - n_3\g)^2\right)^{\frac{\f{1}{2}-i\tilde{\nu}-\Delta_{\phi_2}+m+\Delta'}{2}}} \frac{1}{\left((x_2+n_2 \g - \tilde{x} + n_3 \g)^2\right)^{\frac{\Delta_{\phi_2}+\f{1}{2}-i\tilde{\nu}-m-\Delta'}{2}}}
\eeqa

which is again a conformal partial and thus gives
\beqa
&& \int  \frac{d\tilde{x}}{((\tilde{x}-x_1-n_1 \g)^2)^{\f{\Delta_{\phi_1}+\tilde{\Delta}-m-\Delta'}{2}} ((\tilde{x}-x_2-n_2 \g)^2)^{\f{\Delta_{\phi_1}+\tilde{\Delta}+m+\Delta'}{2}}} \nonumber \\
& \times & \frac{1}{\left((x_1+n_1 \g - x_2 - n_3\g)^2\right)^{\frac{1-\tilde{\Delta}-\Delta_{\phi_2}+m+\Delta'}{2}}} \frac{1}{\left((x_2+n_2 \g - \tilde{x} + n_3 \g)^2\right)^{\frac{\Delta_{\phi_2}+1-\tilde{\Delta}-m-\Delta'}{2}}}\nonumber\\
&=& G_{\f{1}{2}+i\tilde{\nu}}^{ m+\f{1}{2}+i\nu', \, \Delta_{\phi_2},\Delta_{\phi_1}, \, m+\f{1}{2}+i\nu'}(  x_1 + n_1\g - n_3\g ~,~ x_2 + n_2\g - n_3\g~,~x_1+n_1\g~,~x_2+n_2\g) 
\eeqa
and then equation (\ref{partialw2}) can then be expanded as

\beqa
& =& \frac{1}{((x_1 + n_1 - x_2 - n_2)^2)^{\f{\Delta_{\phi_1} + \Delta_{\phi_2} - 1}{2}}} \sum_{m=0}^{\infty} \frac{(+i\tilde{\nu} -i\nu' - \Delta_1)_m \, (1-i\tilde{\nu} +i\nu'- \Delta_{\phi_2})_m}{(2\Delta')_m \, m!} \\
& \times&  \frac{((x_1 + n_1\g - x_2 - n_2\g)^2)^{\f{-\Delta_{\phi_1} - m - \f{d}{2}-i\nu'}{2}}}{((x_1 + n_1\g - x_2 - n_2\g )^2)^{\f{m + \f{1}{2}+i\nu'+ \Delta_{\phi_2}}{2}}} \left( \frac{(x_1 + n_1\g - n_3\g - n_2\g - x_2)^2}{(n_3\g)^2} \right)^{\f{-\Delta_{\phi_2} + m + \f{1}{2}+i\nu'}{2}} \nonumber\\
&\times&\left( \frac{(n_3\g)^2}{(x_1 + n_1\g - x_2 - n_2\g - n_3\g)^2} \right)^{\f{-m - \f{1}{2}-i\nu' + \Delta_{\phi_1}}{2}} \nonumber \\
& \times & \sum_{k=0}^{\infty} \frac{(-\D_{\phi_2}+ 1+i\tilde{\nu} + m + \nu')_k \, (-\D_{\phi_1} +1+ \tilde{\nu} + m + \nu')_k}{(1+2\tilde{\nu})_k \, k!} \nonumber\\
&\times& \left( \frac{ (x_1 + n_1  - x_2 - n_2 \g)^2  (x_1 + n_1\g - x_2 - n_2\g)^2}{(n_3\g)^2 \, (n_3\g)^2} \right)^{\f{k+\f{1}{2}+i\tilde{\nu}}{2}} \nonumber
\eeqa

thus we finally have for this diagram

\beqa
\mathcal{W}&=& \sum_{n_1, n_3=-\infty}^{\infty} \sum_{m, k=0}^{\infty} \int_{-\infty}^{\infty}\f{g^2}{32 \pi^{\f{d+8}{2}}}\f{~\nu'^2\sqrt{\mathcal{C}_{\f{d}{2}+i\nu'}\mathcal{C}_{\f{d}{2}-i\nu'}}~d\nu'}{\left[ \nu'^2+(\Delta'-d/2)^2\right]}\int_{-\infty}^{\infty}\f{\tilde{\nu}^2\sqrt{C_{\f{d}{2}+i\tilde{\nu}}C_{\f{d}{2}-i\tilde{\nu}'}}~d\tilde{\nu}}{\left[ \tilde{\nu}^2+(\tilde{\Delta}-d/2)^2\right]}  \\
&\times & \frac{(\tilde{\nu} - \nu' - \Delta_{\phi_1})_m \, (- \tilde{\nu} - \Delta_{\phi_2} - \nu')_m}{(1+2\nu')_m \, m!} \frac{(-\D_{\phi_2} + \tilde{\nu} + m + \nu'+1)_k \, (-\D_{\phi_1} + \tilde{\nu} + m + \nu'+1)_k}{(1+2\tilde{\nu})_k \, k!} \nonumber \\
& \times & \frac{B_{\D_{\phi_1},\f{d}{2}+i\nu',\f{d}{2}-i\tilde{\nu}}B_{\D_{\phi_2},\f{d}{2}-i\nu',\f{d}{2}+i\tilde{\nu}}}{((n_3\g)^2)^{\frac{2m+2k + 2+2\nu'  + 2\tilde{\nu} - \Delta_{\phi_1} - \Delta_{\phi_2}}{2}} \left( (x_1 + n_1\g  - x_2 )^2 \right)^{\frac{ 2\nu' - 2\tilde{\nu} +2m-2k}{2}}\left( (x_1 + n_1\g - x_2  - n_3\g)^2 \right)^{\frac{-2m -1- 2\nu' + \Delta_{\phi_2} + \Delta_{\phi_1}}{2}}}\nonumber
\eeqa

where we have absorbed the sums over $n_1,n_2$ into a single $n_1$ sum. Now to write the following in terms of Epstein Hurwitz zeta functions 

 \beqa \label{d=1eq}
 \sum_{n_1, n_3=-\infty}^\infty   \frac{1}{((n_3\g)^2)^{\frac{2m+2k + 2+2\nu'  + 2\tilde{\nu} - \Delta_{\phi_1} - \Delta_{\phi_2}}{2}} \left( (x_1 + n_1\g  - x_2 )^2 \right)^{\frac{ 2\nu' - 2\tilde{\nu} +2m-2k}{2}}\left( (x_1 + n_1\g - x_2  - n_3\g)^2 \right)^{\frac{-2m -1- 2\nu' + \Delta_{\phi_2} + \Delta_{\phi_1}}{2}}} \nonumber
  \eeqa
  
we expand the last term as
  
  \beqa
 && \frac{1}{\left( (x_1 + n_1\g - x_2  - n_3\g)^2 \right)^{\frac{-2m -1- 2\nu' + \Delta_{\phi_2} + \Delta_{\phi_1}}{2}}}\nonumber\\
 &=&((n_3)^2)^{\frac{2m + 1+2\nu' - \Delta_{\phi_2} - \Delta_{\phi_1}}{2}}\sum_{j=0}^{\infty} \f{\left({\frac{2m+1+ 2\nu' - \Delta_{\phi_2} - \Delta_{\phi_1}}{2}}\right)_j}{j!}\left(\f{(x_1-x_2-n_1\g)^2}{n_3^2}\right)^{j}
 \eeqa
 giving
 \beqa
\mathcal{W}&=& \sum_{n_1, n_3=-\infty}^{\infty} \sum_{m, k=0}^{\infty} \int_{-\infty}^{\infty}\f{g^2}{32 \pi^{\f{d+8}{2}}}\f{~\nu'^2\sqrt{\mathcal{C}_{\f{d}{2}+i\nu'}\mathcal{C}_{\f{d}{2}-i\nu'}}~d\nu'}{\left[ \nu'^2+(\Delta'-d/2)^2\right]}\int_{-\infty}^{\infty}\f{\tilde{\nu}^2\sqrt{C_{\f{d}{2}+i\tilde{\nu}}C_{\f{d}{2}-i\tilde{\nu}'}}~d\tilde{\nu}}{\left[ \tilde{\nu}^2+(\tilde{\Delta}-d/2)^2\right]}  \\
&\times & \frac{(\tilde{\nu} - \nu' - \Delta_{\phi_1})_m \, (- \tilde{\nu} - \Delta_{\phi_2} - \nu')_m}{(1+2\nu')_m \, m!} \frac{(-\D_{\phi_2} + \tilde{\nu} + m + \nu'+1)_k \, (-\D_{\phi_1} + \tilde{\nu} + m + \nu'+1)_k}{(1+2\tilde{\nu})_k \, k!} \nonumber \\
& \times & \frac{B_{\D_{\phi_1},\f{d}{2}+i\nu',\f{d}{2}-i\tilde{\nu}}B_{\D_{\phi_2},\f{d}{2}-i\nu',\f{d}{2}+i\tilde{\nu}}}{((n_3\g)^2)^{\frac{2k + 2+2\nu'  + j}{2}} \left( (x_1 + n_1\g  - x_2 )^2 \right)^{\frac{1-2\tilde{\nu} -j-2k+\D_{\phi_1}+\D_{\phi_2}}{2}}}\nonumber
\eeqa
 
 which can then be simplified to write the coordinate dependence in eq (\ref{d=1eq}) as
 
 \beqa
&&\sum_{n_1,n_3} \sum_{j=0}^{\infty}\f{\left({\frac{-2m -1- 2\nu' + \Delta_{\phi_2} + \Delta_{\phi_1}}{2}}\right)_j}{j!}\f{1}{((n_3)^2)^{\frac{2m+2k + 2+2\nu'  + 2\tilde{\nu} - \Delta_{\phi_1} - \Delta_{\phi_2}-j}{2}}}\f{1}{((x_1-x_2-n_1\g)^2)^{\f{2\D_{\phi_1}+2\D_{\phi_1}-2-2k-1-2\tilde{\nu}+j}{2}}}\nonumber\\
 &=& \sum_{j=0}^{\infty}\f{\left({\frac{-2m - 1-2\nu' + \Delta_{\phi_2} + \Delta_{\phi_1}}{2}}\right)_j}{j!}\zeta\left(\frac{2m+2k + 2+2\nu'  + 2\tilde{\nu} - \Delta_{\phi_1} -\Delta_{\phi_2}+2j}{2} \right)\nonumber\\
 &\times &\zeta_{\text{EH}}\left(2\D_{\phi_1}+2\D_{\phi_2}-2-2k-1-2\tilde{\nu}+2j,\t_{12},\mathbf{x}_{12}\right)
 \eeqa

Thus the diagram finally becomes
 
 \beqa
\mathcal{W}&=&\sum_{m, k,j=0}^{\infty} \int_{-\infty}^{\infty}\f{g^2}{32 \pi^{\f{d+8}{2}}}\f{~\nu'^2\sqrt{\mathcal{C}_{\f{d}{2}+i\nu'}\mathcal{C}_{\f{d}{2}-i\nu'}}~d\nu'}{\left[ \nu'^2+(\Delta'-\f{d}{2})^2\right]}\int_{-\infty}^{\infty}\f{\tilde{\nu}^2\sqrt{\mathcal{C}_{\f{d}{2}+i\tilde{\nu}}\mathcal{C}_{\f{d}{2}-i\tilde{\nu}'}}~d\tilde{\nu}}{\left[ \tilde{\nu}^2+(\tilde{\Delta}-\f{d}{2})^2\right]}  B_{\D_{\phi_1},\f{d}{2}+i\nu',\f{d}{2}-i\tilde{\nu}}B_{\D_{\phi_2},\f{d}{2}-i\nu',\f{d}{2}+i\tilde{\nu}}\nonumber\\
&\times & \frac{(\tilde{\nu} - \nu' - \Delta_{\phi_1})_m \, (- \tilde{\nu} - \Delta_{\phi_2} - \nu')_m}{(1+2\nu')_m \, m!} \frac{(-\D_{\phi_2} + 1+\tilde{\nu} + m + \nu')_k \, (-\D_{\phi_1} +1+ \tilde{\nu} + m + \nu')_k}{(1+2\tilde{\nu})_k \, k!} \nonumber \\
&\times& \f{({\frac{-2m -1- 2\nu' + \Delta_{\phi_2} + \Delta_{\phi_1}}{2}})_j}{j!}\f{1}{((n_3\g)^2)^{\frac{2m+2k +2+ 2\nu'  + 2\tilde{\nu} - \Delta_{\phi_1} - \Delta_{\phi_2}-j}{2}}}\f{1}{((x_1-x_2-n_1\g)^2)^{\f{2\D_{\phi_1}+2\D_{\phi_2}-2-2k-2\tilde{\D}+j}{2}}}\nonumber\\
 &\times& \f{({\frac{-2m - 1-2\nu' + \Delta_{\phi_2} + \Delta_{\phi_1}}{2}})_j}{j!}\zeta\left(\frac{2m+2k +2+ 2\nu'  + 2\tilde{\nu} - \Delta_{\phi_1} -\Delta_{\phi_2}+j}{2} \right)\nonumber\\
 &\times &\zeta_{\text{EH}}\left(2\D_{\phi_1}+2\D_{\phi_2}-2-2k-1-2\tilde{\nu}+j,\t_{12},\mathbf{x}_{12}\right).
\eeqa
where again we get a KMS invariant correlator.

\vspace{0.5cm}
\textbf{\underline{AdS$_{d+1}$ with $d>1$}:}\\

We first do the $x'$ integral here which is a four point integral arranged as a partial wave, again discarding the shadow here, thus giving $K_{d-\D'}^{\D_{\phi_2},\tilde{\D}}G_{\D'}^{\D_i}(x_i)$ with the general definition \cite{Rosenhaus:2018zqn}
\begin{equation}
K_{\tilde{\Delta}}^{\Delta_3, \Delta_4} = \frac{\pi^{\frac{d}{2}} \Gamma\left(\frac{d}{2} - \Delta\right) \Gamma\left(\frac{\Delta - \Delta_{34}}{2}\right) \Gamma\left(\frac{\Delta + \Delta_{34}}{2}\right)}{\Gamma(\Delta) \Gamma\left(\frac{\tilde{\Delta} - \Delta_{34}}{2}\right) \Gamma\left(\frac{\tilde{\Delta} + \Delta_{34}}{2}\right)} 
\end{equation}

and the conformal block in this case is,
\begin{equation}
G_{\Delta}^{\D_i}(x_i) = \frac{1}{\left(x_{12}^2\right)^{\frac{\Delta_1 + \Delta_2}{2}} \left(x_{34}^2\right)^{\frac{\Delta_3 + \Delta_4}{2}}} \left(\frac{x_{24}^2}{x_{14}^2}\right)^{\frac{\Delta_{12}}{2}} \left(\frac{x_{14}^2}{x_{13}^2}\right)^{\frac{\Delta_{34}}{2}} g_{\Delta}^i(u, v) \,,
\end{equation}

where
\begin{equation}
g_{\Delta}^{\D_i}(u, v) = u^{\frac{\Delta}{2}} \sum_{n,m=0}^{\infty} \frac{u^m}{m!} \frac{(1-v)^n}{n!} \frac{\left(\frac{\Delta+\Delta_{34}}{2}\right)_{n+m} \left(\frac{\Delta-\Delta_{12}}{2}\right)_{n+m} \left(\frac{\Delta+\Delta_{12}}{2}\right)_m \left(\frac{\Delta-\Delta_{34}}{2}\right)_m}{\left(\Delta+1-\frac{d}{2}\right)_m (\Delta)_{2m+n}} \,,
\end{equation}

and $u$ and $v$ are the two conformal cross ratios
\begin{equation}
u = \frac{x_{12}^2 x_{34}^2}{x_{13}^2 x_{24}^2} \,, \qquad v = \frac{x_{14}^2 x_{23}^2}{x_{13}^2 x_{24}^2} \,.
\end{equation}

Thus for our diagram we get after the $x'$ integral (which gives an exchange of $\D'$)

\beqa\label{d=2int}
&&\sum_{n,m=0}^\infty \mathcal{A}\int \frac{d\tilde{x}}{\left((x_1 - \tilde{x} + n_1 \g)^2\right)^{\frac{\Delta_{\phi_1} + d/2-i\tilde{\nu}}{2}}\left((x_2 + n_2 \g - \tilde{x} - n_2 \g)^2\right)^{\frac{\Delta_{\phi_2} +d/2 +i\tilde{\nu}}{2}}}  \nonumber\\
& \times & \left( \frac{(x_1 + n_1 \g - \tilde{x} - n_3 \g)^2}{(x_1 + n_1 \g - x_2 - n_2 \g)^2} \right)^{\frac{\Delta_{\phi_2}-d/2-i\tilde{\nu}}{2}} \left( \frac{(x_1 + n_1 \g - \tilde{x})^2 (x_2 + n_2 \g - \tilde{x} - n_3 \g)^2}{(x_1 + n_1 \g - x_2 - n_2 \g)^2 (n_3\g)^2} \right)^{\frac{\Delta'}{2} + m}\nonumber \\
& \times & \left(\frac{(n_3\g)^2}{(x_1 + n_1 \g - \tilde{x} - n_3 \g)^2} \right)^{\frac{\Delta_{\phi_1}-d/2+i\tilde{\nu}}{2}}\left( 1 - \frac{(x_1 + n_1 \g - \tilde{x} - n_3 \g)^2 (x_2 + n_2 \g - \tilde{x})^2}{(x_1 + n_1 \g - x_2 - n_2 \g)^2 (n_3\g)^2} \right)^n
\eeqa
where
\beqa
\mathcal{A}=\f{K_{d-\D'}^{\D_{\phi_2},\tilde{\D}}}{n!m!}\frac{\left(\frac{\Delta'+\Delta_{\phi_2}-d/2-i\tilde{\nu}}{2}\right)_{n+m} \left(\frac{\Delta'-\Delta_{\phi_1}+d/2-i\tilde{\nu}}{2}\right)_{n+m} \left(\frac{\Delta'+\Delta_{\phi_1}-d/2+i\tilde{\nu}}{2}\right)_m \left(\frac{\Delta'-\Delta_{\phi_2}+d/2+i\tilde{\nu}}{2}\right)_m}{\left(\Delta'+1-\frac{d}{2}\right)_m (\Delta')_{2m+n}} 
\eeqa
Now we need to do integral over $\tilde{x}$. To be able to do this we first expand the last term in the last line of eq (\ref{d=2int}) as

\[
\left(\frac{1}{XY}\right)^n \left[ (XY) - (UV) \right]^n = \frac{1}{(XY)^n} \sum_{k=0}^n \frac{(-1)^{n-k} n! \, (XY)^k (UV)^{n-k}}{k! (n-k)!}
\]

This gives us the following integral

\beqa
&&\int  \frac{d \tilde{x}}{\left( (x_1 + n_1 \g - \tilde{x})^2 \right)^{\frac{\Delta_{\phi_1} - \frac{d}{2} - i\nu' - 2m + \frac{d}{2} - i\tilde{\nu}}{2}}\left( (x_1 + n_1 \g - \tilde{x} - n_3 \g)^2 \right)^{\frac{\Delta_{\phi_1} - \frac{d}{2} + 2i\tilde{\nu} - \Delta_{\phi_2} + \frac{d}{2} - 2k}{2}}} \nonumber \\
&\times& \frac{1}{\left( (x_2 + n_2 \g - \tilde{x})^2 \right)^{-k}\left( (x_2 + n_2 \g - \tilde{x} - n_3 \g)^2 \right)^{\frac{\Delta_{\phi_2}  + i\tilde{\nu} - i\nu' - 2m}{2}}}
\eeqa

This however does not organize itself into a partial wave again unlike the previous case. We thus proceed as follows by writing the integral as
\begin{equation}
I(x_1,x_2) = \int d^d\tilde{x} \frac{\left( (x_2 + n_2 r - \tilde{x})^2 \right)^k}{\left((x_1 + n_1 r - \tilde{x})^2\right)^{a_1} \left((x_1 + n_1 r - n_3 r - \tilde{x})^2\right)^{a_2} \left((x_2 + n_2 r - n_3 r - \tilde{x})^2\right)^{a_4}}
\end{equation}

We evaluate this by using the techniques of Feynman parameters and polynomial expansions, as shown in detail in appendix [], to finally get for the boundary integral
\begin{equation}
I(x_1,x_2) = \sum_{\text{all indices}} \frac{\mathcal{M}}{\left[ (n_3 \g)^2 \right]^{\Omega_{n_3}}  \left[ (x_1 + n_1 \g - x_2 - n_2 \g)^2 \right]^{\Omega_{12}}  \left[ (x_1 + n_1 \g - x_2 - n_2 \g + n_3 \g)^2 \right]^{\Omega_{14}}}
\end{equation}
where $\Omega_{n_3},\Omega_{12},\Omega_{14}$ and $\mathcal{M}$ depend on the indices introduced and are defined in appendix \ref{appb}. We may again expand the last term in the denominator above as
\beqa
\f{1}{\left[ (x_1 + n_1 \g - x_2 - n_2 \g + n_3 \g)^2 \right]^{\Omega_{14}}}=(n_3^2)^{-\Omega_{14}}\sum_{\omega=0}^{\infty}\f{(\Omega_{14})_\omega}{\omega!}\left(\f{n_3^2}{(x_1+n_1\g-x_2)^2}\right)^{-\omega}
\eeqa
Thus
\beqa
I(x_1,x_2) = \sum_{\text{all indices},\omega} \frac{\mathcal{M} ~(\Omega_{14})_j}{\omega!}\f{1}{\left[ (n_3 \g)^2 \right]^{\Omega_{n_3}+\Omega_{14}+\omega}  \left[ (x_1 + n_1 \g - x_2)^2 \right]^{\Omega_{12}+\omega}}
\eeqa

The Witten diagram thus finally becomes

\beqa
\mathcal{W}&=&\int_{-\infty}^{\infty}~\f{g^2}{32 \pi^{\f{d+8}{2}}}~\f{\sqrt{\mathcal{C}_{\f{d}{2}+i\nu}\mathcal{C}_{\f{d}{2}-i\nu'}}~d\nu'}{\left[ \nu'^2+(\Delta'-d/2)^2\right]}\int_{-\infty}^{\infty}\f{\sqrt{\mathcal{C}_{\f{d}{2}+i\tilde{\nu}}\mathcal{C}_{\f{d}{2}-i\tilde{\nu}'}}~d\tilde{\nu}}{\left[ \tilde{\nu}^2+(\tilde{\Delta}-d/2)^2\right]} \f{(-1)^{n-k}}{m!n!(n-k)!}K_{\D'}^{\D_2,\tilde{\D}}\nonumber\\
&\times&\sum_{n,m=0}^{\infty} \sum_{k=0}^{n}\sum_{n_1,n_3}\frac{\left(\frac{\Delta'+(\Delta_{2}-\tilde{\D})}{2}\right)_{n+m} \left(\frac{\Delta'-(\Delta_{1}-(d-\tilde{\D})}{2}\right)_{n+m} \left(\frac{\Delta'+(\Delta'_{1}-(d-\tilde{\D})}{2}\right)_m \left(\frac{\Delta'-(\Delta_{2}-\tilde{\D}}{2}\right)_m}{\left(\Delta'+1-\frac{d}{2}\right)_m (\Delta')_{2m+n}} I(x_1,x_2) \nonumber
\eeqa

Further we can write this in terms of the Epstein Hurwitz zeta function as

\beqa
\sum_{n_1,n_3}I(x_1,x_2)=\xi_{EH}(\Omega_{12}+\omega)\zeta(\Omega_{13}+\Omega_{14}+\omega)
\eeqa\\

\noindent
\textbf{Anomalous dimension:}\\

\noindent
We do the analysis to find the anomalous dimension for the simplest case of the zero relative winding sector $n_3=0$. 
 For this we look at the contribution from the $n_1=0$ term as this is the only contribution that can produce $\log(x_{12}^2)$ term that is needed to read off the anomalous dimensions. We proceed by using the dimensionally regularized boundary integral as shown in \cite{Giombi:2017hpr} to read the the anomalous dimension
\beqa \label{bench0}
\mathcal{W}&=&\f{~g^2}{32 \pi^{\f{d+8}{2}}}\f{\mathcal{C}(\D_{\phi_1})\mathcal{C}(\D_{\phi_2})}{\G(\D_{\phi_1})\G(\D_{\phi_2})}\f{\G\left(\f{\D_{\phi_1}+\D_{\phi_2}-d}{2}\right)}{\G\left(d-\f{\D_{\phi_1}+\D_{\phi_1}}{2}\right)}\f{\G\left(\f{\D_{\phi_1}-\D_{\phi_2}}{2}\right)\G\left(\f{\D_{\phi_1}-\D_{\phi_2}}{2}\right)\G\left(\f{d}{2}\right)}{\G\left(\f{d+\D_{\phi_1}-\D_{\phi_2}}{2}\right)\G\left(\f{d-\D_{\phi_1}+\D_{\phi_2}}{2}\right)} \f{1}{((x_1-x_2)^2)^{\f{\D_{\phi_1}+\D_{\phi_2}}{2}}}
\nonumber\\
&\times& \delta_{\Delta_{\phi_1}, \Delta_{\phi_2}} \frac{2\pi^{\frac{d}{2}}}{\Gamma\left(\frac{d}{2}\right)}  \left[ \frac{2}{\epsilon} + \log(\pi) - \psi \left( \frac{d}{2} \right) + \log \left( (x_{1}+n_1\g-x_2)^2 \right) \right]
\int d\nu' ~ d\tilde{\nu}~\mathcal{F}(\nu',\tilde{\nu})
\eeqa
with
\begin{align}
\mathcal{F}(\nu',\tilde{\nu})
&= \f{\nu'\tilde{\nu}\sinh(\pi\nu')\sinh(\pi\tilde{\nu})}{
\left[\left(\frac{d}{2}-\tilde{\Delta}\right)^2+\tilde{\nu}^2\right]\left[\left(\frac{d}{2}-\Delta'\right)^2+\nu'^2
\right]}
\\
&\quad\times
\Gamma\left(\frac{d-\Delta_{\phi_1}-i(\nu'-\tilde{\nu})}{2}\right)
\Gamma\left(\frac{d-\Delta_{\phi_1}+i(\nu'+\tilde{\nu})}{2}\right)
\Gamma\left(\frac{\Delta_{\phi_1}-i(\nu'-\tilde{\nu})}{2}\right)
\Gamma\left(\frac{\Delta_{\phi_1}+i(\nu+\bar{\nu})}{2}\right)\nonumber
\\
&\quad\times
\Gamma\left(\frac{d-\Delta_{\phi_2}+i(\nu'-\tilde{\nu})}{2}\right)
\Gamma\left(\frac{d-\Delta_{\phi_2}-i(\nu'+\tilde{\nu})}{2}\right)
\Gamma\left(\frac{\Delta_{\phi_2}+i(\nu'-\tilde{\nu})}{2}\right)
\Gamma\left(\frac{\Delta_{\phi_2}-i(\nu'+\tilde{\nu})}{2}\right)\nonumber
\end{align}

Beyond this the spectral analysis to find anomalous dimension is already done in \cite{Giombi:2017hpr} that computes $\g_{\phi}$ to be the following
\beqa
\gamma_\phi &=& -g^2~\delta_{\Delta_{\phi_1}\Delta_{\phi_1}} \frac{\Gamma\left(\frac{\Delta_{\phi_1}+\Delta_{\phi_2}-d}{2}\right)}{64\pi^{\frac{d+8}{2}} \Gamma\left(\frac{d}{2}\right) \Gamma\left(d - \frac{\Delta_{\phi_1}+\Delta_{\phi_1}}{2}\right)\sqrt{\Gamma(\Delta_{\phi_1})\Gamma\left(-\frac{d}{2} + \Delta_{\phi_1} + 1\right)} \sqrt{\Gamma(\Delta_{\phi_2})\Gamma\left(-\frac{d}{2} + \Delta_{\phi_2} + 1\right)}} \nonumber \\
&\times & \int_{-\infty}^{\infty} d\nu ~d\bar{\nu} ~ \mathcal{F}(\nu,\bar{\nu}) 
\eeqa
The authors of \cite{Giombi:2017hpr} give this in terms of the Harmonic numbers to be (with $\D_{\phi_1}=\D_{\phi_2}=\D$ and as written in our notations)
\begin{equation}
\gamma_\phi = -\frac{g^2}{16(\Delta - 1)^2} \left[ \frac{\sin(\pi\Delta)}{\cos(\pi\Delta) - \cos(\pi(\Delta' + \Delta_2))} + \frac{1}{2\pi} \left( H_{\frac{\Delta+\Delta'+\tilde{\Delta}-4}{2}} + H_{\frac{2-\Delta-\Delta'-\tilde{\Delta}}{2}} - H_{\frac{\Delta-\Delta'-\tilde{\Delta}}{2}} - H_{\frac{-\Delta+\Delta'+\tilde{\Delta}-2}{2}} \right) \right].
\end{equation}
\textbf{Full correlator:}\\

\noindent
The full diagram, as we evaluated and on re-introducing the sum over the thermal images has, the following form
\begin{equation}
\mathcal{W} = \sum_{n_1} ((x_1+n_1\g-x_2)^2)^{-\Delta_\phi} [a_1 z_\phi + a_1 \g_\phi \log (x_1+n_1\g-x_2)^2]
\end{equation}
 The $\frac{2}{\epsilon}$ goes into the wave-function renormalization $a_1 z_\phi$ and the $\log (x_{12})^2$ coefficient gives the anomalous dimension where $a_1$ is the $\mathcal{O}(g^0)$ coefficient.

\subsection{Operator families allowed at order $g^2$}

The thermal OPE of $\mathcal{O}_\phi\mathcal{O}_\phi$ is even under $\phi\mapsto-\phi$.  At the order of the Witten diagrams studied her it can thus contain the following:
\begin{enumerate}
 \item The identity;
 \item The single-trace scalar $\mathcal{O}_\chi$, with $f_{\phi\phi\chi}=O(g)$ and $b_\chi=O(g)$;
 \item Even-spin double traces $[\mathcal{O}_\phi\mathcal{O}_\phi]_{n,J}$, whose mean-field thermal coefficients are $O(1)$ and whose dimensions and coefficients receive $O(g^2)$ corrections;
 \item  Also it was noted in \cite{Iliesiu:2018fao} that the $[\mathcal{O}_\phi\mathcal{O}_\phi]_{n,J}$, family does not analytically continue down to $J=0$ and $\chi$ instead plays the role of $\phi_i\phi_i$ in the IR. Thus, the most basic scalar operators are powers of $\chi$.
 \item We also get mixing between scalar operators $[\mathcal{O}_\phi\mathcal{O}_\phi]$ and $[\mathcal{O}_{\tilde{\D}}\mathcal{O}_{\tilde{\D}}]$ as we describe later in the tadpole diagram analysis in section ().
\end{enumerate}

\section{Contact Diagrams}\label{section4}
In this section we evaluate scalar quartic contact diagrams both with and without derivative interactions with the coupling $g_1$.
\subsection{Contact Diagram with loop}
\vspace{1cm}

\[\begin{wittendiagram} 
 \draw (-1.15,-1.6) node[vertex] -- (-0.1,-0.5) node[vertex];
 \draw (1.15,-1.6) node[vertex] -- (-0.1,-0.5) node[vertex];
 \draw(-0.1,0.1) circle (0.6);
  \node at (-2.1, -1.7) {$x_1+n_1\g$};
  \node at (2.1, -1.7) {$x_2+n_2\g$};
  \node at (0.7, -0.2) {$X$};
   \node at (-1.2, -1) {$\D_{\phi_1}$};
  \node at (1.2, -1) {$\D_{\phi_2}$};
    \node at (0, 1) {$\D$};
  \end{wittendiagram}\hspace{0.3cm}\quad
\bm{\xrightarrow{\hspace{1cm}} }
\quad
\begin{wittendiagram}
 \draw (-1.15,-1.6) node[vertex] -- (-0.1,-0.5) node[vertex];
 \draw (1.15,-1.6) node[vertex] -- (-0.1,-0.5) node[vertex];
   \draw[dashed] (-0.1,-0.5) node[vertex] -- (-1.7,1) node[vertex];
       \draw[dashed] (-0.1,-0.5) node[vertex] -- (1.7,1) node[vertex];
  \node at (-2.1, -1.7) {$x_1+n_1\g$};
  \node at (2.1, -1.7) {$x_2+n_2\g$};
  \node at (0.7, -0.6) {$X$};
  \node at (2.9, 1.1) {$x+n_3\g$};
  \node at (-2.9, 1.1) {$x+n_4\g$};
\end{wittendiagram}\]

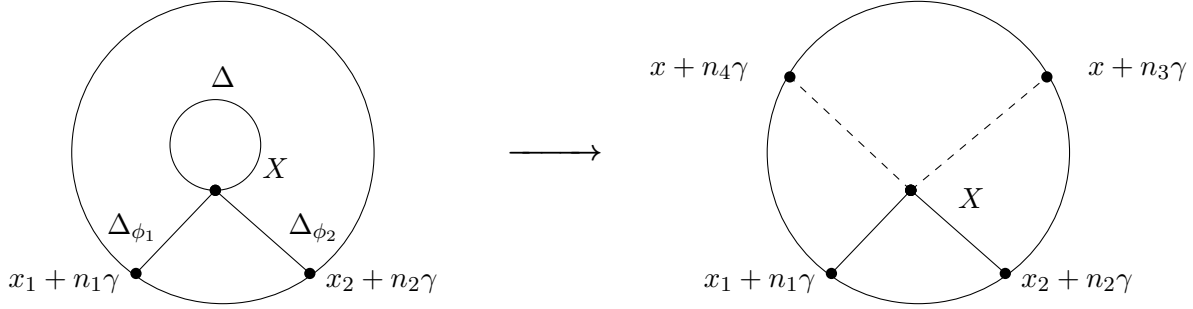
\captionof{figure}{The two-point contact diagram in thermal AdS with details of the split representation and the thermal arrangement of points. Thermally identified points are shown in black, non identified points are shown in blue.}\label{fig12}

\vspace{1cm}
We evaluate the contact diagram as follows.
\beqa
\mathcal{W}&=&\f{g_1}{\mathcal{N}^3}\sum_{\substack{a,b,c,d\neq 0\\n_1,n_2}}^{}\int dX K_{\D_{\phi_1}}(X^{a\g},x_1+n_1\g)K_{\D_{\phi_2}}(X^{b\g},x_2+n_2\g)G_{\D}(X^{c\g},X^{d\g})\nonumber\\
&=&\f{1}{\mathcal{N}^2}\sum_{\substack{a,c,d\neq 0\\n_1,n_2}}^{}\int_{-\infty}^{\infty}\f{2\nu^2 d\nu \sqrt{\mathcal{C}_{\f{d}{2}-i\nu}\mathcal{C}_{\f{d}{2}+i\nu}}}{2\pi\left[ \nu^2+(\Delta-d/2)^2\right]}\int dX \int d x K_{\D_{\phi_1}}(X^{a\g},x_1+n_1\g)K_{\D_{\phi_2}}(X^{a\g},x_2+n_2\g)\nonumber\\
&\times& K_{\f{d}{2}+i\nu}(X^{c\g},x)K_{\f{d}{2}-i\nu}(X^{d\g},x)\nonumber\\
&=&\f{1}{\mathcal{N}}\sum_{\substack{a\neq 0\\n_1,n_2,n_3,n_4}}^{}\int_{-\infty}^{\infty}\f{\nu^2 d\nu\sqrt{\mathcal{C}_{\f{d}{2}-i\nu}\mathcal{C}_{\f{d}{2}+i\nu}}}{\left[ \nu^2+(\Delta-d/2)^2\right]}\int dX \int d x K_{\D_{\phi_1}}(X^{a\g},x_1+n_1\g)K_{\D_{\phi_2}}(X^{a\g},x_1+n_2\g)\nonumber\\
&\times& K_{\f{d}{2}+i\nu}(X^{a\g},x+n_3\g)K_{\f{d}{2}-i\nu}(X^{a\g},x+n_4\g)\nonumber\\
\eeqa
We split the diagram as shown in figure \ref{fig13} to simplify the boundary integrals while also keeping intact our channel.  The computation reduces to that of a four point function which we evaluate as follows. To do the above computation we use the following relations \cite{Carmi:2019ocp}
\beqa\label{specprop1}
\Omega_{\nu}(y_1,y_2)=\f{\nu'^2 \sqrt{\mathcal{C}_{\f{d}{2}+i\nu'}\mathcal{C}_{\f{d}{2}-i\nu'}}}{\pi}\int d x K_{\f{d}{2}+i\nu}(x,y_1)K_{\f{d}{2}-i\nu}(x,y_2)
\eeqa
and
\beqa\label{specprop2}
F(X,Y)=\int_{-\infty}^{\infty}d \nu \tilde{F} \ \Omega_{\nu}(X,Y)
\eeqa
with the property
\beqa\label{specprop3}
\int_{-\infty}^{\infty}\Omega_{\nu}(y_1,y_2)=\d(y_1,y_2)
\eeqa

where, for us $F(X,Y)=\delta(X,Y)$ and $\tilde{F}=1$. This converts the above diagram to two 3 point integrals as given below and indicated in going from diagram \ref{fig12} to \ref{fig13} and can be written as

\vspace{1cm}

\[\begin{wittendiagram} 
 \draw (-1.15,-1.6) node[vertex] -- (-0.1,-0.7) node[vertex];
 \draw (1.15,-1.6) node[vertex] -- (-0.1,-0.7) node[vertex];
   \draw (-0.1,0.3) node[vertex] -- (-1.3,1.5) node[vertex];
       \draw (-0.1,0.3) node[vertex] -- (1.15,1.6) node[vertex];
       \draw [dashed] (-0.1,0.3) -- (-0.1,-0.7) ;
  \node at (-1.7, -1.7) {$x_1$};
  \node at (1.7, -1.7) {$x_2$};
  \node at (0.7, -0.6) {$X$};
   \node at (0.7, 0.2) {$Y$};
  \node at (2.7, 1.6) {$x+n_3\g$};
  \node at (-2.7, 1.6) {$x+n_4\g$};
\end{wittendiagram}\hspace{0.3cm}\quad
\bm{\xrightarrow{\hspace{1cm}} }
\quad\begin{wittendiagram} 
 \draw (-1.15,-1.6) node[vertex] -- (-0.1,-0.7) node[vertex];
 \draw (1.15,-1.6) node[vertex] -- (-0.1,-0.7) node[vertex];
   \draw (-0.1,0.3) node[vertex] -- (-1.3,1.5) node[vertex];
       \draw (-0.1,0.3) node[vertex] -- (1.15,1.6) node[vertex];
       \draw [dashed] (-0.1,0.3) -- (-2,-0.1) ;
       \draw [dashed] (-0.1,-0.7) -- (-2,-0.1) node[vertex, color=blue];
  \node at (-1.7, -1.7) {$x_1$};
  \node at (1.7, -1.7) {$x_2$};
  \node at (0.7, -0.6) {$X$};
   \node at (0.7, 0.2) {$Y$};
  \node at (2.7, 1.6) {$x+n_3\g$};
  \node at (-2.7, 1.6) {$x+n_4\g$};
  \node at (-2.4, -0.1) {$x$};
   \node at (-0.45, -1.5) {$\D_{\phi_1}$};
  \node at (0.45, -1.5) {$\D_{\phi_2}$};    \node at (-1.4, -0.55) {{\tiny $\f{d}{2}-i\nu'$}};
    \node at (-1.4, 0.25) {{\tiny $\f{d}{2}+i\nu'$}};
        \node at (1.1, 1) {{\tiny $\f{d}{2}-i\nu$}};
    \node at (-1.3, 1) {{\tiny $\f{d}{2}+i\nu$}};
\end{wittendiagram}\] 

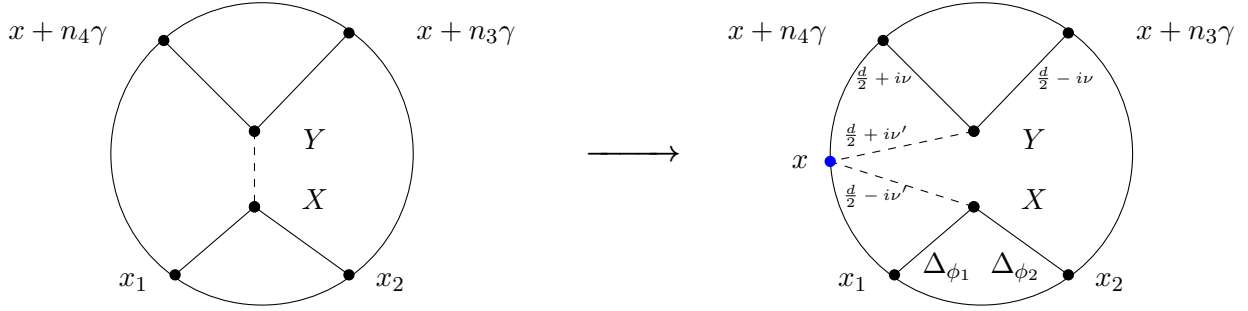
\captionof{figure}{Thermally identified points are given in black, non identified points are in blue.} \label{fig13}

\vspace{0.4cm}

\beqa
\mathcal{W}&=&\f{1}{\mathcal{N}}\sum_{\substack{a\neq 0\\n_1,n_2,n_3,n_4}}^{}\f{g_1}{32 \pi^{\f{d+8}{2}}}\int_{-\infty}^{\infty}\f{\nu^2 d\nu \sqrt{\mathcal{C}_{\f{d}{2}-i\nu}\mathcal{C}_{\f{d}{2}+i\nu}}}{\pi\left[ \nu^2+\left(\Delta-\f{d}{2}\right)^2\right]}\int_{-\infty}^{\infty} \f{\nu'^2}{\pi} \sqrt{\mathcal{C}_{\f{d}{2}-i\nu'}\mathcal{C}_{\f{d}{2}+i\nu'}} ~d\nu'\nonumber\\
&\times& \int dX ~ d Y \int d x' ~d x~ K_{\D_{\phi_1}}(X^{a\g},x_1+n_1\g) K_{\D_{\phi_2}}(X^{a\g},x_1+n_2\g) K_{\f{d}{2}+i\nu'}(X^{a\g},x')\nonumber\\
&\times& K_{\f{d}{2}-i\nu'}(Y^{a\g},x') K_{\f{d}{2}+i\nu}(Y^{a\g},x+n_3\g) K_{\f{d}{2}-i\nu}(Y^{a\g},x+n_4\g)
\eeqa
We notice that here $Y^{a\g}$ will be replaced by $Y^{b\g}$ to have the correct $\mathcal{N}$ dependence as split rep of $\delta(X,Y)$ also introduces a factor of $1/\mathcal{N}$. The standard bulk three point integrals for $X$ and $Y$ as before give

\beqa
&&\f{1}{\mathcal{N}}\int d X K_{\D_{\phi_1}}(X^{a\g},x_1+n_1\g) K_{\D_{\phi_2}}(X^{a\g},x_2+n_2\g) K_{\f{d}{2}+i\nu'}(X^{a\g},x')\nonumber\\
&=&\f{1}{\mathcal{N}}\f{B_{\D_{\phi_1},\D_{\phi_2},\f{d}{2}+i\nu'}}{((x_1+n_1-x_2-n_2)^2)^{\f{\D_{\phi_1}+\D_{\phi_2}-\f{d}{2}-i\nu'}{2}}((x_2+n_2-x')^2)^{\f{-\D_{\phi_1}+\D_{\phi_2}+\f{d}{2}+i\nu'}{2}}((x_1+n_1-x')^2)^{\f{\D_{\phi_1}-\D_{\phi_2}+\f{d}{2}+i\nu'}{2}}} \nonumber
\eeqa
and

\beqa
&&\f{1}{\mathcal{N}}\int d Y K_{\f{d}{2}-i\nu'}(Y^{a\g},x') K_{\f{d}{2}+i\nu}(Y^{a\g},x+n_3\g) K_{\f{d}{2}-i\nu}(Y^{a\g},x+n_4\g)
\nonumber\\
&=&\f{1}{\mathcal{N}}\f{B_{\f{d}{2}-i\nu',\f{d}{2}+i\nu,\f{d}{2}-i\nu}}{((x'-x-n_3)^2)^{\f{\f{d}{2}-i\nu'+2i\nu}{2}}((n_3-n_4)^2)^{\f{\f{d}{2}+i\nu'}{2}}((x'-x-n_4)^2)^{\f{\f{d}{2}-i\nu'-2i\nu}{2}}} \nonumber
\eeqa
Next, the $x'$ and $x$ integrals need to be performed

\beqa
&&\f{1}{((x_1+n_1-x_2-n_2)^2)^{\f{\D_{\phi_1}+\D_{\phi_2}-\f{d}{2}-i\nu'}{2}}((n_3-n_4)^2)^{\f{\f{d}{2}+i\nu'}{2}}}
\int  \f{dx'}{((x_2+n_2-x')^2)^{\f{-\D_{\phi_1}+\D_{\phi_2}+\f{d}{2}+i\nu'}{2}}}\nonumber\\
&\times& \f{d\tilde{x}}{((x_1+n_1-x')^2)^{\f{\D_{\phi_1}-\D_{\phi_2}+\f{d}{2}+i\nu'}{2}}((x'-x-n_3)^2)^{\f{\f{d}{2}-i\nu'+2i\nu'}{2}}((n_3-n_4)^2)^{\f{\f{d}{2}+i\nu'}{2}}((x'-x-n_4)^2)^{\f{\f{d}{2}-i\nu'-2i\nu}{2}}}\nonumber
\eeqa

We first do the $x$ integral. This is a two point integral and also converts the susequent $x'$ integral to a two point integral. This gives

\beqa
\int  \f{dx}{((x'-x-n_3)^2)^{\f{\f{d}{2}-i\nu'+2i\nu}{2}}((x'-x-n_4)^2)^{\f{\f{d}{2}-i\nu'-2i\nu}{2}}}=\f{\pi^{\f{d}{2}}\G\left( \f{2 i \nu+\f{d}{2}+i\nu'}{2} \right)\G\left( \f{-2 i \nu + \f{d}{2}+i\nu'}{2} \right)\G(-i\nu')}{\G\left(\f{\f{d}{2}-i\nu'-2 i \nu}{2} \right)\G\left(\f{\f{d}{2}-i\nu'+2 i \nu}{2} \right)\G\left(\f{d}{2}+i\nu'\right)}\f{1}{((n_3-n_4)^2)^{-i\nu'}}\nonumber\\
\eeqa

Similarly, the $x'$ integral gives

\beqa
&&\int  \f{dx'}{((x_2+n_2-x')^2)^{\f{-\D_{\phi_1}+\D_{\phi_2}+\f{d}{2}+i\nu'}{2}}((x_1+n_1-x')^2)^{\f{\D_{\phi_1}-\D_{\phi_2}+\f{d}{2}+i\nu'}{2}}}\nonumber\\
&=&\f{\pi^{\f{d}{2}}\G\left( \f{-\D_{\phi_1}+\D_{\phi_2}+\f{d}{2}-i\nu'}{2} \right)\G\left( \f{\D_{\phi_2}-\D_{\phi_2}+\f{d}{2}-i\nu'}{2} \right)\G(i\nu')}{\G\left(\f{\f{d}{2}+i\nu'-\D_{\phi_2}+\D_{\phi_2} }{2}\right)\G\left(\f{\f{d}{2}+i\nu'+\D_{\phi_2}-\D_{\phi_1}}{2} \right)\G\left(\f{d}{2}-i\nu'\right)}\f{1}{((x_2-x_1+n_2-n_1)^2)^{-i\nu'}}
\eeqa

Next, we need to perform the sums over $n_i$. The sums over $n_3, n_4$ cancel factors of $\mathcal{N}^2$, while

\beqa
\sum_{n_1}\f{1}{((x_2-x_1+n_2)^2)^{\f{\D_{\phi_1}+\D_{\phi_2}-\f{d}{2}+i\nu'}{2}}}&=&\zeta_{\text{EH}}\left(\D_{\phi_1}+\D_{\phi_2}-\f{d}{2}+i\nu',\t_{12},\mathbf{x}_{12}\right)
\eeqa
For the sum over $n_3$ we may use

\beqa
\sum_{n_3=-\infty}^{\infty}\f{1}{(n_3^2)^{\f{d}{2}-i\nu'}}=\left[1+(-1)^{i\nu'-\f{d}{2}} \right]\zeta\left(\f{d}{2}-i\nu'\right)
\eeqa

The expression thus finalizes to

\beqa
\mathcal{W}&=&\f{g_1}{32 \pi^{\f{d+8}{2}}}\int_{-\infty}^{\infty}d\nu'  \int_{-\infty}^{\infty}\f{d \nu}{\left[ \nu^2+\left(\Delta-\f{d}{2}\right)^2\right]}\Gamma\left(\f{\Delta_{\phi_1}+\Delta_{\phi_2}-\f{d}{2}\pm i\nu'}{2}\right)\Gamma\left(\f{\f{d}{2}\pm i\nu'}{2}\right)\\
&\times&\f{\Gamma\left(\f{\D_{\phi_2}-\D_{\phi_1}+\f{d}{2}-i\nu'}{2}\right)\Gamma\left(\f{-\D_{\phi_2}+\D_{\phi_1}+\f{d}{2}-i\nu'}{2}\right)\Gamma\left(\f{d}{4}+\f{i\nu'\pm 2i\nu}{2}\right)}{\sqrt{\G(\D_{\phi_1})\G(\D_{\phi_2})\G(\D_{\phi_1}-\f{d}{2}+1)\G(\D_{\phi_2}-\f{d}{2}+1)}\G( i \nu')\G(-i\nu)\G(\f{d}{2}\pm i\nu')}\nonumber\\
&\times&  \zeta_{\text{EH}}\left(\D_{\phi_1}+\D_{\phi_2}-\f{d}{2}+i\nu',\t_{12},\mathbf{x}_{12}\right)\zeta\left(\f{d}{2}-i\nu'\right)\nonumber
\eeqa
We notice the following pole structure for this in the UHP:\\

\begin{enumerate}
    \item  \textbf{Double trace poles} from $\Gamma\left(\frac{\D_{\phi_1}+\D_{\phi_2}-h+i\nu'}{2}\right)$

       These poles occur when $\frac{\D_{\phi_1}+\D_{\phi_2}-h+i\nu'}{2}=-2n \rightarrow \nu'=\D_{\phi_1}+\D_{\phi_2}+2n$\\

       \item \textbf{Physical poles} from $\nu=\D-d/2$ of the loop, these further affect the next set of poles in point (3)\\
    
    \item Poles from $\Gamma\left(\f{i\nu+2\tilde{\D}}{2}\right)$
    
   These poles occur when $\f{i\nu+2\tilde{\D}}{2}=-2m \rightarrow \nu'=2\tilde{\D}+2m$\\

    \item There is a cancellation of poles between $\Gamma\left(\f{\f{d}{2}\pm i\nu'}{2}\right)$ and $\G(\f{d}{2}\pm i\nu')$.

\end{enumerate}

Double poles occur from the collision of poles in (1) and (3) when $\D_{\phi_1}+\D_{\phi_2}+2n=2\tilde{\D}+2m$ and as before we instead find the anomalous dimension from the effective vertex diagram in section , having shown that the loop reduces to effective vertices in appendix \ref{effverapp}.

\subsection{Contact Diagram with derivatives}
\vspace{0.4cm}

\[\begin{wittendiagram} 
 \draw (-1.15,-1.6) node[vertex] -- (-0.1,-0.5) node[vertex];
 \draw (1.15,-1.6) node[vertex] -- (-0.1,-0.5) node[vertex];
 \draw(-0.1,0.1) circle (0.6);
  \node at (-2.1, -1.7) {$x_1+n_1\g$};
  \node at (2.1, -1.7) {$x_2+n_2\g$};
  \node at (-0.1, -0.9) {$X$};
     \node at (-1.2, -1) {$\D_{\phi_1}$};
  \node at (1.2, -1) {$\D_{\phi_2}$};
    \node at (0, 1) {$\D$};
     \node at (-0.4, -0.6) { \footnotesize{1}};
      \node at (-0.4, -0.2) {\footnotesize{2}};
       \node at (0.25, -0.6) {\footnotesize{4}};
        \node at (0.25, -0.2) {\footnotesize{3}};
  \end{wittendiagram}\]
  
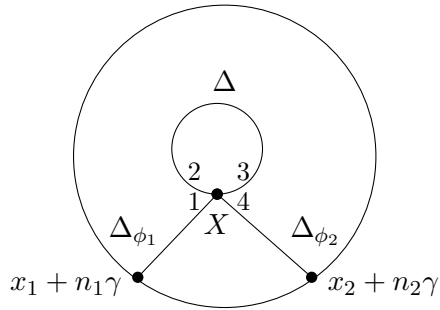
\captionof{figure}{Contact diagram with derivative interactions in thermal AdS} \label{fig14}

  \vspace{1cm}
We now evaluate the one-loop contact diagram generated by the quartic vertex with derivatives. We again follow the conventions set in \cite{Giombi:2017hpr} to write down the scalar vertex with derivative terms as

\beqa
\mathcal{V}_{k,m}(X)=\frac{g_1}{(k+m)\mbox{!}}\left[\phi_1(X)(\del_U\del_X)^k \phi_2(X)\right](\del_U\del_X)^m\phi_3(X)(U.\del_X)^{k+m}\phi_4(X)
\eeqa

We also use the following relation
\beqa
(U.\del_X)^n K_{\D,0}(X,P)=2^n \left(\D+1-\f{d}{2} \right)_n (U.P)^n K_{\D+n,0}(X,P)
\eeqa

For this diagram we have 4 different configurations that contribute, labeled using the point splitting representation mentioned above and written as $\mathcal{W}_{1234},\mathcal{W}_{1342},\mathcal{W}_{3142}$ and $\mathcal{W}_{4132}$. These are defined below
\beqa
\mathcal{W}_{1234}&=&\f{1}{\mathcal{N}^3}\sum_{\substack{a,b,c,d\neq 0\\n_1,n_2}}^{}g_1\int dX K_{\D_{\phi_1}}(X^{a\g},x_1+n_1\g)(\del_U\del_{X_1})^k (\del_U \del X_2)^m G(X_1^{c\g},X_2^{d\g})\nonumber\\
&\times&(U.\del_X)^{k+m}
K_{\D_{\phi_2}}(X^{b\g},x_2+n_2\g)|_{X_i=X}\\
\mathcal{W}_{3142}&=&\f{1}{\mathcal{N}^3}\sum_{\substack{a,b,c,d\neq 0\\n_1,n_2}}^{}\int dX (\del_U \del X_2)^m K_{\D_{\phi_1}}(X^{a\g},x_1+n_1\g)(\del_U\del_{X_1})^k  K_{\D_{\phi_2}}(X^{b\g},x_2+n_2\g)\nonumber\\
&\times&(U.\del_X)^{k+m} G(X_1^{c\g},X_2^{d\g})
|_{X_i=X}\\
\mathcal{W}_{4132}&=&\f{1}{\mathcal{N}^3}\sum_{\substack{a,b,c,d\neq 0\\n_1,n_2}}^{}\int dX (\del_U\del_{X_1})^k K_{\D_{\phi_1}}(X^{a\g},x_1+n_1\g) (\del_U \del X_2)^m G(X_1^{c\g},X_2^{d\g})\nonumber\\
&\times&(U.\del_X)^{k+m}
K_{\D_{\phi_2}}(X^{b\g},x_2+n_2\g)|_{X_i=X}\\
\mathcal{W}_{1342}&=&\f{1}{\mathcal{N}^3}\sum_{\substack{a,b,c,d\neq 0\\n_1,n_2}}^{}\int dX (U.\del_X)^{k+m} K_{\D_{\phi_1}}(X^{a\g},x_1+n_1\g)(\del_U\del_{X_1})^k (\del_U \del X_2)^m G(X_1^{c\g},X_2^{d\g})\nonumber\\
&\times&
K_{\D_{\phi_2}}(X^{b\g},x_2+n_2\g)|_{X_i=X}
\eeqa

Here we evaluate the contribution $\mathcal{W}_{1234}$  as follows and as illustrated in figure \ref{fig14}
\beqa
\mathcal{W}_{1234}&=&\f{1}{\mathcal{N}^3}\sum_{\substack{a,b,c,d\neq 0\\n_1,n_2}}^{}g_1\int dX K_{\D_{\phi_1}}(X^{a\g},x_1+n_1\g)(\del_U\del_{X_1})^k (\del_U \del X_2)^m G(X_1^{c\g},X_2^{d\g})\nonumber\\
&\times&(U.\del_X)^{k+m}
K_{\D_{\phi_2}}(X^{b\g},x_2+n_2\g)|_{X_i=X}\nonumber\\
&=&\f{1}{\mathcal{N}}\sum_{\substack{a\neq 0\\n_1,n_2,n_3,n_4}}^{}g_1\int_{-\infty}^{\infty}\f{2\nu^2 d\nu \sqrt{\mathcal{C}_{\f{d}{2}-i\nu}\mathcal{C}_{\f{d}{2}+i\nu}}}{2\pi\left[ \nu^2+(\Delta-d/2)^2\right]}\int_{-\infty}^{\infty}\f{\nu'^2}{\pi}  d\nu' \sqrt{\mathcal{C}_{\f{d}{2}+i\nu'}\mathcal{C}_{\f{d}{2}-i\nu'}}\int dX \int d Y \nonumber\\
&\times& \int d x' \int d x \ K_{\D_{\phi_1}}(X^{a\g},x_1+n_1\g) K_{\f{d}{2}+i\nu}(Y^{a\g},x+n_3)(\del_U\del_{X_1})^k (\del_U\del_{X_1})^m K_{\f{d}{2}+i\nu'}(X^{a\g},x')\nonumber\\
&\times&  K_{\f{d}{2}-i\nu'}(Y^{a\g},x')  K_{\f{d}{2}-i\nu}(Y^{a\g},x+n_4)(U.\del_X)^{k+m}
K_{\D_{\phi_2}}(X^{a\g},x_2+n_2\g)|_{X_i=Y=X}
\eeqa

 We act
\beqa
(\del_U\del_{X_1})^k (\del_U\del_{X_1})^m K_{\f{d}{2}+i\nu'}(X^{a\g},x') = 2^{k+m}\left(1+i\nu'\right)_{k+m}(\del_U.x')^{k+m} K_{\f{d}{2}+i\nu'+k+m}(X^{a\g},x') 
\eeqa
and
\beqa
(U.\del_X)^{k+m}
K_{\D_{\phi_2}}(X^{a\g},x_2+n_2\g)=2^{k+m}\left(\D_{\phi_2}+1-\f{d}{2}\right)_{k+m}(U.(x_2+n_2\g))^{k+m}K_{\D_{\phi_2}+k+m}(X^{a\g},x_2+n_2\g)\nonumber\\
\eeqa
Here too $Y^{a\g}$ will be replaced by $Y^{b\g}$ to have the correct $\mathcal{N}$ dependence from the $\d(X,Y)$ function as before.
Thus our integrals from the case of no derivatives get modified as the standard bulk three point integrals for $X$ and $Y$ giving

\beqa
&&\int d X K_{\D_{\phi_1}}(X^{a\g},x_1+n_1\g) K_{\D_{\phi_2}+k+m}(X^{a\g},x_2+n_2\g) K_{\f{d}{2}+i\nu'+k+m}(X^{a\g},x')\nonumber\\
&=&\f{B_{\D_{\phi_1},\D_{\phi_2}+k+m,\f{d}{2}+i\nu'+k+m}}{((x_1+n_1\g-x_2-n_2\g)^2)^{\f{\D_{\phi_1}+\D_{\phi_2}-\f{d}{2}-i\nu'}{2}}((x_2+n_2\g-x')^2)^{\f{-\D_{\phi_2}+\D_{\phi_2}+\f{d}{2}+i\nu'}{2}}((x_1+n_1\g-x')^2)^{\f{\D_{\phi_1}-\D_{\phi_2}+\f{d}{2}+i\nu'}{2}}} \nonumber
\eeqa
and

\beqa
&&\int d Y K_{\f{d}{2}-i\nu'}(Y^{a\g},x') K_{\f{d}{2}+i\nu}(Y^{a\g},x+n_3\g) K_{\f{d}{2}-i\nu}(Y^{a\g},x+n_4\g)
\nonumber\\
&=&\f{B_{\f{d}{2}-i\nu',\f{d}{2}+i\nu,\f{d}{2}-i\nu}}{((x'-x-n_3\g)^2)^{\f{\f{d}{2}-i\nu'+2i\nu}{2}}((n_3\g-n_4\g)^2)^{\f{\f{d}{2}+i\nu'}{2}}((x'-x-n_4\g)^2)^{\f{\f{d}{2}-i\nu'-2i\nu}{2}}} \nonumber
\eeqa
Next, the $x'$ and $x$ integrals need to be performed 

\beqa
&&\f{1}{((x_1+n_1-x_2-n_2)^2)^{\f{\D_1+\D_2-d/2-i\nu'}{2}}((n_3-n_4)^2)^{\f{\f{d}{2}-i\nu'}{2}}}
\int  \f{dx'}{((x_2+n_2\g-x')^2)^{\f{-\D_1+\D_2+\f{d}{2}+i\nu'}{2}}}\nonumber\\
&\times& \f{dx}{((x_1+n_1\g-x')^2)^{\f{\D_1-\D_2+\f{d}{2}+i\nu'}{2}}((x'-x-n_3\g)^2)^{\f{\f{d}{2}-i\nu'+2i\nu'}{2}}((n_3-n_4\g)^2)^{\f{\f{d}{2}+i\nu'}{2}}((x'-x-n_4\g)^2)^{\f{\f{d}{2}-i\nu'-2i\nu}{2}}}\nonumber
\eeqa

We first do the $x$ integral. This is a two point integral and also converts the $x'$ integral to a two point integral. 

\beqa
&&\int \f{dx}{((x'-x-n_3\g)^2)^{\f{\f{d}{2}-i\nu'+2i\nu}{2}}((x'-x-n_4\g)^2)^{\f{\f{d}{2}-i\nu'-2i\nu}{2}}}\nonumber\\
&=& \f{\pi^{\f{d}{2}}\G\left( \f{2 i \nu+\f{d}{2}+i\nu'}{2} \right)\G\left( \f{-2 i \nu + \f{d}{2}+i\nu'}{2} \right)\G(-i\nu')}{\G\left(\f{\f{d}{2}-i\nu'-2 i \nu}{2} \right)\G\left(\f{\f{d}{2}-i\nu'+2 i \nu}{2} \right)\G\left(\f{d}{2}+i\nu'\right)}\f{1}{((n_3-n_4\g)^2)^{-i\nu'}}
\eeqa

Similarly, the $x'$ integral gives

\beqa
&&\int   \f{dx'}{((x_2+n_2\g-x')^2)^{\f{-\D_{\phi_1}+\D_{\phi_2}+\f{d}{2}+i\nu'}{2}}}\f{1}{((x_1+n_1\g-x')^2)^{\f{\D_{\phi_1}-\D_{\phi_2}+\f{d}{2}+i\nu'}{2}}}\nonumber\\
&=&\f{\pi^{\f{d}{2}}\G\left( \f{-\D_{\phi_1}+\D_{\phi_2}+\f{d}{2}-i\nu'}{2} \right)\G\left( \f{\D_{\phi_1}-\D_{\phi_2}+\f{d}{2}-i\nu'}{2} \right)\G(i\nu')}{\G\left(\f{\f{d}{2}+i\nu'-\D_{\phi_2}+\D_{\phi_1} }{2}\right)\G\left(\f{\f{d}{2}+i\nu'+\D_{\phi_2}-\D_{\phi_1}}{2} \right)\G\left(\f{d}{2}-i\nu'\right)}\f{1}{((x_2-x_1+n_2\g-n_2\g)^2)^{-i\nu'}}
\eeqa

Next, we need to perform the sums over $n_i$. The sums over $n_3, n_4$ give $\mathcal{N}^2$, while

\beqa
\sum_{n_1}\f{1}{((x_2-x_1+n_2\g)^2)^{\f{\D_1+\D_2-\f{d}{2}+i\nu'}{2}}}&=&\zeta_{\text{EH}}\left(\D_{\phi_1}+\D_{\phi_2}-d/2+i\nu',\t_{12},x_{12}\right)
\eeqa
For the sum over $n_3$ we may use

\beqa
\sum_{n_3=-\infty}^{\infty}\f{1}{(n_3^2)^{\f{d}{2}-i\nu'}}=\left[1+(-1)^{i\nu'-\f{d}{2}} \right]\zeta\left(\f{d}{2}-i\nu'\right)
\eeqa

The final expression is

\beqa
\mathcal{W}_{1234}&=&g_1\f{2^{k+m}}{32\pi^{\f{d+8}{2}}}\left(1+i\nu'\right)_{k+m}\left(\D_{\phi_2}+1-\f{d}{2}\right)_{k+m}\int_{-\infty}^{\infty}\nu'^2d\nu'  \int_{-\infty}^{\infty}\f{d \nu}{\left[ \nu^2+(\Delta-\f{d}{2})^2\right]}\Gamma\left(\f{\Delta_{\phi_1}+\Delta_{\phi_2}-\f{d}{2}\pm i\nu'}{2}\right)\nonumber\\
&\times&\f{\Gamma\left(\f{\D_{\phi_2}-\D_{\phi_1}+\f{d}{2}-i\nu'}{2}\right)\Gamma\left(\f{-\D_{\phi_2}+\D_{\phi_1}+\f{d}{2}-i\nu'}{2}\right)\Gamma\left(\f{d}{4}+\f{i\nu'\pm 2i\nu}{2}\right)}{\sqrt{\G(\D_{\phi_1})\G(\D_{\phi_2}+k+m)\G(\D_{\phi_1}-\f{d}{2}+1)\G(\D_{\phi_1}-\f{d}{2}+1+k+m)}}\f{\G( -i \nu')}{\G(-i\nu)\G(\f{d}{2}\pm i\nu')}\nonumber\\
&\times& \sqrt{\f{\G(\f{d}{2}+i\nu')\G(\f{d}{2}+i\nu'+k+m)}{\G(1+i\nu')\G(1+i\nu'+k+m)}} \f{\Gamma\left(\f{\f{d}{2}\pm i\nu'}{2}\right)}{\G(1-i\nu')\G(\f{d}{2}+i\nu'+k+m)} \nonumber\\
&\times&\zeta\left(\f{d}{2}-i\nu'\right) \zeta_{\text{EH}}\left(\D_{\phi_1}+\D_{\phi_2}-\f{d}{2}+i\nu',x_2-x_1\right)
\eeqa
This matches with no derivatives case when $k=m=0$. Following a similar procedure we can also evaluate the other three contributions.

\section{Spinning Exchange Diagram }\label{section5}
In this section we extend our analysis to compute the Witten diagram with exchange of a spinning particle. The the cubic vertex, considered in the de Donder gauge, here is written as $g~(\del_X \del_{U_{X}})^J$. Further for the evaluation in this scetion we take the one point interaction vertex to be $g'=g~ \phi_{t_1,t_2,t_3,\cdots, t_s}$ where we have contracted all the indices of the higher spin field $\phi_{\mu_1,\mu_2,\mu_3,\cdots, \mu_s}$ with the thermal AdS Killing vector as using any other directions will break the isometry of thermal AdS. 

\vspace{0.5cm}
\begin{center}
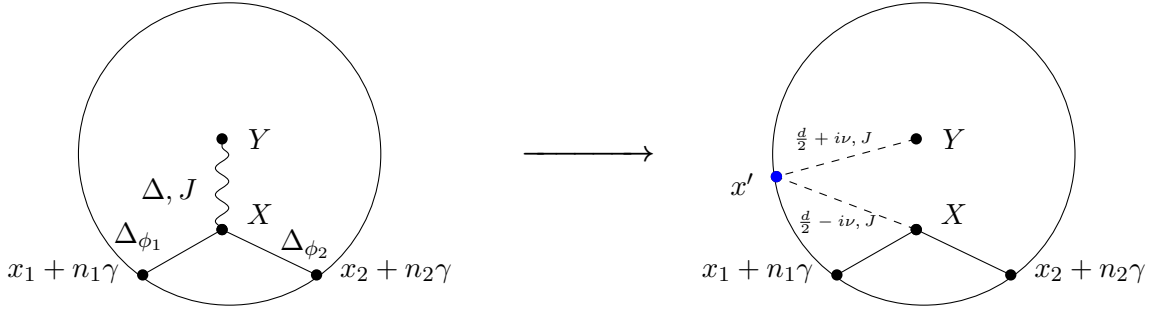

\[\begin{wittendiagram}

  \draw (-1.15,-1.6) node[vertex] -- (-0.1,-1) node[vertex];
 \draw (1.15,-1.6) node[vertex] -- (-0.1,-1) node[vertex];

 \draw [snake it] (-0.1,-1)  -- (-0.1,0.2) node[vertex] ;
  \node at (-2.2, -1.55) {$x_1+n_1\g$};
  \node at (2.2, -1.55) {$x_2+n_2\g$};
  \node at (0.4, 0.2) {$Y$};
  \node at (0.4, -0.8) {$X$};
  \node at (-0.8, -0.5) {$\D,J$};
  \node at (1, -1.2) {$\D_{\phi_2}$};
  \node at (-1.2, -1.1) {$\D_{\phi_1}$};
  
\end{wittendiagram}\hspace{0.3cm}\quad
\bm{\xrightarrow{\hspace{1.5cm}} }\quad
\begin{wittendiagram}
 
  \draw (-1.15,-1.6) node[vertex] -- (-0.1,-1) node[vertex];
 \draw (1.15,-1.6) node[vertex] -- (-0.1,-1) node[vertex];
  \draw[dashed] (-0.1,0.2) node[vertex] -- (-1.95,-0.3) node[vertex];
    \draw[dashed] (-0.1,-1) -- (-1.95,-0.3) node[vertex, color=blue];

   \node at (-2.2, -1.55) {$x_1+n_1\g$};
  \node at (2.2, -1.55) {$x_2+n_2\g$};
  \node at (0.4, 0.2) {$Y$};
  \node at (0.4, -0.8) {$X$};
  \node at (-2.4,-0.4) {$x'$};

    \node at (-1.15, -0.9) {{\tiny $\f{d}{2}-i\nu, J$}};
    \node at (-1.2, 0.2) {{\tiny $\f{d}{2}+i\nu, J$}};
  
\end{wittendiagram}\]
\captionof{figure}{The spin $J$ exchange diagram in thermal AdS with details of the split representation and the thermal arrangement of points. Thermally identified points are shown in black, non identified points are shown in blue.}\label{fig15}
\end{center}

\vspace{0.5cm}

We will use the relations presented in section \ref{section2} and our diagram then becomes, where we have already done the thermal re-arrangement like the previous cases,

\beqa
\mathcal{W}&=&\f{-g g'}{\mathcal{N}}\int_{TAdS}dX dY \sum_{\substack{a,b \neq 0\\n_1,n_2}}G_{\D,J}(X^{a\g},U_{X^{a\g}},Y^{b\g},U_{Y^{b\g}})K_{\Delta_{\phi_1},0}(X^{a\gamma},x_1+n_1\gamma)\nonumber\\
&\times&(\del_{X^{a\g}} \del_{U_{X^{a\g}}})^J K_{\Delta_{\phi_2},0}(X^{a\gamma},x_2+n_2\gamma)\\
&=&\f{-g g'}{\mathcal{N}}\int_{-\infty}^{\infty}\f{2\nu^2 d\nu}{2\pi\left[ \nu^2+(\Delta-d/2)^2\right]}\int_{TAdS}dX dY\int d x \sum_{\substack{a,d \neq 0\\n_1,n_2}}K_{\Delta_{\phi_1},0}(X^{a\gamma},x_1+n_1\gamma)\nonumber\\
&\times&(\del_{X^{a\g}} \del_{U_{X^{a\g}}})^J K_{\Delta_{\phi_2},0}(X^{a\gamma},x_2+n_2\gamma)K_{\f{d}{2}+i\nu',J}(Y^{b\gamma},U_{Y^{b\g}},x,\hat{\del}_Z)K_{\f{d}{2}-i\nu,J}(X^{a\gamma},U_{X^{a\g}},x,Z) \nonumber
\eeqa

Next use

\beqa
K_{\f{d}{2}+i\nu,J}(X^{a\g},x)=\f{1}{\left(\f{d}{2}+i\nu-1\right)_J}\mathcal{D}_x(Z,U_{X^{a\g}})^J K_{\f{d}{2}+i\nu,0}(X^{a\g},x)
\eeqa
and

\beqa
(\partial_{X^{a\g}}.\partial_{U_{X^{a\g}}})^J K_{\D_{\phi_2}}(X^{a\g},x_2+n_2\g)K_{\f{d}{2}+i\nu,J}(X^{a\g},x) &=&2^J~\mathcal{D}_x(Z,x_2+n_2\g)^J\f{\left(1-\f{d}{2}+\D_{\phi_2}\right)_J}{\left(\f{d}{2}+i\nu-1\right)_J}\nonumber\\
&\times&  K_{\D_{\phi_2+J}}(X^{a\g},x_2+n_2\g)K_{\f{d}{2}+i\nu}(X^{a\g},x)\nonumber
\eeqa
The $Y$ integral can actually be done following the normalization procedure of the bulk to boundary propagator similar to the scalar case  and following\cite{Giombi:2017hpr}\cite{Aharony:2020omh}\cite{Aharony:2021ovo} to write

\beqa\label{spinJY}
 \frac{1}{\mathcal{N}}\int dY  K_{\f{d}{2}-i\nu, J}(Y^{b\g},U_{Y^{b\g}},x',\hat{\partial}_Z)& =& \frac{1}{\mathcal{N}} \frac{(Z'.\hat{\del}_Z)^J}{\sqrt{C_{\D,J}}} \frac{1}{(-i \nu)}\int_0^{\infty}z^{(i\nu -\f{d}{2}-J-1)}\nonumber\\
 &=& \frac{1}{\mathcal{N}}\frac{(Z'.\hat{\del}_Z)^J}{\sqrt{\mathcal{C}_{\D,J}}} \frac{1}{(-i \nu)} \frac{4 d}{(d-2J+2i\nu)(d-2J-2i\nu)}
\eeqa

where $Z'$ is the boundary value of $U_Y$ and the $z$ integral again has been done by splitting the integral into two parts and analytical continuation as in the scalar case. Also, $\mathcal{C}_{\D,J}$ is the corresponding normalization for the bulk to boundary propagator. As a check, if we put s=0 in the above expression we retrieve the scalar  (with $\a=0$):

\beqa
\frac{(Z'.\hat{\del}_Z)^J}{\sqrt{\mathcal{C}_{\D,J}}} \frac{1}{(-i \nu)} \frac{4 d}{(d-2J+2i\nu)(d-2J-2i\nu)} \rightarrow \frac{1}{\sqrt{\mathcal{C}_{\D}}} \frac{1}{(-i \nu)} \frac{4 d}{d^2 +4 \nu^2}
\eeqa

The second component of the diagram is

\beqa
 \int dX \int dx~ \mathcal{D}_{x}(x_2+n_2\g,Z)^JK_{\Delta_{\phi_2+J,0}}(X^{a\gamma},x_2+n_2\gamma)
 K_{\Delta_{\phi_1,0}}(X^{a\gamma},x_1+n_1\gamma) K_{\f{d}{2}+i\nu}(X^{a\gamma},x)
\eeqa \\

Next we can do the $X$ bulk integral as before (except that $\D_{\phi_2} \rightarrow \D_{\phi_2}+J$ now). The boundary integrals should be done after acting with $\mathcal{D}_x(x_2,Z)^J$. Thus we get  

\beqa
\mathcal{W}&=&gg'\sum_{a,c,n_1,n_2=-\infty}^{\infty}\int_{-\infty}^{\infty}\f{1}{\mathcal{N}^3}\f{ d\nu}{\left[ \nu^2+(\Delta-d/2)^2\right]}\f{4 d}{(d-2J+2i\nu)(d-2J-2i\nu)} \ B_{\Delta_{\phi_1},\Delta_{\phi_2+J},\f{d}{2}+i\nu}2^{J}\left(\D_2+1-\f{d}{2} \right)_J\nonumber\\
&\times&  \frac{-g g'}{\left(\f{d}{2}+i\nu-1 \right)_J}(\mathcal{D}_x(x_2+n_2\g,Z)^J)\left(\f{1}{((x_1-x_2+n_1\g-n_2\g)^2)^{\f{\Delta_{\phi_1}+\Delta_{\phi_2}+J-\f{d}{2}-i\nu}{2}}} \right.\nonumber\\
&\times& \left. \int  \f{dx}{((x_2-x+n_2\g)^2)^{\f{-\Delta_{\phi_1}+\Delta_{\phi_2}+\f{d}{2}+i\nu}{2}}((-x+x_1+n_1\g)^2)^{\f{\Delta_{\phi_1}-\Delta_{\phi_2}+\f{d}{2}+i\nu}{2}}}\right)
\eeqa \\

and again here too the expressions reduce to the scalar case when we put $s=0$. Next, we act with $(\mathcal{D}_x(x_2,Z)^J)$ on the coordinate dependent part to get

\beqa \label{2ndint}
&&\left(\f{\D_{\phi_1}-\D_{\phi_2}-s+\f{d}{2}+i\nu}{2}\right)_J\nonumber\\
&\times&\int d x \f{\left((Z.(x_1+n_1))((x_2+n_2)(x))-(Z.(x_2+n_2))((x_1+n_1)(x))\right)^J}{((x_1-x_2+n_1\g-n_2\g)^2)^{\f{\Delta_{\phi_1}+\Delta_{\phi_2}+J-\f{d}{2}-i\nu}{2}}((x_2-x+n_2\g)^2)^{\f{-\Delta_{\phi_1}+\Delta_{\phi_2}+\f{d}{2}+i\nu}{2}}((-x+x_1+n_1\g)^2)^{\f{\Delta_{\phi_1}-\Delta_{\phi_1}+\f{d}{2}+i\nu}{2}}} \nonumber
\eeqa \\

Now we need to evaluate the boundary integral

\beqa
I=\int  \f{\left((Z.(x_1+n_1\g))((x_2+n_2).x)-(Z.(x_2\g+n_2))((x_1+n_1\g).x))\right)^J dx}{((x_1-x_2+n_1\g-n_2\g)^2)^{\f{\Delta_{\phi_1}+\Delta_{\phi_2}+J-\f{d}{2}-i\nu}{2}}((x_2-x+n_2\g)^2)^{\f{-\Delta_{\phi_1}+\Delta_{\phi_2}+\f{d}{2}+i\nu}{2}}((-x+x_1+n_1\g)^2)^{\f{\Delta_{\phi_1}-\Delta_{\phi_1}+\f{d}{2}+i\nu}{2}}} \nonumber
\eeqa \\

At this stage act with $(Z'.\hat{\partial}_Z)^s$ from equation \ref{spinJY} on the numerator following \cite{Giombi:2017hpr}
\begin{equation}
(a \cdot \hat{\partial}_z)^k (b \cdot z)^k
=
\frac{k!}{2^k \left(\frac{d}{2}-1\right)_k}
\left(a^2 b^2\right)^{k/2}
C_k^{\left(\frac{d}{2}-1\right)}
\left(
\frac{a\cdot b}{\sqrt{a^2 b^2}}
\right),
\end{equation}
which for us (using $P_i^2=0$) on denoting
\beqa
a&=&Z' \nonumber\\
b&=&(((x_1+n_1\g))((x_2+n_2\g).x)-((x_2+n_2\g))((x_1+n_1\g).x))^J
\eeqa
gives
\beqa
(a \cdot \hat{\partial}_z)^J (b \cdot z)^J
&=&
\frac{J!}{2^J \left(\frac{d}{2}-1\right)_J}
\left(Z'^2((x_1+n_1\g).x)((x_2+n_2\g).x)((x_1+n_1\g).(x_2+n_2\g))^2\right)^{J/2}\nonumber\\
&\times& C_J^{\left(\frac{d}{2}-1\right)}
\left(
\frac{(Z'.(x_1+n_1\g))((x_2+n_2\g).x)-(Z'.(x_2+n_2\g))((x_1+n_1\g).x)}{\sqrt{Z'^2 (((x_1+n_1\g).x)((x_2+n_2\g).x)((x_1+n_1\g).(x_2+n_2\g))) ^2}}
\right)
\eeqa
At this stage we can define
\beqa
x_4=(Z'.(x_1+n_1\g))(x_2+n_2\g)-(Z'.(x_2+n_2\g))(x_1+n_1\g)
\eeqa
and then expand the  Gegenbauer polynomial as 
\begin{equation}
C_n^{(\alpha)}(z)
=
\sum_{k=0}^{\lfloor n/2 \rfloor}
(-1)^k
\frac{\Gamma(n-k+\alpha)}
{\Gamma(\alpha)\,k!\,(n-2k)!}
(2z)^{\,n-2k}.
\end{equation}
Thus in our case we get the following series
\beqa
&&C_J^{\left(\frac{d}{2}-1\right)}
\left(
\frac{((Z'.(x_1+n_1))((x_2+n_2).x)-(Z'.(x_2+n_2))((x_1+n_1).x))}{\sqrt{Z'^2 (((x_1+n_1).x)((x_2+n_2).x)((x_1+n_1).(x_2+n_2))) ^2}}\right)\nonumber\\
&=&\sum_{k=0}^{\lfloor J/2 \rfloor} (-1)^k
\frac{\Gamma(J-k+d/2-1)}
{\Gamma(\f{d}{2}-1)\,k!\,(J-2k)!}(x_4.x)^{J-2k} (Z'^2)^{2k-J} \left(\f{2}{Z'}\right)^{J-2k} \nonumber\\
&\times&(((x_1+n_1\g).x)((x_2+n_2\g).x)((x_1+n_1\g).(x_2+n_2\g)))^{2k-J}
\eeqa

Finally, following \cite{Liu:2018jhs}, the above can be written and evaluated as, treating $J$ as a parameter for now,

\beqa
\int \f{\left((Z.(x_1+n_1))((x_2+n_2).x)-(Z.(x_2+n_2))((x_1+n_1).x))\right)^J~dx}{(((x_2+n_2\g).x)^2)^{\f{-\Delta_{\phi_1}+\Delta_{\phi_2}+J+\f{d}{2}+i\nu-2k}{2}}((x.(x_1+n_1\g))^2)^{\f{\Delta_{\phi_1}-\Delta_{\phi_2}-J+\f{d}{2}+i\nu-2k}{2}}} \nonumber
\eeqa \\

We can write the numerator as $x.x_{4}$ and note the property $x_4^2=0$. Thus we can evaluate this three point and club all the factors to get
 \beqa
\mathcal{W} &=&\sum_{n_1}\int\f{d\nu}{\nu^2+\left(\D-\f{d}{2}\right)^2}\f{gg'}{\sqrt{\mathcal{C}_{\f{d}{2}-i\nu}}}\f{4 B_{\D_{\phi_1},\D_{\phi_2}+J,\f{d}{2}+i\nu}}{(d-2J+2i\nu)(d-2J-2i\nu)} \nonumber\\
 &\times& \f{\left(\f{\D_{\phi_1}-\D_{\phi_2}-J-\f{d}{2}+i\nu}{2}\right)_J}{\left(\f{d}{2}+i\nu-1\right)_J(\f{d}{2}-1)_J} \f{\G\left(\f{4J+2i\nu-4k}{2}\right)\G\left(\f{-\D_{\phi_1}+\D_{\phi_2}+i\nu-d/2+2J}{2}\right)\G\left(\f{\D_{\phi_2}-\D_{\phi_1}+\f{d}{2}-i\nu+J-2k}{2}\right)}{\G(J-2k)\G\left(\f{-\D_{\phi_1}+\D_{\phi_2}+J+d/2+i\nu}{2}\right)\G\left(\f{\D_{\phi_1}-\D_{\phi_2}+J+\f{d}{2}+i\nu}{2}\right)}\nonumber\\
 &\times& \sum_{k=0}^{\lfloor J/2 \rfloor}\f{J!(-1)^k\Gamma(J-k+d/2-1)}{\G(d/2-1)k!(J-2k)!} \f{(Z'.(x_1+n_1)^{\f{\D_{\phi_2}-\D_{\phi_1}+\f{d}{2}-i\nu-J-2k}{2}}(Z'.(x
 _2+n_2))^{\f{\D_{\phi_2}-\D_{\phi_1}+i\nu-\f{d}{2}+J+2k}{2}}}{((x_1+n_1\g-x_2-n_2\g)^2)^{\f{\D_{\phi_1}+\D_{\phi_2}+6J+i\nu-8k}{2}}} \nonumber
 \eeqa

\section{Effective bulk vertices from scalar loops}\label{section6}

As discussed in section (\ref{section2}) the coincident point thermal bulk to bulk propagator can be written as
\beqa
G^\b_{\D}(X,\g^n X)=\sum_{n \neq 0}\mathcal{C}_\D \left(\f{n}{z}\right)^{-2\D} ~{}_2F_1\left(\D,\D-\f{d}{2}+\f{1}{2},2\D-d+1,-\f{4 z^2}{n^2}\right)
\eeqa
This hypergeometric function can then be expanded in a series allowing us to write
\beqa\label{effver}
G_{\Delta}^{\beta}(X, \gamma^n X) &=& \mathcal{C}_{\Delta} \sum_{n \neq 0} \sum_{k=0}^{\infty} \frac{(\Delta)_k \left(\Delta - \frac{d}{2} + \frac{1}{2}\right)_k}{(2\Delta - d + 1)_k k!} (-4)^k n^{-2\Delta -2 k} z^{2\Delta + 2k} \nonumber\\
 &=& \mathcal{C}_{\Delta}  \sum_{k=0}^{\infty} \frac{(\Delta)_k \left(\Delta - \frac{d}{2} + \frac{1}{2}\right)_k}{(2\Delta - d + 1)_k k!} (-4)^k \zeta(2\D+2k)z^{2\Delta + 2k} \nonumber\\
&=& \mathcal{A} ~ z^\a
\eeqa

This coincident point bulk to bulk propagator thus has an effect of introducing a modified effective vertex $\mathcal{A}~z^\alpha$ into a Witten diagram. This we also verify using the split representation applied to the bulk loop in appendix \ref{effverapp}. In the sections that follow we investigate how this affects our diagrams.

\subsection{Scalar loop and $a_\chi$}\label{chi1}

We evaluate the one-point Witten diagram shown in figure \ref{figure7} which helps us to evaluate the thermal one point coefficient of $\mathcal{O}_\chi$.

\[\begin{wittendiagram}
\draw[color=black, fill=white,](-0.1,0) circle (0.5);
   \draw[dashed] (-0.1,-0.5) node[vertex] -- (-0.1,-2) node[vertex];
  \node at (0.3, -0.8) {$Y$};
  \node at (-0.2,-2.4) {$x'$};
  \node at (-0.5, -1.3) {$\D_\chi$};
  \node at (-0.2,0.9) {$\D$};
\end{wittendiagram}\]

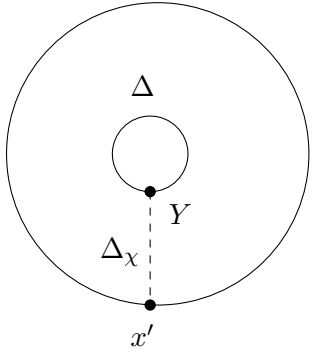
\captionof{figure}{The tadpole diagram that generates $a_\chi$.} \label{figure7}

\vspace{0.5cm}

The $\mathcal{O}(g)$ boundary one point function is evaluated in Poincare coordinates as follows
\beqa
\mathcal{W}=\langle\mathcal{O}_\chi\rangle_\b &=&\f{1}{\mathcal{N}}\sum_{m,n \neq 0}\int dY K_{\D_\chi}(Y^{m\g},x')G_{\D}(Y^{n\g},Y)\nonumber\\
&=&\f{1}{\mathcal{N}}\sum_{n,m \neq 0}\sqrt{\mathcal{C}_{\D_\chi}}\mathcal{C}_{\D}\int_0^\infty \f{d z}{z^{d+1}}\int d^d x~ \left(\f{z}{n}\right)^{2\D} \left(\f{z}{z^2+|x_m-x'|^2}\right)^{\Delta_\chi}\nonumber\\
&\times& ~{}_2F_1\left(\D,\D-\f{d}{2}+\f{1}{2},2\D-d+1,-4 \f{z^2}{n^2}\right)\nonumber\\
\eeqa
Now the bulk to bulk propagator does not involve any $x$ so we can do this integral separately by shifting $z$ by $x'$. The factor of $\mathcal{N}$ then cancels the sum over $m$ and we have
\beqa
\int d^d x \left(\f{z}{z^2+|x-x'|^2}\right)^{\Delta_\chi}= (z)^{d-\D_1}\pi^{\f{d}{2}}\f{\Gamma\left(\D_\chi-\f{d}{2}\right)}{\Gamma(\D_\chi)}
\eeqa
Next we do the integral over $z$ to write
\beqa
\mathcal{M}&=&\int_0^\infty \f{dz}{z^{d+1}}z^{d-\D_\chi} \left(\f{z}{n}\right)^{2\D}  ~{}_2F_1\left(\D,\D-\f{d}{2}+\f{1}{2},2\D-d+1,-4 \f{z^2}{n^2}\right) \nonumber \\
&=&  \f{2^{\D_\chi-d-1}\Gamma\left(\f{\D_\chi}{2}\right)\Gamma\left(\f{2\D-\D_\chi}{2}\right)\Gamma\left(\f{1+\D_\chi-d}{2}\right)\G\left(1+\D-\f{d}{2}\right)}{\sqrt{\pi}\Gamma(\D)\Gamma\left(1+\D+\f{\D_\chi}{2}-d\right)}
\eeqa
while the sum over $n$ gives $\zeta(\D)$. Thus collecting all terms we have the diagram being evaluated to
\beqa
\mathcal{W}=\langle\mathcal{O}_\chi\rangle^{(1)}=b_{\chi}^{(1)}=\pi^{\f{d}{2}}\zeta(\D)\sqrt{\mathcal{C}_{\D_{\chi}}}\mathcal{C}_{\D}\f{\Gamma\left(\D_\chi-\f{d}{2}\right)}{\Gamma(\D_\chi)}\mathcal{M}
\eeqa
Thus we may read off the thermal OPE coefficient as
\beqa
a_\chi=f_{\phi\phi\chi}^{(1)}b_{\chi}^{(1)}
\eeqa
where we have from the zero temperature relation
\beqa
f_{\phi\phi\chi}=B(\D_\phi,\D_\phi,\D_\chi)
\eeqa

\subsection{Scalar Exchange Diagram with modified vertex}
\vspace{1cm}
\[\begin{wittendiagram} 
 \draw (-1.15,-1.6) node[vertex] -- (-0.1,-1) node[vertex];
 \draw (1.15,-1.6) node[vertex] -- (-0.1,-1) node[vertex];
 \draw (-0.1,-1) node[vertex] -- (-0.1,0) node[vertex]node[vertex, red];
  \node at (-2.1, -1.7) {$x_1+n_1\g$};
  \node at (2.1, -1.7) {$x_2+n_2\g$};
  \node at (0.4, -0.2) {$Y$};
  \node at (0.4, -0.8) {$X$};
      \node at (-1.1, -1.2) {$\D_{\phi_1}$};
  \node at (1, -1.2) {$\D_{\phi_2}$};
    \node at (-0.5, -0.5) {$\D$};
\end{wittendiagram}\hspace{0.3cm}\quad
\bm{\xrightarrow{\hspace{1cm}} }
\quad
\hspace{0.8cm}
\begin{wittendiagram}
 \draw (-1.15,-1.6) node[vertex] -- (-0.1,-1) node[vertex];
 \draw (1.15,-1.6) node[vertex] -- (-0.1,-1) node[vertex];
 \draw[dashed] (-0.1,0) node[vertex, red] -- (-1.95,-0.45) node[vertex];
    \draw[dashed] (-0.1,-1) node[vertex] -- (-1.95,-0.45) node[vertex, color=blue];
  \node at (-2.1, -1.7) {$x_1+n_1\g$};
  \node at (2.1, -1.7) {$x_2+n_2\g$};
  \node at (0.4, -0.1) {$Y$};
  \node at (0.4, -0.8) {$X$};
  \node at (-2.4,-0.4) {$x'$};
\end{wittendiagram}\]

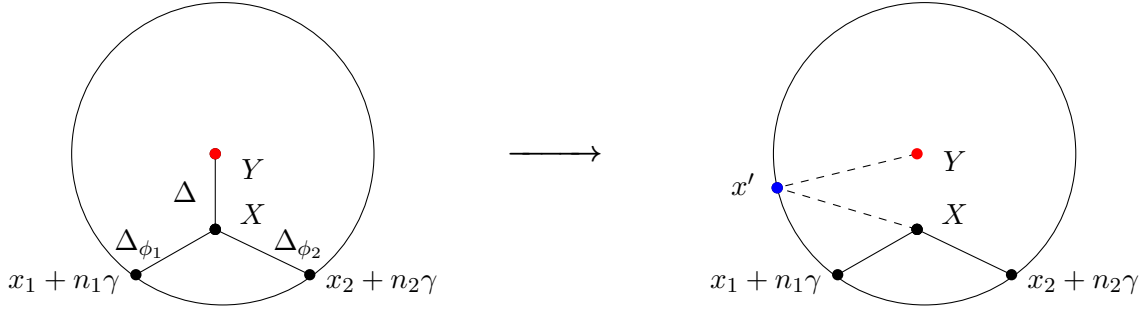
\captionof{figure}{The scalar exchange diagram with the effective vertex, shown in red, in thermal AdS. The thermally identified points are given in black, non identified points are in blue and the effective vertex in red.}

\vspace{0.5cm}

Using the description in equation \ref{effver} the diagram becomes

\beqa \label{Y0}
\mathcal{W}&=&\f{1}{\mathcal{N}^3}\sum_{\substack{a,b,c,d \neq 0\\n_1,n_2}}^{}~g^2~\int_{TAdS} dX dY \mathcal{A}~z^\alpha ~ K_{\Delta_{\phi_1}}(X^{a\gamma},x_1+n_1\gamma)K_{\Delta_{\phi_2}}(X^{b\gamma},x_2+n_2\gamma)G_{\Delta}(X^{c\gamma},Y^{d\gamma})\nonumber\\
&=&\f{1}{\mathcal{N}^3}\sum_{\substack{a,b,c,d \neq 0\\n_1,n_2}}^{}~g^2~\int_{TAdS}dX dY \mathcal{A}z^\alpha ~ K_{\Delta_{\phi_1}}(X^{a\gamma},x_1+n_1\gamma)K_{\Delta_{\phi_2}}(X^{a\gamma},x_2+(n_2+b-a)\gamma)G_{\Delta}(X^{a\gamma},Y^{(d+c-a)\gamma})\nonumber\\
&=&\f{1}{\mathcal{N}}\sum_{\substack{a,c \neq 0\\n_1,n_2}}^{}~g^2~\int_{TAdS}dX dY \mathcal{A}~z^\alpha ~ K_{\Delta_{\phi_1}}(X^{a\gamma},x_1+n_1\gamma)K_{\Delta_{\phi_2}}(X^{a\gamma},x_2+n_2\gamma)G_{\Delta}(X^{a\gamma},Y^{c\gamma})\nonumber\\
&=&\f{1}{\mathcal{N}}\sum_{\substack{a,c \neq 0\\n_1,n_2}}^{}~g^2~\int_{TAdS}dX dY \mathcal{A}~z^\alpha ~ \int d x'\int_{-\infty}^{\infty}\f{\nu^2 d\nu \sqrt{\mathcal{C}_{\f{d}{2}-i\nu}\mathcal{C}_{\f{d}{2}+i\nu}}}{\left[ \nu^2+(\Delta-d/2)^2\right]}K_{\Delta_{\phi_1}}(X^{a\gamma},x_1+n_1\gamma)\nonumber\\
&\times&K_{\Delta_{\phi_2}}(X^{a\gamma},x_2+n_2\gamma)K_{\f{d}{2}+i \nu}(X^{a\gamma},x')K_{\f{d}{2}-i \nu}(Y^{c\gamma},x')\nonumber\\
\eeqa

where $\a=2\D+2k$. To do the bulk integrals we follow the same set of steps as in section \ref{opetadpole}. For the $Y$ integral we have

\beqa
&&\f{1}{\mathcal{N}}\int dY z^\alpha ~ K_{\f{d}{2}-i \nu}(Y,x')\nonumber\\
&=&\f{1}{\mathcal{N}}\left(\f{\Gamma\left(\f{d}{2}-i\nu\right)}{2\pi^{\f{d}{2}}\Gamma(1-i\nu)}\right)^{\f{1}{2}} \int_0^{\infty}\f{d z ~ z^\alpha}{(z)^{d+1}}\int d^{d} x \left(\f{z}{z^2+|x-x'|^2}\right)^{\f{d}{2}-i\nu}\nonumber\\
&=&\f{1}{\mathcal{N}} \left(\f{\Gamma\left(\f{d}{2}-i\nu\right)}{\pi^{\f{d}{2}}}\right)^{-\f{1}{2}}\frac{\G(-i\nu)}{\sqrt{\G(1-i\nu)}} \int_{0}^{\infty}d z (z)^{\alpha+i\nu-d/2-1} \nonumber \\
& =& \f{1}{\mathcal{N}} \left(\f{2\Gamma\left(\f{d}{2}-i\nu\right)}{\pi^{\f{d}{2}}}\right)^{-\f{1}{2}}\frac{\G(-i\nu)}{\sqrt{\G(1-i\nu)}} \f{4 d}{(d+2\a+2i\nu)(d-2\a-2i\nu)}
\eeqa
Where again we first did the $x$ integral by shifting $x$ by $x'$  and then do the $z$ integral by splitting the integral in two parts and then analytically continuing in $d$. The $X$ integral whereas is similar to that in the previous case and gives

\beqa
&&\f{1}{\mathcal{N}}\int d X ~ K_{\Delta_{\phi_1}}(X,x_1+n_1) K_{\Delta_{\phi_2}}(X,x_2+n_2)K_{\f{d}{2}+i\nu}(X,x')\\
&=&\f{1}{\mathcal{N}}\f{B_{\Delta_{\phi_1},\Delta_{\phi_2},\f{d}{2}+i\nu}}{((x_1-x_2+n_1\g-n_2\g)^2)^{\f{\Delta_{\phi_1}+\Delta_{\phi_2}-\f{d}{2}-i\nu}{2}}((x_2-x'+n_2\g)^2)^{\f{-\Delta_{\phi_1}+\Delta_{\phi_2}+\f{d}{2}+i\nu}{2}}((x'-x_1-n_1\g)^2)^{\f{\Delta_{\phi_1}-\Delta_{\phi_2}+\f{d}{2}+i\nu}{2}}}\nonumber
\eeqa

This leaves us with the following expression

\beqa
\mathcal{W}&=&\sum_{a,c,n_1,n_2=-\infty}^{\infty}~g^2~\int_{-\infty}^{\infty}\f{1}{\mathcal{N}^3}\f{d\nu}{\left[ \nu^2+(\Delta-\f{d}{2})^2\right]}\f{4 d }{d^2+4(\alpha+i \nu)^2}\f{ \mathcal{A} ~B_{\Delta_{\phi_1},\Delta_{\phi_2},\f{d}{2}+i\nu}}{((x_1-x_2+n_1\g-n_2\g)^2)^{\f{\Delta_{\phi_1}+\Delta_{\phi_2}-\f{d}{2}-i\nu}{2}}} \nonumber\\
&\times&\int dx' \f{1}{((x_2-x'+n_2\g)^2)^{\f{-\Delta_{\phi_1}+\Delta_{\phi_2}+\f{d}{2}+i\nu}{2}}((-x'+x_1+n_1\g)^2)^{\f{\Delta_{\phi_1}-\Delta_{\phi_2}+\f{d}{2}+i\nu}{2}}}
\eeqa

At this stage we again observe that the sums over $a,c$ cancel the factors of $\mathcal{N}^2$ in the denominator as $\mathcal{N}=|\mathbb{Z}|.$ Also the 2 point boundary integral can be evaluated to give

\beqa
\mathcal{W}&=&\f{g^2}{8\pi^{\f{d+2}{2}}}\sum_{n_1,n_2=-\infty}^{\infty}\f{1}{\mathcal{N}}\int_{-\infty}^{\infty}\f{ d\nu}{\left[ \nu^2+(\Delta-\f{d}{2})^2\right]}\f{ d~ \mathcal{A}}{(d+2\a+2i\nu)(d+2\a-2i\nu)} \nonumber\\
&\times&\f{\Gamma\left(\f{\D_{\phi_1}+\D_{\phi_2}-\f{d}{2}+i\nu}{2}\right)\Gamma\left(\f{\D_{\phi_1}-\D_{\phi_2}+d/2-i\nu}{2}\right)\Gamma\left(\f{-\D_{\phi_1}+\D_{\phi_2}+d/2-i\nu}{2}\right)\Gamma\left(\f{\D_{\phi_1}+\D_{\phi_2}-d/2-i\nu}{2}\right)}{\sqrt{\Gamma(\D_{\phi_1}-\f{d}{2}+1)\Gamma(\D_{\phi_2}-\f{d}{2}+1)\Gamma(\D_{\phi_1})\Gamma(\D_{\phi_2})}\G\left(\f{d}{2}-i\nu\right)} \nonumber \\
&\times&\f{1}{(((x_1-x_2+n_1\g-n_2\g))^2)^{\f{\D_{\phi_1}+\D_{\phi_2}-\f{d}{2}+i\nu}{2}}}
\eeqa

Again by shifting the $n_2$ sum we can write

\beqa
\f{1}{\mathcal{N}}\sum_{n_1,n_2=-\infty}^{\infty}\f{1}{((x_1-x_2+n_1\g-n_2\g)^2)^{\f{\D_{\phi_1}+\D_{\phi_2}-\f{d}{2}+ i\nu}{2}}} =\zeta_{\text{EH}}\left(\D_{\phi_1}+\D_{\phi_2}-\f{d}{2}+ i \nu, \t_{12},\mathbf{x}_{12}\right)
\eeqa

This finally gives the two-point diagram to be

\beqa \label{finitet}
\mathcal{W}&&\f{g^2}{8\pi^{\f{d+2}{2}}}\int_{-\infty}^{\infty}\f{ d\nu}{\left[ \nu^2+(\Delta-\f{d}{2})^2\right]}\f{ \mathcal{A}~d}{ (d+2\a+2i\nu)(d+2\a-2i\nu)} \nonumber \\
&\times&\f{\Gamma\left(\f{\D_{\phi_1}+\D_{\phi_2}-\f{d}{2}+i\nu}{2}\right)\Gamma\left(\f{\D_{\phi_1}-\D_{\phi_2}+d/2-i\nu}{2}\right)\Gamma\left(\f{-\D_{\phi_1}+\D_{\phi_2}+\f{d}{2}-i\nu}{2}\right)\Gamma\left(\f{\D_{\phi_1}+\D_{\phi_2}-\f{d}{2}-i\nu}{2}\right)}{\sqrt{\Gamma(\D_{\phi_1}-d/2+1)\Gamma(\D_{\phi_2}-d/2+1)\Gamma(\D_{\phi_1})\Gamma(\D_{\phi_2})}\G\left(\f{d}{2}-i\nu\right)}\nonumber \\
&\times&\zeta_{\text{EH}}\left(\D_{\phi_1}+\D_{\phi_2}-\f{d}{2}+ i \nu, \t_{12},\mathbf{x}_{12}\right)
\eeqa

Thus we get a KMS invariant expression. We now need to do the spectral integral and for this we close the integration contour in the UHP.\\

\noindent
\textbf{Pole Structure in the Upper Half-Plane:}\\

\noindent
We get three distinct families of poles in the UHP:

\begin{enumerate}
    \item \textbf{Double-Trace Pole Towers ($\nu_1$):} That come from the argument of $\Gamma\left(\frac{\D_{\phi_1}+\D_{\phi_2} - \f{d}{2} + i\nu}{2}\right)$. These poles occur when
    \begin{equation}
   \f{ \D_{\phi_1}+\D_{\phi_2} - \f{d}{2} + i\nu}{2} = -m \implies \nu_1 = i\left(\D_{\phi_1}+\D_{\phi_2} - \f{d}{2}  + 2m\right), \quad m \in \mathbb{N}
    \end{equation}
    
    \item \textbf{Tadpole Operator Pole ($\nu_2$):} These come from the polynomial term $d+2\a+2i\nu$ vanishing in the denominator and occur when
    \begin{equation}
    \nu_2 = i\left(\alpha - \frac{d}{2}\right)
    \end{equation}
    
    \item \textbf{Physical poles ($\nu_3$):} These are Dictated by the exchange of $\Delta$ occuring when
    \begin{equation}
    \nu_3 = i\left(\Delta - \frac{d}{2}\right)
    \end{equation}
\end{enumerate}
We notice that when the parameters align such that the double-trace tower merges with the tadpole mixing factor, a double pole forms.
\begin{equation}
\nu_1 = \nu_2 \implies \D_{\phi_1}+\D_{\phi_2} - \f{d}{2} + 2m = \alpha -\f{d}{2} \implies \D_{\phi_1}+\D_{\phi_2} + 2m = \alpha
\end{equation}

Going ahead we use the notation $\D_{12}=\D_{\phi_1}-\D_{\phi_2}$ and also define the $\nu$ independent terms to be clubbed in
\beqa
\mathcal{A'}=\f{-g^2}{4\pi^{\f{d+2}{2}}}\f{\mathcal{A}}{\sqrt{\Gamma\left(\D_{\phi_1}-\f{d}{2}+1\right)\Gamma\left(\D_{\phi_2}-\f{d}{2}+1)\Gamma(\D_{\phi_1}\right)\Gamma(\D_{\phi_2})}}
\eeqa
We first compute the residues at the above poles when $\alpha \neq S + 2m$, i.e when the poles are isolated and simple. Evaluating the contour integration using the standard Residue Theorem $\oint d\nu f(\nu) = 2\pi i \sum \text{Res} f(\nu)$ gives the following residues

\begin{enumerate}

\item \textbf{Double-Trace Sector}
Evaluating at $\nu_1=\D_{\phi_1}+\D_{\phi_2}-\f{d}{2}+2m$ we get
\beqa
&&\mathcal{W}_{\text{double-trace}}^{(m)} = \frac{\pi d \mathcal{A}'(-1)^m \Gamma(\Delta_1+m)\Gamma(\Delta_2+m)\Gamma(\D_{\phi_1}+\D_{\phi_2}+m-\frac{d}{2})}{m! \, \Gamma(\D_{\phi_1}+\D_{\phi_2}+2m) (d+\alpha-\D_{\phi_1}+\D_{\phi_2}-2m)(\D_{\phi_1}+\D_{\phi_2}+2m-\alpha)}\nonumber\\
&\times& \f{1}{(\Delta-\D_{\phi_1}+\D_{\phi_2}-2m)(\Delta+\D_{\phi_1}+\D_{\phi_2}+2m-d)} |x|^{2m}
\eeqa

\item \textbf{Tadpole Sector}
Evaluating the residue at $\nu_2 = i(\alpha - h)$ gives
\begin{equation}
\mathcal{W}_{\text{tadpole}} = -\frac{\pi}{2}\mathcal{A}' \frac{\Gamma\left(\frac{\alpha \pm \Delta_{12}}{2}\right)\Gamma\left(\frac{\D_{\phi_1}+\D_{\phi_2}+\alpha-d}{2}\right)\Gamma\left(\frac{\D_{\phi_1}+\D_{\phi_2}-\alpha}{2}\right)}{\Gamma(\alpha) (\Delta-\alpha)(\Delta+\alpha-d)} |x|^{\alpha-\D_{\phi_1}+\D_{\phi_2}}
\end{equation}

\item \textbf{Physical exchange}
Evaluating at $\nu_3 = i(\Delta - h)$, the the residue yields
\begin{equation}
\mathcal{W}_{\text{single-trace}}=\frac{\G(\f{\D_{\phi_1}+\D_{\phi_2}-d+\D}{2})\G(\f{\D_{\phi_1}+\D_{\phi_2}-\D}{2})\G(\f{\pm \D_{12}+\D}{2})}{\G(\D)(\Delta-\alpha)(d+\alpha-\Delta)} |x|^{\Delta-\D_{\phi_1}-\D_{\phi_2}}
\end{equation}

\end{enumerate}

\textbf{Anomalous Dimension Computation:}

The two set of poles collide producing a double pole when $\nu_2 = \nu_1 + \epsilon,~~\e\rightarrow 0$. Factoring out the intrinsic simple pole residue of the Gamma function, the double pole has the form
\begin{equation}
\frac{(-1)^m}{m!} \frac{F(\nu)}{(\nu - \nu_1)(\nu - \nu_1 - \epsilon)} |x|^{-(\D_{\phi_1}+\D_{\phi_2}- h + i\nu)}
\end{equation}
and we need to find the coefficient of $\e^{-1}$ where
\beqa
F(\nu)&=&\f{ 1}{\left[ \nu^2+(\Delta-\f{d}{2})^2\right]}\f{ d~ \mathcal{A}}{(d+2\a+2i\nu)(d+2\a-2i\nu)} \nonumber\\
&\times&\f{\Gamma\left(\f{\D_{\phi_1}+\D_{\phi_2}-\f{d}{2}+i\nu}{2}\right)\Gamma\left(\f{\D_{\phi_1}-\D_{\phi_2}+d/2-i\nu}{2}\right)\Gamma\left(\f{-\D_{\phi_1}+\D_{\phi_2}+d/2-i\nu}{2}\right)\Gamma\left(\f{\D_{\phi_1}+\D_{\phi_2}-d/2-i\nu}{2}\right)}{\sqrt{\Gamma(\D_{\phi_1}-\f{d}{2}+1)\Gamma(\D_{\phi_2}-\f{d}{2}+1)\Gamma(\D_{\phi_1})\Gamma(\D_{\phi_2})}\G\left(\f{d}{2}-i\nu\right)} \nonumber \\
&\times&\f{1}{(((x_1-x_2+n_1\g))^2)^{\f{\D_{\phi_1}+\D_{\phi_2}-\f{d}{2}+i\nu}{2}}}
\eeqa
Evaluating the derivative using the product rule yields the following structure
\begin{equation}
\frac{d}{d\nu}\left[ F(\nu) |x_{12}|^{-(\D_{\phi_1}+\D_{\phi_2}-h+i\nu)} \right] = |x|^{-(\D_{\phi_1}+\D_{\phi_2}-h+i\nu)} \left[ F'(\nu) - i F(\nu)\ln|x_{12}| \right]
\end{equation}
We further note that only the $n_1=0$ term can give the $\ln |x_{12}|$ term and hence the anomalous dimension.
At the common pole $\nu_0 = i(\alpha - \f{d}{2})$, this simplifies since $-(\D_{\phi_1}+\D_{\phi_2}-h+i\nu_0) = 2m$ to give
\begin{equation}
\int_{\text{double-pole}} d\nu ~ F(\nu) = 2\pi \mathcal{A}' \frac{(-1)^m}{m!} F(\nu_0) \left[ \ln|x_{12}| + F_0 \right] |x_{12}|^{2m}
\end{equation}
We evaluate the factors appearing above to be
\begin{equation}
F(\nu_0) = \frac{\Gamma\left(\frac{\alpha \pm \Delta_{12}}{2}\right)\Gamma\left(\frac{(\D_{\phi_1}+\D_{\phi_2}+\alpha-d}{2}\right)}{2\Gamma(\alpha)(\Delta - \alpha)(\Delta + \alpha - d)}
\end{equation}

While $F_0 = \frac{d}{d(-i\nu)} \ln F(\nu)\Big|_{\nu_0}$ is given by
\beqa
F_0 &=& \frac{2(\alpha - \frac{d}{2})}{(\Delta - \alpha)(\Delta + \alpha - d)} + \frac{1}{d} - \psi(\alpha) + \frac{1}{2}\psi\left(\frac{\alpha+\Delta_{12}}{2}\right) + \frac{1}{2}\psi\left(\frac{\alpha-\Delta_{12}}{2}\right) \nonumber\\
&+& \frac{1}{2}\psi\left(\frac{(\D_{\phi_1}+\D_{\phi_2}+\alpha-d}{2}\right) - \frac{1}{2}\psi(m+1)
\eeqa
and  we have clubbed the $\nu$ independent terms in
\beqa
\mathcal{A}'=\f{g^2\mathcal{A}}{8\pi^{\f{d+2}{2}}\sqrt{\Gamma(\D_{\phi_1}-\f{d}{2}+1)\Gamma(\D_{\phi_2}-\f{d}{2}+1)\Gamma(\D_{\phi_1})\Gamma(\D_{\phi_2})}\G\left(\f{d}{2}-i\nu\right)}
\eeqa

To extract the anomalous dimension, we read the the coefficient of the $\ln\vert x_{12} \vert$ term in the double-pole residue and get

\begin{equation}
\gamma^{(m)} = - \f{\pi \mathcal{A}'}{a_0} \frac{(-1)^m}{m!} \frac{\Gamma\left(\frac{\alpha \pm \Delta_{12}}{2}\right)\Gamma\left(\frac{\D_{\phi_1}+\D_{\phi_2}+\alpha-d}{2}\right)}{2\Gamma(\alpha)(\Delta - \alpha)(\Delta + \alpha - d)}~,~ a^{(1)}=2\pi F(\nu_0)F_0
\end{equation}

Here $a_{n,0}^{(0)}$ are the $\mathcal{O}(g^0)$ coefficients. Only the scalar operators here pick up an anomalous dimension. The coefficient of the non-logarithmic term $F_0$ give the leading order correction to the OPE coefficients $a^{(1)}_{n,0}$.\\

\noindent
\textbf{Full correlator:}\\

\noindent
To construct the full correlator we sum over the thermal images. We do so for the case of the collison of the two set of the spurious poles. The full correlator is then split into

\begin{equation}
\mathcal{W} = \mathcal{W}_{\text{physical}} + \mathcal{W}_{\text{double pole}~(n=0)} + \mathcal{W}_{\text{double pole}~(n \neq 0)}
\end{equation}

where
    \begin{equation}
    \mathcal{W}_{\text{double pole}~(n=0)} = 2\pi \mathcal{A}' \frac{(-1)^m}{m!} F(\nu_0) {\left[ \ln|x_{12}| + F_0 \right]} |x_{12}|^{2m}
    \end{equation}
and
\begin{equation}
    \mathcal{W}_{\text{double pole}~(n_1 \neq 0)} = \sum_{n_1\neq 0} 2\pi \mathcal{A}' \frac{(-1)^m}{m!} F(\nu_0) \left[ \ln|x_{12} + n_1| + F_0 \right] |x_{12} + n_1|^{2m}
    \end{equation}

\vspace{1cm}
\subsection{Scalar Contact Diagram with modified vertex}\label{effY}
In this section we see how the bulk loop gives rise to effective vertices for the contact diagram as shown in figure \ref{effcontact}.
\[\begin{wittendiagram} 
 \draw (-1.15,-1.6) node[vertex] -- (-0.1,-0.5) node[vertex];
 \draw (1.15,-1.6) node[vertex] -- (-0.1,-0.5) node[vertex];
 \draw(-0.1,0.1) circle (0.6);
  \node at (-2.1, -1.7) {$x_1+n_1\g$};
  \node at (2.1, -1.7) {$x_2+n_2\g$};
  \node  at (0.7, -0.4) {$Y$};
     \node at (-1.2, -1) {$\D_{\phi_1}$};
  \node at (1.2, -1) {$\D_{\phi_2}$};
    \node at (0, 1) {$\D$};
\end{wittendiagram}\hspace{0.3cm}\quad
\bm{\xrightarrow{\hspace{1cm}} }
\quad
\hspace{0.8cm}
\begin{wittendiagram} \draw (-1.15,-1.6) node[vertex] -- (-0.1,-0.5) node[vertex];
 \draw (1.15,-1.6) node[vertex] -- (-0.1,-0.5) node[vertex, red];
  \node at (-2.1, -1.7) {$x_1+n_1\g$};
  \node at (2.1, -1.7) {$x_2+n_2\g$};
  \node at (0.4, -0.2) {$Y$};
\end{wittendiagram}\]

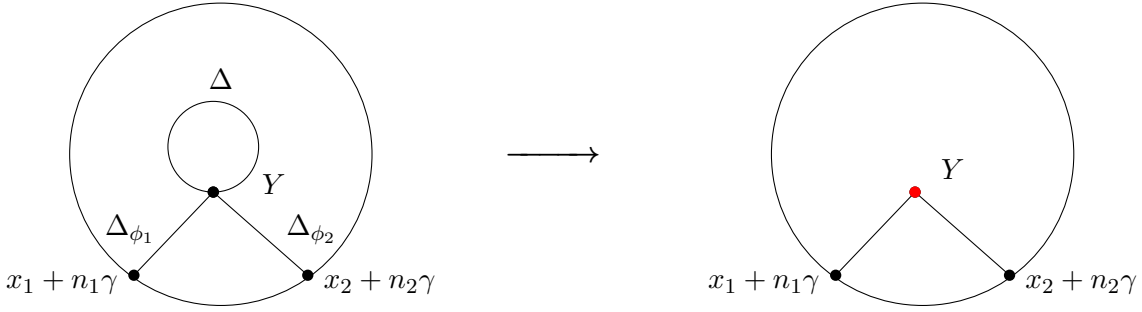
\captionof{figure}{The contact diagram with the effective vertex depicted by the red point}\label{effcontact}

\vspace{0.5cm}
The diagram is given by the two bulk to boundary propagators and an effective vertex $\mathcal{A}~z^\a$ at the point of contact.
\beqa
\mathcal{W}&=&\sum_{n_1,n_2}\int_0^\infty dz_0
\int d^d\vec{z}\,
\frac{\mathcal{A}~z^{\a+\D_{\phi_1}+\D_{\phi_2}-d-1}}
{\left[z^2+(x-x_1-n_1\g)^2\right]^{\D_{\phi_1}}
 \left[z^2+(x-x_2-n_2\g)^2\right]^{\D_{\phi_2}}}
\eeqa

This integral may be performed using \ref{3pt-integral} and our diagram simply is
\beqa
\mathcal{W}&=&\f{g}{4}\sum_n \f{\mathcal{A}\Gamma(\alpha-d-1+\D_{\phi_1}+\D_{\phi_2})\G\left(\f{\D_{\phi_1}+\D_{\phi_2}-\a}{2}\right)\G\left(\f{\a+\D_{\phi_2}-\D_{\phi_1}}{2}\right)\G\left(\f{\a-\D_{\phi_2}+\D_{\phi_1}}{2}\right)}{\G(\alpha)\sqrt{\Gamma\left(\D_{\phi_1}-\f{d}{2}+1\right)\Gamma\left(\D_{\phi_2}-\f{d}{2}+1\right)\G(\D_{\phi_1})\G(\D_{\phi_2})}}\nonumber\\
&\times& \f{1}{((x_1-x_2-n\g)^2)^{\f{\D_{\phi_1}+\D_{\phi_2}-\a}{2}}}
\eeqa
where $n=n_1-n_2$ and again
\beqa
\sum_{n}\f{1}{((x_1-x_2-n\g)^2)^{\f{\D_{\phi_1}+\D_{\phi_2}-\a}{2}}}=\xi_{EH}(\D_{\phi_1}+\D_{\phi_2}-\a,\t_{12},\mathbf{x}_{12})
\eeqa
We can thus keep increasing the power $\a$ above to get the non local factor from the additional loop.\\

\noindent
\textbf{Anomalous dimensions:}\\

\noindent
We notice that $\G(\f{\D_{\phi_1}+\D_{\phi_2}-\a}{2})$ has poles when $\D_{\phi_1}+\D_{\phi_2}-\a=-2 j$. This is precisely where $\D_{\phi_1}+\D_{\phi_2}+2 n=2\tilde{\D}+2m$ or when $[\phi\phi]$ and $[\tilde{\D}\tilde{\D}]$ become degenerate. Focusing on the $n=0$ term as before, this gives for $\a=\D_{\phi_1}+\D_{\phi_2}+\e$
\beqa
\G(\f{\D_{\phi_1}+\D_{\phi_2}-\a}{2})\sim\f{2(-1)^j}{n!\e},~~|x_{12}|^{2j+\e}\sim |x_{12}|^{2j}(1-\e \ln |x_{12}|)
\eeqa
and thus
\beqa
\g_\phi=\mathcal{A}\f{g}{2}\f{(-1)^j}{n!} \f{\Gamma(2\D_{\phi_1}+2\D_{\phi_2}+2j-d-1)\G\left(\D_{\phi_2}+j\right)\G\left(\D_{\phi_1}+j\right)}{\G(\D_{\phi_1}+\D_{\phi_2}+\j)\sqrt{\Gamma\left(\D_{\phi_1}-\f{d}{2}+1\right)\Gamma\left(\D_{\phi_2}-\f{d}{2}+1\right)\G(\D_{\phi_1})\G(\D_{\phi_2})}}
\eeqa

\subsection{Spinning loop effective vertix}
In this section we show that a spinning loop too gives rise to effective vertices in the bulk. We first collect a few results for the spinning propagators form \cite{Costa:2014kfa} to use in our analysis. The bulk to bulk propagator for a spinning field with spin $J$ and with polarization vectors $W_1$ and $W_2$ can be written as
\begin{equation}
\Pi_{\Delta,J}(X_1,X_2,W_1,W_2)
=
\sum_{k=0}^{J}
(W_{12})^{J-k}
\bigl[(W_1\cdot X_2)(W_2\cdot X_1)\bigr]^k
\,g_k(u),
\tag{28}
\end{equation}
where we have
\[
W_{12}=W_1\cdot W_2.
\]
Further for a spinning bulk to bulk propagator the split representation reads
\begin{equation}
\frac{\nu^2}{\pi J! \left(\f{d}{2}-1\right)_J}
\int \;
\Pi_{\f{d}{2}+i\nu,J}(X_1,P;W_1,D_Z)\,
\Pi_{\f{d}{2}-i\nu,J}(X_2,P;W_2,Z).
\tag{201}
\end{equation}

The boundary contraction can be done resulting in
\begin{equation}
\frac{2^J J! \, \mathcal{C}_{\f{d}{2}+i\nu,J}\, \mathcal{C}_{\f{d}{2}-i\nu,J}\, \nu^2}
{\pi \left(\f{d}{2}-1\right)_J}
\int dP \,
\frac{(P\!\cdot\! W_1)^J (P\!\cdot\! W_2)^J}
{(-2P\!\cdot\! X_1)^{\f{d}{2}+i\nu+J}
(-2P\!\cdot\! X_2)^{\f{d}{2}-i\nu+J}}
\,C_J^{\,h-1}(t),
\tag{202}
\end{equation}

where $C_J^{\,\f{d}{2}-1}(t)$ is the Gegenbauer polynomial and \(t\) is
\begin{equation}
t
=
X_1\!\cdot\! X_2
+
\frac{(P\!\cdot\! X_1)(P\!\cdot\! X_2)}
{(P\!\cdot\! W_1)(P\!\cdot\! W_2)}
\,W_1\!\cdot\! W_2
-
\frac{P\!\cdot\! X_1}{P\!\cdot\! W_1}
\,W_1\!\cdot\! X_2
-
\frac{P\!\cdot\! X_2}{P\!\cdot\! W_2}
\,W_2\!\cdot\! X_1.
\tag{203}
\end{equation}
For simplicity, we choose polarizations such that $W_1 ·X_2$ and $W_2 ·X_1$ vanish in the above analysis. Thus the propagator only has the $g_0$ term surviving.
Thus, from the term proportional to $(W_{12})^J$ in the propagator we obtain
\begin{equation}
\begin{aligned}
g_0(u)
&=
\frac{2\pi^{\f{d}{2}} J!\, \mathcal{C}_{\Delta,J}\, \mathcal{C}_{d-\Delta,J}\,
\Gamma\left(1+\f{d}{2}-\Delta\right)}
{\Gamma(d-\Delta)}
(2u)^{-\Delta}
\\
&\quad\times
\sum_{l=0}^{J}
\frac{(-1)^l}
{(J-l)!\,(d-\Delta)_l}
\,{}_2F_1\!\left(
-l,\,
\frac{3}{2}-\f{d}{2}-J;\,
3-d-2J;\,
-\frac{2}{u}
\right)
\\
&\qquad\qquad\times
{}_2F_1\!\left(
\frac{1}{2}+\Delta-\f{d}{2},\,
\Delta+l;\,
1+2\Delta-d;\,
-\frac{2}{u}
\right).
\end{aligned}
\end{equation}

Using the above we essentially have for a coincident thermal bulk to bulk propagator the following
\begin{equation}
\begin{aligned}
g_0(u)
&=\sum_{n\neq 0}
\frac{2\pi^{\f{d}{2}} J!\, \mathcal{C}_{\Delta,J}\, \mathcal{C}_{2h-\Delta,J}\,
\Gamma\left(1+\f{d}{2}-\Delta\right)}
{\Gamma(d-\Delta)}
\left(\f{z^2}{n^2}\right)^{\Delta}
\\
&\quad\times
\sum_{l=0}^{J}
\frac{(-1)^l}
{(J-l)!\,(d-\Delta)_l}
\,{}_2F_1\!\left(
-l,\,
\frac{3}{2}-\f{d}{2}-J;\,
3-2h-2J;\,
-\frac{4 z^2}{n^2}
\right)
\\
&\qquad\qquad\times
{}_2F_1\!\left(
\frac{1}{2}+\Delta-\f{d}{2},\,
\Delta+l;\,
1+2\Delta-d;\,
-\frac{4 z^2}{n^2}
\right).
\end{aligned}
\end{equation}
The two hypergeometric functions in the above expression can be combined using the formulae in appendix \ref{2f1} to write $g_0$ as
\beqa
g_0=\frac{2\pi^{\f{d}{2}} J!\, \mathcal{C}_{\Delta,J}\, \mathcal{C}_{d-\Delta,J}\,
\Gamma\left(1+\f{d}{2}-\Delta\right)}
{\Gamma(d-\Delta)}
\left(\f{z_0^2}{n^2}\right)^{\Delta}\sum_{l=0}^{J}\frac{(-1)^l}
{(J-l)!\,(2h-\Delta)_l} \sum_{k=0}^{\infty}c_k~z^{k}
\eeqa
with
\beqa
c_k&=&\left(-\f{4}{n^2}\right)^\D\f{\left(\f{1}{2}+\D-\f{d}{2}\right)_k(\D+l)_k}{(1+2\D-d)_k}\nonumber\\
&\times&{}_4F_3\left(-k,-2\D+\f{d}{2},-l,\f{3}{2}-\f{d}{2}-J;\f{1}{2}+\D-\f{d}{2}-k,\f{1}{2}+\D-\f{d}{2},\D+l;1\right).
\eeqa
 Further following \cite{Giombi:2017hpr}, the corresponding vertex factor for the coincident bulk to bulk propagtor is
\[
\frac{(\partial_{W_1}\cdot D_{W_2})^J}
{(J!)^2\left(\frac{d-1}{2}\right)_J}
.\]
This contacts with the $(W_1.W_2)^J$ part of the previous expression of the bulk to bulk propagator to give
\begin{equation}
\frac{\left(\partial_{W_1}\cdot D_{W_2}\right)^J}{(J!)^2}
\,(W_1\cdot W_2)^J
=
\frac{(d+2J-1)(d+J-2)!}{(d-1)!\,J!}.
\end{equation}
Thus effectively we again have additional powers of $z^\a$ appearing from the spinning bulk loop and giving rise to effective vertices as
\beqa
\mathcal{A}_J~z^\a=\sum_{n\neq 0}\sum_{l=0}^{J}\sum_{k=0}^{\infty}\frac{(-1)^l2\pi^{\f{d}{2}}(d+2J-1)(d+J-2)!\, \Gamma\left(1+\f{d}{2}-\Delta\right)\mathcal{C}_{\Delta,J}\, \mathcal{C}_{d-\Delta,J}\,
} {(d-1)!(J-l)!\,(d-\Delta)_l ~\Gamma(d-\Delta) ~n^{2\D}} c_k~z^{k+\D}
\eeqa

\section{Witten diagrams and relation to thermal bootstrap $g_{dr}$ and $g_{arcs}$}\label{drarc}

In this section we compare the results of two-point correlators as established by bulk Witten diagram computations and the ones built by dispersion relations following \cite{Barrat:2025nvu}. We do this both for the bench (for the zero relative winding sector) and the tadpole with the effective vertex. In the bootstrap approach $g_{dr}$ denotes the part of the correlator that can be reconstructed from its discontinuities and is given for the zero spatial separation upto a constant $\k$ 
 while $g_{arcs}$ contains theory dependent low spin data that is not captured by the discontinuity. 
 We thus write at zero spatial separation
\beqa
g(\tau)=\sum_{\D}\zeta(2\D_\phi-\D,\t)+\zeta(2\D_\phi-\D,1-\t)+\kappa
\eeqa
while at non zero spatial separation can write
\beqa
g_{dr}(z,\bar{z})=\f{1}{(z-1)^{\D}(\bar{z}-1)^{\D}}+\f{1}{(z+1)^{\D}(\bar{z}+1)^{\D}}
\eeqa
and as constructed from its discontinuity in the $r-w$ plane
\begin{equation}
g_{\mathrm{dr}}(r, w) = \int_{0}^{r} \frac{\mathrm{d}w'}{2\pi i} \frac{w^2(1 - w'^4)}{w'(w' - w)(w' + w)(1 - w^2 w'^2)} \operatorname{Disc} g(r, w') \,,
\end{equation}
with the definition
\begin{equation}
\text{Disc } g(r,w') = g(r, w'+i0) - g(r,w'-i0).
\label{eq:2.18}
\end{equation}
The full correlator is then given by
\begin{equation}
g(z, \bar{z}) = g_{\text{dr}}(z, \bar{z}) + g_{\text{arcs}}(z, \bar{z}).
\label{eq:2.19}
\end{equation}
Since $g_{\text{dr}}(z, \bar{z})$ is KMS-symmetric on its own, note that $g_{\text{arcs}}(z, \bar{z})$ should also be KMS-symmetric.

\subsection{The exchange diagram with effective vertex}

At order $\mathcal{O}(\e)$ we only have Identity and $n=0$ of the single trace tower ($\chi$) contributing as for this theory $g\sim\sqrt{\e}$ for the fixed point in $d=6$ \cite{Fei:2014xta}.\\

\textbf{The complex $\t$ plane}:\\
We need to apply the above discussed relations to the spectrum of our tadpole diagram. We can thus write the contributions coming from the identity and $\chi$ as
\beqa
g^{\text{tad}}(\t)=g^{\text{tad}(\mathbf{1})}(\t)+g^{(\chi)}(\t)
\eeqa
We specialize to identical external scalars here with $\D
_{\phi_1}=\D_{\phi_2}=\D_\phi$. For the identity we have
\beqa
g_{dr}^{\text{tad}(\mathbf{1})}(\t)=\zeta(2\D_\phi,\t)+\zeta(2\D_\phi,1-\t)=\sum_{n=-\infty}^{\infty}|\t+n|^{\D-2\D_\phi}
\eeqa
Now for operator $\chi$
\beqa
\D_{\chi}=2\D_\phi+2m
\eeqa
thus
\beqa
g^{\text{tad}(\chi)}_{dr}(\t)-\kappa=\sum_{n=-\infty}^{\infty}\sum_{m=0}^{\infty}a^{(0)}_{\chi^m}|\t+n|^{2m}
\eeqa
But we have the identity
\beqa
\zeta(-2m,\t)+\zeta(-2m,1-\t)=0
\eeqa
so no these receive contribution at this order. Let us now introduce anomalous dimensions thus at $\mathcal{O}(\e)$
\beqa
\D_\phi=2-\f{\epsilon}{2}~~,~~\D=2\D_\phi+2m+\e\g
\eeqa
Thus we have for the identity
\beqa
g_{dr}^{\text{tad}(\mathbf{1})}=\zeta\left(4-\epsilon\g,\t\right)+\zeta\left(4-\epsilon\g,1-\t\right)
\eeqa
while for $\chi$ we write
\beqa
g_{dr}^{\text{tad}(\chi)}=\sum_m\zeta\left(-\epsilon\g-2m,\t\right)+\zeta\left(-\epsilon\g-2m,1-\t\right)=\sum_{n=-\infty}^{\infty}\sum_{m=0}^{\infty}a_\D|\t+n|^{2m+\epsilon\g}
\eeqa
we can expand the RHS using Taylor Series
\beqa
\f{1}{|\t+n|^{\epsilon\g-2m}}\sim |\t+n|^{2m}+\epsilon\g|\t+n|^{2m}\log(\t+n)
\eeqa
and the second term on the LHS above can be written as
\beqa
-\f{\partial}{\partial s}\left[\zeta(s,\t)+\zeta(s,1-\t)\right]|_{s=-2m}
\eeqa
as we can expand zeta and write
\beqa
\zeta\left(-\epsilon\g-2m,\t\right)=\zeta(-2m,\t)-\epsilon\g\zeta'(-2m,\t)~,~\zeta\left(-\epsilon\g-2m,1-\t\right)=\zeta(-2m,1-\t-\epsilon\g\zeta'(-2m,1-\t)\nonumber
\eeqa
so at this order we get the contribution
\beqa
g_{dr}^{\text{tad}(\chi)}=\epsilon\g\left(\zeta'(-2m,1-\t)+\zeta'(-2m,1-\t)\right)
\eeqa
However, operators $[\phi\phi]_{m,J}$ with $m>1$ appear at a higher order than $\mathcal{O}(\e)$, thus at this order we only have
\beqa
g_{dr}^{\text{tad}(\chi)}=\epsilon\g\left(\zeta'(-\e\g_\phi,1-\t)+\zeta'(-\e\g_\phi,1-\t)\right)
\eeqa\\

\textbf{Non-zero spatial separation}:\\

Here at $\mathcal{O}(\e)$ for the identity we have \cite{Barrat:2025nvu}
\begin{equation}
g_{\mathrm{dr}}^{\text{tad}(\mathbf{1})} = \frac{\varepsilon(z\bar{z} + 1)}{(z^2 - 1)(\bar{z}^2 - 1)} \log \left[ \frac{(z - 1)(\bar{z} - 1)}{z\bar{z}} \right] + \frac{\varepsilon}{(z + 1)(\bar{z} + 1)} \left[ \tanh^{-1}(z) + \tanh^{-1}(\bar{z}) \right] \,.
\end{equation}

The new contribution is for $\chi$
\beqa
g_{\text{dr}}^{\text{tad}(\phi^2)} = \varepsilon a_{\D}^{(0)}\gamma_{\phi} \left[ \log(1 - z^2) + \log(1 - \bar{z}^2) \right] ,
\eeqa

We can now compare our Witten diagram result to write
\beqa
g_{arcs}^{\text{tad}}=g_{WD}^{\text{tad}}-g_{dr}^{\text{tad}}
\eeqa
which gives
\beqa
g_{\text{arcs}}^{\text{tad}}(\tau, x) = \sum_{m}\f{(-1)^m}{m!} 2 \pi \mathcal{A}'F(\nu_0)F_0 \left(x^2 + (\tau+n)^2\right)^{\f{\e}{2}+m} 
\eeqa
as expected from

\beqa
g_{\text{arcs}}(\tau, x) = \sum_{\Delta} \sum_{J < J_\star} a_{\mathcal{O}}^{(\text{arcs})} \left(x^2 + \tau^2\right)^{\frac{\Delta}{2} - \Delta_\phi} C_J^{(\nu)} \left( \frac{\tau}{\sqrt{x^2 + \tau^2}} \right)
\eeqa
with $C^{(\nu)}_{J=0}(x)=1$. The total discontinuity sees the non-analytic terms of the full correlator. Our correlator is a complete image sum, so it contains no non-thermal term. $g_{\text{arcs}} \neq 0$ means that the discontinuity data do not reproduce the full $g^{\text{WD}}$: it contains, besides the non-analytic terms, analytic terms as well and these have no discontinuity.
We may now perform the generalized method of images to write
 \beqa
 g_{\text{arcs}}^{\text{tad}} = g^{\text{tad}}_{\text{WD}} - \frac{1}{2}\sum_m g_{\text{dr}}^{\text{tad}}(z - m, \bar{z} - m)
 \eeqa
We observe from the analytically continued expression for $\xi_{EH}$ that at the value $s=-2m$ the $1/\G(-m)$ goes to zero and hence the arcs of the generalized method of images vanish.

\subsection{The bubble diagram}

For this diagram (zero relative winding case) we get corrections to the conformal dimension and normalization as $\g_{\phi}$ and $z_{\phi}$ as at $\mathcal{O}(\e)$

\beqa
\D_\phi=\D_\phi+\e\g_\phi ~~,~~Z_\phi=1+\e z_\phi
\eeqa

\textbf{The complex $\t$ plane:}\\
With the above $\mathcal{O}(\e)$ corrections we can write
\beqa
g^{\text{bench}}(\t)=\zeta(\D_\phi,\t)+\zeta(\D_\phi,1-\t)-\e\g_\phi[\zeta'(\D_\phi,\t)+\zeta'(\D_\phi,1-\t)]
\eeqa
thus at this order we get the contribution
\beqa
g^{\text{bench}}_{dr}=-\e\g_\phi[\zeta'(\D_\phi,\t)+\zeta'(\D_\phi,1-\t)]
\eeqa
or alternately using the generalized method of images and combining the two contributions
\beqa
g_{dr}^{\text{bench}}=\sum_{n_1}\f{1}{(\t+n_1\g)^{\D_\phi+\e \g_\phi}}=\sum_{n_1}\left(\f{1}{(\t+n_1\g)^{\D_\phi}}-\f{\e\g_\phi}{(\t+n_1\g)^{\D_\phi+\e \g_\phi}}\log(\t+n_1\g)\right)
\eeqa

\textbf{Non-zero spatial separation:}\\
We have at $\mathcal{O}(1)$
\beqa
g_{dr}^{\text{bench}}=\sum_{n_1}\f{1}{(x_1+n_1\g-x_2)^{\D_\phi}}
\eeqa
The new contribution at $\mathcal{O}(\e)$ is
\beqa
g_{\text{dr}}^{\text{bench}} =\varepsilon a_1\gamma_{\phi} \left[ \log(1 - z^2) + \log(1 - \bar{z}^2) \right] =a_1 \g_\phi\sum_{n_1}\log(x_1+n_1\g-x_2)^2
\eeqa

where $a_1=1$ and $\gamma_{\phi}$ can be read from \ref{bench0}. To find $g_{arcs}^{\text{bench}}$ we compare the correlator constructed here with the full Witten diagram contribution as given in \ref{bench0}.

\beqa
g_{arcs}^{\text{bench}}=g^{\text{bench}}_{WD}-\f{1}{2}\sum_{n_1}g_{dr}^{\text{bench}}=\sum a_1z_\phi (x_1+n_1-x_2)^{-\D_\phi}
\eeqa
thus we see that does not vanish even after the generalized method of images.

\section{Discussion and Conclusions}\label{section7}
In this paper we studied two-point Witten diagrams in thermal AdS defined with the conformal boundary $S_\b^1\times \mathbb{R}^{d-1}$.
In this pursuit we give a general prescription that successfully reduces thermal AdS Witten diagrams to the usual vacuum integrals that get dressed by thermal sums, and obtained manifestly KMS invariant spectral representations for the contact, exchange, and one-loop tadpole and bubble diagrams. The thermal OPE data were also read off from thermal Witten diagrams. The spectral poles, apart from the physical exchange, fall into the following two families: (a) the double-trace tower of operators $[O_{\phi}O_\phi]_{n,J}$ of the external fields, and, (b) for the loop diagrams double traces built from the fields running in the bulk loop. At leading order in the coupling, collisions between these families produce double poles with resulting $\log|x_{12}|$ terms giving the anomalous dimensions of the double-trace operators. \\

As an interesting result from the analysis of coincident point bulk to bulk propagators for both scalar and spinning fields in thermal AdS we established that these act as effective vertices from which the one-loop tadpole diagrams can be reconstructed. For the tadpole we found that the coincident point thermal bulk-to-bulk propagator can be expanded as a sum of powers of the radial coordinate $z^{2\tilde\Delta+2k}$, where each power acts as an effective bulk vertex. Each term from the series corresponds to a thermal one-point function of a double-trace operator $[O_{\phi}O_{\phi}]_{n,J}$ of the loop field. This works for both scalar and spinning loop fields. The bench diagram does not have this property since its internal propagators are not at coincident points, so no local effective vertices arise. We also generalized our Witten diagram computations to a spin exchange diagram with external scalars. Finally, we compared the Witten diagram expressions with the correlators constructed using the dispersion relations where the image sum of the part of the Witten diagram determined by the discontinuity reproduces the dispersive reconstruction, and the remainder
(the “arcs”) is computed explicitly from the bulk diagram. The bulk computations could therefore supply independent thermal data and a way to reinterpret the CFT results. We computed the anomalous dimensions only for scalar double traces; the spinning ones receive contributions from other diagrams, which we did not compute.\\

 A natural extension of the work done in this paper would be to consider Witten diagrams for two-point functions by generalizing the boundary scalars to boundary spinning operators. We also aim to extend our analysis to the analysis for the $S^1_\beta\times S^{d-1}$ boundary. There the Hawking–Page transition appears, and the thermal-AdS Witten diagrams computed here would describe the low-temperature phase, with the black-hole phase requiring a different saddle. We could also study the case of the alternate quantization and related double-trace flows by repeating the loop analysis with the Neumann boundary condition as compared to the Dirichlet boundary condition studied in this paper. This tests how the effective vertices change along a double-trace RG flow at finite temperature. We could also perform a correlator-based exploration of the phases of scalar field theories studied in thermal AdS in \cite{Kakkar:2022hub} 
\cite{Bishwakarma:2026xhq} by studying the long-range behavior of correlators.

\vspace{5.5mm}
\noindent
{\bf Acknowledgements}
\vspace{0.5cm}

\noindent
I thank Shota Komatsu for several insightful discussions,
 extremely helpful advice and comments on the draft.
 I am also grateful to Apratim Kaviraj for various helpful discussions and also thank Arjun Bagchi, Diptarka Das and Swarnendu Sarkar for related discussions.
I also acknowledge the support and financial assistance of Council of Scientific \& Industrial Research (CSIR), Government of India through the CSIR Research Associateship scheme.

\appendix
\section{Useful formulae}
In this appendix we collect the useful formulae that we use in the evaluation of our Witten daigrams.
\subsection{AdS integrals}
\subsubsection{Bulk integral}
 The three point bulk point integral in embedding coordinates is
\beqa \label{bulkint}
\int dX K_a(P_1, X)K_b(P_2, X)K_c(P_3, X) = \frac{B_{a,b,c}}{(P_{12})^{\frac{a+b-c}{2}} (P_{23})^{\frac{b+c-a}{2}} (P_{31})^{\frac{c+a-b}{2}}} ,
\eeqa
where
\beqa
B_{a,b,c}=\f{\pi^{d/2}}{2}\sqrt{\mathcal{C}_a\mathcal{C}_b\mathcal{C}_c}\Gamma\left(-\f{d}{2}+\f{a+b+c}{2}\right)\f{\Gamma(\f{a+b-c}{2})\Gamma(\f{a-b+c}{2})\Gamma(\f{-a+b+c}{2})}{\Gamma(a)\Gamma(b)\Gamma(c)}
\eeqa
\subsubsection{Boundary integrals}
The boundary integrals can be written in Poincar\'e coordinates as follows. The two point boundary integral is given by \cite{Giombi:2017hpr}
\beqa\label{2ptbint}
\int \f{d^d y}{[(y_1-y)^2]^{a_1}[(y_2-y)^2]^{a_2}}=\f{\pi^{d/2}\Gamma(\f{d}{2}-a_1)\Gamma(\f{d}{2}-a_2)\Gamma(a_1+a_2-\f{d}{2})(y_{12}^2)^{\f{d}{2}-a_1-a_2}}{\Gamma(a_1)\Gamma(a_2)\Gamma(d-a_1-a_2)}
\eeqa
A particularly useful version of the above evaluated using dimensional regularisation, with $a_1+a_2=d$, gives
\beqa
\int \f{d^d y}{[(y_1-y)^2]^{a_1}[(y_2-y)^2]^{a_2}}= \delta_{a_1,a_2} \frac{2\pi^{d/2} (y_{12}^2)^{-\frac{d}{2}} \left( \log\left(\pi(y_{12}^2)\right) - \psi^{(0)} \left( \frac{d}{2} \right) \right)}{\Gamma \left( \frac{d}{2} \right)}
\eeqa

We may also write the three point integral as
 \beqa\label{3pt-integral}
 \int \f{d^d x}{[(x_2-x_1)^2]^{a_1} [(x_3-x_1)^2]^{a_2} [(x_4-x_1)^2]^{a_3}}=\frac{\pi ^{\f{d}{2}}\G\left(a_1+a_2-\frac{d}{2}\right) \G\left(a_1+a_3-\frac{d}{2}\right) \G\left(\frac{d}{2}-a_1\right)}{\G(a_1)\G(a_2)\G(a_3)(x_{34}^2)^{\f{d}{2}-a_1} (x_{24}^2)^{a_1+a_3-\f{d}{2}} (x_{23}^2)^{a_1+a_2-\f{d}{2}}}\nonumber\\
 \eeqa 
We also have the following useful integral \cite{Freedman:1998tz} 
\beqa
\int_0^\infty dz
\int d^dx_0\,
\frac{z^{a}}
{\left[z^2+(x_0-x_1)^2\right]^b
 \left[z^2+(x_0-x_2)^2\right]^c}=
I_{a,b,c,d}
(x_{12}^2)^{\f{1+a+d-2b-2c}{2}}
\eeqa
where
\beqa
I_{a,b,c,d}
=
\frac{\pi^{d/2}}{2}
\frac{
\Gamma\!\left(\frac{a+1}{2}\right)
\Gamma\!\left(b+c-\frac{a+d+1}{2}\right)
\Gamma\!\left(\frac{a+d+1}{2}-b\right)
\Gamma\!\left(\frac{a+d+1}{2}-c\right)
}
{
\Gamma(b)\Gamma(c)
\Gamma(1+a+d-b-c)
}.
\eeqa
\subsection{Hypergeometric function identities}\label{2f1}
The product of the two hypergeometric functions can be combined into a single hypergeometric function using \cite{wolfram}
\beqa
_{p}F_{q}(a_1, \dots, a_p; b_1, \dots, b_q; c z) \, _{r}F_{s}(\alpha_1, \dots, \alpha_r; \beta_1, \dots, \beta_s; d z) = \sum_{k=0}^{\infty} c_k z^k
\eeqa
where
\beqa
c_k & = & \frac{d^k \prod_{j=1}^r (\alpha_j)_k}{k! \prod_{j=1}^s (\beta_j)_k} \, _{p+s+1}F_{q+r} \Big( -k, 1-\beta_1-k, \dots, 1-\beta_s-k, a_1, \dots, \nonumber \\
& & a_p; 1-\alpha_1-k, \dots, 1-\alpha_r-k, b_1, \dots, b_q; \frac{(-1)^{r+s-1} c}{d} \Big)  \nonumber \\
\text{or}~~~~ c_k &=& \frac{c^k \prod_{j=1}^p (a_j)_k}{k! \prod_{j=1}^q (b_j)_k} _{q+r+1}F_{p+s} \Big( -k, 1-b_1-k, \dots, 1-b_q-k, \alpha_1, \dots, \alpha_r; \nonumber\\
& & 1-a_1-k, \dots, 1-a_p-k, \beta_1, \dots, \beta_s; \frac{(-1)^{p+q-1} d}{c} \Big)
\eeqa

This may also be seen directly from \cite{harry}
\begin{equation}
{}_2F_1(a,b;c;pz)\,
{}_2F_1(a',b';c';qz)
=
\sum_{n=0}^{\infty}
\frac{(a)_n(b)_n(pz)^n}{n!\,(c)_n}
\,{}_4F_3\!\left(
\begin{matrix}
a',\,b',\,1-c-n,\,-n\\
c',\,1-a-n,\,1-b-n
\end{matrix};
\frac{q}{p}
\right).
\end{equation}

\section{Boundary integral for the bubble with $n_3\neq 0$ and $d>1$}\label{appb}
We show the detailed computation of the following boundary integral in this appendix
\begin{equation}
I(x_1,x_2) = \int d^d\tilde{x} \frac{\left( (x_2 + n_2 r - \tilde{x})^2 \right)^k}{\left((x_1 + n_1 r - \tilde{x})^2\right)^{a_1} \left((x_1 + n_1 r - n_3 r - \tilde{x})^2\right)^{a_2} \left((x_2 + n_2 r - n_3 r - \tilde{x})^2\right)^{a_4}}
\end{equation}
where we have defined
\begin{align}
a_1 = \f{\D_{\phi_1}-i\nu'-i\tilde{\nu}-2m}{2} ~,~
a_2 = \f{\D_{\phi_1}-\D_{\phi_2}+2i\tilde{\nu}-2k}{2} ~,~
a_4 = \f{\D_{\phi_2}-i\nu'+i\tilde{\nu}-2m}{2}
\end{align}

To unify the denominators, we introduce three Feynman parameters $z_1, z_2, z_4 \geq 0$ that are bound by the constraint $\delta(1 - z_1 - z_2 - z_4)$. This helps us to write
\begin{equation}
\frac{1}{\prod_{i =1,2,4} ((\tilde{x}-y_i)^2)^{a_i}} = \frac{\Gamma(a_1+a_2+a_4)}{\Gamma(a_1)\Gamma(a_2)\Gamma(a_4)} \int_0^1 dz_1 \int_0^1 dz_2 \int_0^1 dz_4 \, \delta(1-z_1-z_2-z_4) \frac{z_1^{a_1-1} z_2^{a_2-1} z_4^{a_4-1}}{[\mathcal{D}]^{a_1+a_2+a_4}}
\end{equation}
Expanding the denominator terms inside the bracket and grouping by powers of $\tilde{x}$ gives
\beqa
\mathcal{D} = z_1(\tilde{x}-y_1)^2 + z_2(\tilde{x}-y_2)^2 + z_4(\tilde{x}-y_4)^2 = (\tilde{x} - y_0)^2 + \Delta
\eeqa
where we have used the short hand notation $y_1=x_1+n_1\g,y_2=x_1+n_1\g-n_3\g, y_4=x_2+n_2\g-n_3\g$. Using the constraint $\sum z_i = 1$, $y_0$ and $\Delta$ are evaluated to be
\beqa
y_0 &=& z_1 (x_1 + n_1 \g) + z_2 (x_1 + n_1 \g - n_3 \g) + z_4 (x_2 + n_2 \g - n_3 \g) \\
\Delta &=& z_1 z_2 (n_3 \g)^2 + z_1 z_4 (x_1 + n_1 \g - x_2 - n_2 \g + n_3 \g)^2 + z_2 z_4 (x_1 + n_1 \g- x_2 - n_2 \g)^2.
\eeqa

Shifting the coordinate to $l = \tilde{x} - y_0$, the numerator $(\tilde{x}-y_3)$ transforms to $(l + y_0 - y_3)$. Using the identity $1 = z_1 + z_2 + z_4$, we can write the numerator as
\begin{align}
V &= z_1(y_1 - y_3) + z_2(y_2 - y_3) + z_4(y_4 - y_3) \nonumber \\
&= z_1 (x_1 + n_1 \g - x_2 - n_2 \g) + z_2 (x_1 + n_1 \g - x_2 - n_2 \g - n_3 \g) + z_4 (-n_3 \g)
\end{align}
Expanding the total numerator factor $((l+V)^2)^k$ and realizing that only the even terms in the $r$ series survive we can write
\begin{equation}
(l^2 + 2l\cdot V + V^2)^k = \sum_{p+2r+s=k} \frac{k!}{p! \, (2m)! \, s!} 2^{2r} (l^2)^p (l \cdot V)^{2r} (V^2)^s
\end{equation}
The integral can further be simplified as
\begin{equation}
\int d^d l \, f(l^2) \, (l \cdot V)^{2r} = \frac{(2r)!}{4^r r! \left(\frac{d}{2}\right)_r} (V^2)^r \int d^d l \, f(l^2) (l^2)^r
\end{equation}
Substituting this back, the coefficients simplify to yield the simplified loop integrand
\begin{equation}
\sum_{p+2r+s=k} \frac{k!}{p! \, r! \, s!} \frac{1}{\left(\frac{d}{2}\right)_r} (l^2)^{p+r} (V^2)^{s+r}
\end{equation}

Next we compute the integration over $d^d l$ using standard dimensional regularization formulas. Setting the total power of momentum to $j = p+r$ and total denominator exponent to $A = a_1+a_2+a_4$ we get
\begin{equation}
\int d^d l \frac{(l^2)^j}{(l^2 + \Delta)^A} = \pi^{\f{d}{2}} \frac{\Gamma(j + \frac{d}{2})\Gamma(A - j - \frac{d}{2})}{\Gamma(\frac{d}{2})\Gamma(A)} \Delta^{j + \frac{d}{2} - A}
\end{equation}
To perform the parameter integrations, we perform the following transformation
\begin{equation}
z_1 = 1 - x, \quad z_2 = x w, \quad z_4 = x(1-w)
\end{equation}

This helps us to re-write $\D$ as
\beqa
\Delta &=& x(1-w) y_{14}^2 \left[ 1 - x \left( 1 - w \frac{y_{24}^2}{y_{14}^2} \right) + (1-x)\frac{w}{1-w} \frac{y_{12}^2}{y_{14}^2} \right]\nonumber\\
&=& \f{x(1-w)y_{14}^2}{1-x} \left[ 1 +\f{x w}{1-x} u_1  + \f{w}{1-w} u_2 \right]
\eeqa
where we have defined 
\beqa
u_1=\frac{y_{24}^2}{y_{14}^2} ,~~u_2=\frac{y_{12}^2}{y_{14}^2} 
\eeqa
Raising $\Delta$ to the total denominator power $B = a_1 + a_2 + a_4 - p - r - \frac{d}{2}$ splits the components as
\begin{equation}
\frac{1}{\Delta^B} = \frac{1}{(y_{14}^2)^B} \cdot x^{-B} (1-w)^{-B} (1-x)^B\cdot \left[ 1 +\f{x w}{1-x} u_1  + \f{w}{1-w} u_2 \right]^{-B}
\end{equation}
which can be expanded to write
\beqa
=\frac{1}{(y_{14}^2)^B} \cdot x^{-B} (1-w)^{-B} (1-x)^B\cdot \sum_{M,N=0}^{\infty}\f{(-1)^{M+N} (B)_{M+N}}{M! N!}\left(\f{x w}{1-x}u_1\right)^M\left(\f{w}{1-w}u_2\right)^N
\eeqa

Shifting back to the variable $l = \tilde{x} - y_0$, we use the identity $1 = z_1 + z_2 + z_4$ to write $V \equiv y_0 - y_3$ and squaring it as
\begin{equation}
V^2 = z_1^2 y_{13}^2 + z_2^2 y{_23}^2 + z_4^2 y_{43}^2 + 2z_1 z_2 (y_{13} \cdot y_{23}) + 2z_1 z_4 (y_{13} \cdot y_{43}) + 2z_2 z_4 (y_{23} \cdot y_{43})
\end{equation}
Using the coordinate definitions $y_{13} = x_{12}$, $y_{43}= -n_3 r$, and $y_{23} = x_{12} - n_3 r$, the vector cross-terms reduce to pure linear combinations of the distance squares:
\begin{align}
2 (y_{13} \cdot y_{23}) &=y_{24}^2 + y_{14}^2 - y_{12}^2 \nonumber \\
2 (y_{13} \cdot y_{43}) &= y_{14}^2 - y_{24}^2 - y_{12}^2 \nonumber\\
2 (y_{23} \cdot y_{23}) &= y_{14}^2 - y_{24}^2 + y_{12}^2
\end{align}

We then expand $(V^2)^{s+r}$ to write
\begin{equation}
(V^2)^{s+r} = \sum_{j_1 + \dots + j_6 = s+r} \frac{(s+r)!}{j_1! \dots j_6!} \, \left( z_1^i z_2^g z_4^h \right) \, \cdot \, \Pi_{\text{coords}}
\end{equation}
where the powers and the corresponding parameters are
\begin{align}
i &= 2j_1 + j_4 + j_5 \implies (1-x)^i \nonumber \\
g &= 2j_2 + j_4 + j_6 \implies (xw)^g \nonumber \\
h &= 2j_3 + j_5 + j_6 \implies (x(1-w))^h
\end{align}
and the accompanying coordinate factor $\Pi_{\text{coords}}$ is
\begin{equation}\label{coordprod}
\Pi_{\text{coords}} = (y_{24}^2)^{j_1} (y_{14}^2)^{j_2} (y_{12}^2)^{j_3} (y_{24}^2 + y_{14}^2 - y_{12}^2)^{j_4} (y_{14}^2 - y_{24}^2 - y_{12}^2)^{j_5} (y_{14}^2 - y_{24}^2 + y_{12}^2)^{j_6}
\end{equation}
Applying successive binomial expansions to the three multi-term cross-term brackets in eq (\ref{coordprod}) breaks up the polynomials simply into
\begin{equation}
\Pi_{\text{coords}} \longrightarrow \sum_{k_1,k_2,k_3,p_1,p_2,p_3,r_1,r_2,r_3} (y_{12}^2)^{\alpha} (y_{24}^2)^{\beta} (y_{14}^2)^{\delta}
\end{equation}
where the cumulative powers are
\beqa
\a&=&j_3+p_3+k_3+r_3\nonumber\\
\b&=&j_1+p_1+k_1+r_1\nonumber\\
\d&=&j_2+p_2+k_2+r_2
\eeqa

We finally note that the $x$ and $w$ integrals give Beta functions
\beqa
\int_0^1 dx ~x^{A_x-1}(1-x)^{A_{x-1}-1}=B(A_x,A_{x-1})=\f{\G(A_x)\G(A_{x-1})}{\G(A_x+A_{x-1})}
\eeqa
and
\beqa
\int_0^1 dw ~w^{A_w-1}(1-w)^{A_{w-1}-1}=B(A_w,A_{w-1})=\f{\G(A_w)\G(A_{w-1})}{\G(A_w+A_{w-1})}
\eeqa
where collecting all powers we have defined
\beqa
A_x-1&=&-B+M+g+h+a_4+a_2-2\nonumber\\
A_{x-1}-1&=& B-M+i+a_1-1\nonumber\\
A_w-1&=&M+n+g+a_2-1\nonumber\\
A_{w-1}-1&=&-B-N+h+a_4-1
\eeqa

\section{Scalar loop as an effective vertex}\label{effverapp}

In this appendix we show how the bulk loop reduces using the split representation to effective vertices mention in eq (\ref{effver}). For this we look at following the segment of the tadpole diagram in figure \ref{loopY} that connects to the bulk point $Y$.
\beqa
I&=&\sum_{n_3\neq 0}\int dY d\tilde{x}\int d\tilde{\nu} \f{\tilde{\nu}^2}{\tilde{\nu}^2-(\tilde{\D}-d/2)^2} \sqrt{C_{\frac{d}{2}+i\tilde{\nu}}} \, \sqrt{C_{\frac{d}{2}-i\tilde{\nu}}} \, \sqrt{C_{\frac{d}{2}-i\nu'}} \, K_{\frac{d}{2}-i\nu'}(Y,x') \, K_{\frac{d}{2}+i\tilde{\nu}}(Y, \tilde{x}) \, K_{\frac{d}{2}-i\tilde{\nu}}(Y, \tilde{x}-n_3\g)\nonumber\\
&=&\int d\tilde{\nu} \f{\tilde{\nu}^2}{\nu'^2-(\tilde{\D}-d/2)^2} \sqrt{C_{\frac{d}{2}+i\tilde{\nu}}} \, \sqrt{C_{\frac{d}{2}-i\tilde{\nu}}} \, \sqrt{C_{\frac{d}{2}-i\nu'}} \int_0^\infty \frac{dz}{(z)^{d+1}} \int d^d x' \left(\frac{z}{z^2 + |x'-x''|^2}\right)^{\frac{d}{2}-i\nu'}\nonumber\\
&\times & \int d\tilde{x} \left(\frac{z}{z^2 + |x-\tilde{x}|^2}\right)^{\frac{d}{2}+i\tilde{\nu}} \left(\frac{z}{z^2 + |x-\tilde{x}-n_3|^2}\right)^{\frac{d}{2}-i\tilde{\nu}}
\eeqa

We do the $\tilde{x}$ integral first by shifting  $\tilde{x} \rightarrow \tilde{x} + x$ as this removes the $x$ dependence from the following
\begin{equation}
\int d \tilde{x}  \left(\frac{z}{z^2+\tilde{x}^2}\right)^{\frac{d}{2}+i\tilde{\nu}} \left(\frac{z}{z^2+(\tilde{x}-n_3)^2}\right)^{\frac{d}{2}-i\tilde{\nu}}.
\end{equation}

To evaluate this we use Feynman parameter identity 
\begin{equation}
\frac{1}{A^a B^b} = \frac{\Gamma(a+b)}{\Gamma(a)\Gamma(b)} \int_0^1 \frac{t^{a-1}(1-t)^{b-1}}{[tA + (1-t)B]^{a+b}} \, dt
\end{equation}
Setting
\begin{equation}
A = z^2 + \tilde{x}^2 \quad \text{and} \quad B = z^2 + (\tilde{x}-n_3)^2
\end{equation}
the denominator becomes:
\begin{align}
 t(z^2+\tilde{x}^2) + (1-t)(z^2+(\tilde{x}-n_3)^2)=
    &= z^2 + \tilde{x}^2 + n_3^2(1-t) - 2\tilde{x}~n_3(1-t)
\end{align}
Completing the square we write this as
\beqa
 z^2 + t(1-t)n_3^2 + (\tilde{x}-(1-t)n_3)^2
\eeqa
Thus the integral becomes
\begin{equation}
I = \frac{z^d ~ \Gamma(d)}{\Gamma\left(\frac{d}{2}+i\tilde{\nu}\right)\Gamma\left(\frac{d}{2}-i\tilde{\nu}\right)} \int_0^1 dt \, t^{a-1}(1-t)^{b-1} \int \frac{d^d\tilde{x}}{\left[z^2+t(1-t)n_3^2 + (\tilde{x}-(1-t)n_3)^2\right]^d}
\end{equation}

\noindent
Shift integration variable $\tilde{\tilde{x}} = \tilde{x} - (1-t)~n_3 \implies d^d\tilde{\tilde{x}} = d^d\tilde{x}$ allows us to write
\begin{equation}
I=\frac{z^d \Gamma(d)}{\Gamma\left(\frac{d}{2}+i\tilde{\nu}\right)\Gamma\left(\frac{d}{2}-i\tilde{\nu}\right)} \int_0^1 dt \, t^{a-1}(1-t)^{b-1} \int \frac{d^d\tilde{\tilde{x}}}{\left[(z^2+t(1-t)n_3^2) + \tilde{\tilde{x}}^2\right]^d}
\end{equation}
\noindent
Now using standard result
\beqa
\int \frac{d^d w}{(w^2+\Delta)^d}  = \frac{\pi^{d/2}\Gamma(d/2)}{\Gamma(d)} \Delta^{-d/2}
\eeqa
we write the integral as
\begin{align}
I &=\frac{\pi^{d/2}\Gamma(d/2)}{\Gamma(a)\Gamma(b)} \int_0^1 dt \, t^{a-1}(1-t)^{b-1} \left[1 + t(1-t)\frac{n_3^2}{z^2}\right]^{-d/2}
\end{align}
We next substitute $t = \frac{s}{1+s}$ giving

\begin{equation}
I = \frac{\pi^{d/2}\Gamma(d/2)}{\Gamma(a)\Gamma(b)} \int_0^\infty ds \, \frac{s^{a-1}}{(1+s)^d} \left[ \frac{s^2+s\left(2+\frac{n_3^2}{z^2}\right)+1}{(1+s)^2} \right]^{-d/2}
\end{equation}

\noindent
Now let $\cosh\rho = 1 + \frac{n_3^2}{2z^2}$ as this further simplifies the integral as
\begin{equation}
I = \frac{\pi^{d/2}\Gamma(d/2)}{\Gamma(a)\Gamma(b)} \int_0^\infty ds \, s^{a-1} \left[(s+e^\rho)(s+e^{-\rho})\right]^{-d/2}
\end{equation}

\noindent
Now applying Euler's generalized Beta integral
\begin{equation}
\int_0^\infty x^{\nu-1}(\beta+x)^{-\mu}(\gamma+x)^{-\varrho} \, dx = \beta^{-\mu}\gamma^{\nu-\mu-\varrho} B(\nu, \mu+\varrho-\nu) \, {}_2F_1\left(\mu, \nu; \mu+\varrho; 1-\frac{\gamma}{\beta}\right)
\end{equation}
The standard linear transformation formula to change the argument from $1 - w$ to $w$ is written as
\beqa
    {}_2F_1(A, B; C; 1 - w) &=& \frac{\Gamma(C)\Gamma(C - A - B)}{\Gamma(C - A)\Gamma(C - B)} {}_2F_1(A, B; A + B - C + 1; w)\nonumber\\
    &+& \frac{\Gamma(C)\Gamma(A + B - C)}{\Gamma(A)\Gamma(B)} w^{C - A - B} {}_2F_1(C - A, C - B; C - A - B + 1; w)
\eeqa

Using the above we get
\beqa
I=\pi^{d/2}\sum_{\pm}\f{\G(\pm i\tilde{\nu})}{\G(d/2\pm i\tilde{\nu})}e^{-\rho (d/2\pm i\tilde{\nu})}{}_2F_1(d/2,d/2\pm i\tilde{\nu},1\pm i\tilde{\nu},e^{-2\rho})
\eeqa

Now we do the $\nu'$ integral by picking the pole at $i\tilde{\nu}=\tilde{\D}-d/2$ we get
\beqa
I  = \mathcal{C}_{\tilde{\D}} e^{-\rho\tilde{\Delta}} \, {}_2F_1\left(\frac{d}{2}, \tilde{\Delta}; 1-\frac{d}{2}+\tilde{\Delta}; e^{-2\rho}\right)
\eeqa

We now want to expand this as a series in $z$. For this we define
\begin{equation}
\xi = e^{-\rho} \implies \cosh\rho = \frac{e^\rho+e^{-\rho}}{2} = 1 + \frac{n_3^2}{2z^2} = \frac{\xi+\xi^{-1}}{2}
\end{equation}
and
\begin{equation}
u = \frac{n_3^2}{z^2} \rightarrow \xi+\xi^{-1} = 2+u 
\end{equation}
Now we use the following transformation identity of the hypergeometric functions
\beqa
{}_2F_1(b,a; a-b+1; \xi^2) = (1+z_y)^{-2a} {}_2F_1\left(a, a-b+\frac{1}{2}; 2a-2b+1; \frac{4\xi}{(1+\xi)^2}\right)
\eeqa
to write for us
\begin{equation}
\xi^{\tilde{\Delta}}{}_2F_1\left(\frac{d}{2}, \tilde{\Delta}; \tilde{\Delta}-\frac{d}{2}+1; \xi^2\right) = \left(\frac{\xi}{(1+\xi)^2}\right)^{\tilde{\Delta}} {}2F_1\left(\tilde{\Delta}, \tilde{\Delta}-\frac{d}{2}+\frac{1}{2}; 2\tilde{\Delta}-d+1; \frac{4\xi}{(1+\xi)^2}\right)
\end{equation}
further since
\begin{equation}
\frac{\xi}{(1+\xi)^2} = \frac{\xi}{1+\xi^2+2\xi} = \frac{1}{4+u}
\end{equation}
we get
\begin{equation}
I=\mathcal{C}_{\tilde{\Delta}} (4+u)^{-\tilde{\Delta}}{}_2F_1\left(\tilde{\Delta}, \tilde{\Delta}-\frac{d}{2}+\frac{1}{2}; 2\tilde{\Delta}-d+1; \frac{4}{4+u}\right)
\end{equation}
\noindent
Next we use the Pfaff transformation for the hypergeometric function
\beqa
{}2F_1(a,b,c; x) = (1-x)^{-a} {}_2F_1\left(a, c-b, c; \frac{x}{x-1}\right)
\eeqa
to finally write this as the series in $z$ thus matching the expression in eq (\ref{effver})
\beqa
I&=& \sum_{n_3} \mathcal{C}_{\tilde{\Delta}} \sum_k\left(-4\right)^k \frac{(\tilde{\Delta})_k \left(\tilde{\Delta}-\frac{d}{2}+\frac{1}{2}\right)_k}{k!(2\tilde{\Delta}-d+1)_k}(n_3^2)^{-\tilde{\Delta}-k}(z^2)^{\tilde{\D}+k}=\mathcal{A}~z^\a
\eeqa


\begin{thebibliography}{99}

\bibitem{Witten:1998zw}
E.~Witten,
``Anti-de Sitter space, thermal phase transition, and confinement in gauge theories,''
Adv. Theor. Math. Phys. \textbf{2} (1998), 505-532
doi:10.4310/ATMP.1998.v2.n3.a3
[arXiv:hep-th/9803131 [hep-th]].

\bibitem{Hawking:1982dh}
S.~W.~Hawking and D.~N.~Page,
``Thermodynamics of Black Holes in anti-De Sitter Space,''
Commun. Math. Phys. \textbf{87} (1983), 577
doi:10.1007/BF01208266

\bibitem{Iliesiu:2018fao}
L.~Iliesiu, M.~Kolo{\u{g}}lu, R.~Mahajan, E.~Perlmutter and D.~Simmons-Duffin,
``The Conformal Bootstrap at Finite Temperature,''
JHEP \textbf{10} (2018), 070
doi:10.1007/JHEP10(2018)070
[arXiv:1802.10266 [hep-th]].


\bibitem{Iliesiu:2018zlz}
L.~Iliesiu, M.~Kolo{\u{g}}lu and D.~Simmons-Duffin,
``Bootstrapping the 3d Ising model at finite temperature,''
JHEP \textbf{12} (2019), 072
doi:10.1007/JHEP12(2019)072
[arXiv:1811.05451 [hep-th]].

\bibitem{Alday:2020eua}
L.~F.~Alday, M.~Kologlu and A.~Zhiboedov,
``Holographic correlators at finite temperature,''
JHEP \textbf{06} (2021), 082
doi:10.1007/JHEP06(2021)082
[arXiv:2009.10062 [hep-th]].

\bibitem{Barrat:2025nvu}
J.~Barrat, D.~N.~Bozkurt, E.~Marchetto, A.~Miscioscia and E.~Pomoni,
``The analytic bootstrap at finite temperature,''
[arXiv:2506.06422 [hep-th]].



\bibitem{Barrat:2025twb}
J.~Barrat, D.~N.~Bozkurt, E.~Marchetto, A.~Miscioscia and E.~Pomoni,
``Analytic thermal bootstrap meets holography,''
JHEP \textbf{05} (2026), 180
doi:10.1007/JHEP05(2026)180
[arXiv:2510.20894 [hep-th]].

\bibitem{Rodriguez-Gomez:2021pfh}
D.~Rodriguez-Gomez and J.~G.~Russo,
``Correlation functions in finite temperature CFT and black hole singularities,''
JHEP \textbf{06} (2021), 048
doi:10.1007/JHEP06(2021)048
[arXiv:2102.11891 [hep-th]].

\bibitem{Karlsson:2022osn}
R.~Karlsson, A.~Parnachev, V.~Prilepina and S.~Valach,
``Thermal stress tensor correlators, OPE and holography,''
JHEP \textbf{09} (2022), 234
doi:10.1007/JHEP09(2022)234
[arXiv:2206.05544 [hep-th]].

\bibitem{Dodelson:2022yvn}
M.~Dodelson, A.~Grassi, C.~Iossa, D.~Panea Lichtig and A.~Zhiboedov,
``Holographic thermal correlators from supersymmetric instantons,''
SciPost Phys. \textbf{14} (2023) no.5, 116
doi:10.21468/SciPostPhys.14.5.116
[arXiv:2206.07720 [hep-th]].


\bibitem{Kraus:2020nga}
P.~Kraus, S.~Megas and A.~Sivaramakrishnan,
``Anomalous dimensions from thermal AdS partition functions,''
JHEP \textbf{10} (2020), 149
doi:10.1007/JHEP10(2020)149
[arXiv:2004.08635 [hep-th]].


\bibitem{Buric:2026pes}
I.~Buri{\'c}, F.~Mangialardi, F.~Russo, V.~Schomerus and A.~Vichi,
``Thermal One-point Functions and Asymptotic CFT Data: QFT in AdS,''
[arXiv:2606.17167 [hep-th]].

\bibitem{Buric:2024kxo}
I.~Buric, F.~Russo, V.~Schomerus and A.~Vichi,
``Thermal one-point functions and their partial wave decomposition,''
JHEP \textbf{12} (2024), 021
doi:10.1007/JHEP12(2024)021
[arXiv:2408.02747 [hep-th]].

\bibitem{Kraus:2017ezw}
P.~Kraus, A.~Maloney, H.~Maxfield, G.~S.~Ng and J.~q.~Wu,
``Witten Diagrams for Torus Conformal Blocks,''
JHEP \textbf{09} (2017), 149
doi:10.1007/JHEP09(2017)149
[arXiv:1706.00047 [hep-th]].


\bibitem{Gobeil:2018fzy}
Y.~Gobeil, A.~Maloney, G.~S.~Ng and J.~q.~Wu,
``Thermal Conformal Blocks,''
SciPost Phys. \textbf{7} (2019) no.2, 015
doi:10.21468/SciPostPhys.7.2.015
[arXiv:1802.10537 [hep-th]].



\bibitem{Dodelson:2023vrw}
M.~Dodelson, C.~Iossa, R.~Karlsson and A.~Zhiboedov,
``A thermal product formula,''
JHEP \textbf{01} (2024), 036
doi:10.1007/JHEP01(2024)036
[arXiv:2304.12339 [hep-th]].


\bibitem{Arnaudo:2026axe}
P.~Arnaudo, C.~Iossa, R.~Karlsson and B.~Withers,
``OPE = QNM,''
[arXiv:2607.24909 [hep-th]].

\bibitem{Barrat:2026jfg}
J.~Barrat, D.~N.~Bozkurt, E.~Marchetto, A.~Miscioscia and E.~Pomoni,
``Analytic thermal bootstrap in momentum space: From thermal OPE to QNMs,''
[arXiv:2607.24919 [hep-th]].

\bibitem{Guo:2026xyl}
Y.~Guo, Z.~Li and T.~Shen,
``Thermal Polyakov bootstrap,''
[arXiv:2609.28627 [hep-th]].

\bibitem{Costa:2014kfa}
M.~S.~Costa, V.~Gon{$\c${c}}alves and J.~Penedones,
``Spinning AdS Propagators,''
JHEP \textbf{09} (2014), 064
doi:10.1007/JHEP09(2014)064
[arXiv:1404.5625 [hep-th]].

\bibitem{Mikhailov:2002bp}
A.~Mikhailov,
``Notes on higher spin symmetries,''
[arXiv:hep-th/0201019 [hep-th]].

\bibitem{Giombi:2017hpr}
S.~Giombi, C.~Sleight and M.~Taronna,
``Spinning AdS Loop Diagrams: Two Point Functions,''
JHEP \textbf{06} (2018), 030
doi:10.1007/JHEP06(2018)030
[arXiv:1708.08404 [hep-th]].

\bibitem{Sleight:2016gru}
C.~Sleight,
``Interactions in Higher-Spin Gravity: A Holographic Perspective,''
doi:10.5282/edoc.19912

\bibitem{Sleight:2016hyl}
C.~Sleight,
``Interactions in Higher-Spin Gravity: a Holographic Perspective,''
J. Phys. A \textbf{50} (2017) no.38, 383001
doi:10.1088/1751-8121/aa820c
[arXiv:1610.01318 [hep-th]].



\bibitem{Carmi:2019ocp}
D.~Carmi,
``Loops in AdS: From the Spectral Representation to Position Space,''
JHEP \textbf{06} (2020), 049
doi:10.1007/JHEP06(2020)049
[arXiv:1910.14340 [hep-th]].

\bibitem{Carmi:2021dsn}
D.~Carmi,
``Loops in AdS: from the spectral representation to position space. Part II,''
JHEP \textbf{07} (2021), 186
doi:10.1007/JHEP07(2021)186
[arXiv:2104.10500 [hep-th]].


\bibitem{Carmi:2024tzj}
D.~Carmi,
``Loops in AdS: from the spectral representation to position space. Part III,''
JHEP \textbf{08} (2024), 193
doi:10.1007/JHEP08(2024)193
[arXiv:2402.02481 [hep-th]].

\bibitem{Marchetto:2023xap}
E.~Marchetto, A.~Miscioscia and E.~Pomoni,
``Sum rules {\&} Tauberian theorems at finite temperature,''
JHEP \textbf{09} (2024), 044
doi:10.1007/JHEP09(2024)044
[arXiv:2312.13030 [hep-th]].


\bibitem{Carmi:2018qzm}
D.~Carmi, L.~Di Pietro and S.~Komatsu,
``A Study of Quantum Field Theories in AdS at Finite Coupling,''
JHEP \textbf{01} (2019), 200
doi:10.1007/JHEP01(2019)200
[arXiv:1810.04185 [hep-th]].











\bibitem{Liu:2018jhs}
J.~Liu, E.~Perlmutter, V.~Rosenhaus and D.~Simmons-Duffin,
``$d$-dimensional SYK, AdS Loops, and $6j$ Symbols,''
JHEP \textbf{03} (2019), 052
doi:10.1007/JHEP03(2019)052
[arXiv:1808.00612 [hep-th]].


\bibitem{Aharony:2020omh}
O.~Aharony, S.~M.~Chester and E.~Y.~Urbach,
``A Derivation of AdS/CFT for Vector Models,''
JHEP \textbf{03} (2021), 208
doi:10.1007/JHEP03(2021)208
[arXiv:2011.06328 [hep-th]].


\bibitem{Aharony:2021ovo}
O.~Aharony, S.~M.~Chester and E.~Y.~Urbach,
``AdS from CFT for scalar QED,''
Phys. Rev. D \textbf{104} (2021) no.12, 126011
doi:10.1103/PhysRevD.104.126011
[arXiv:2109.05512 [hep-th]].













\bibitem{Freedman:1998tz}
D.~Z.~Freedman, S.~D.~Mathur, A.~Matusis and L.~Rastelli,
``Correlation functions in the CFT(d) / AdS(d+1) correspondence,''
Nucl. Phys. B \textbf{546} (1999), 96-118
doi:10.1016/S0550-3213(99)00053-X
[arXiv:hep-th/9804058 [hep-th]].



\bibitem{Rosenhaus:2018zqn}
V.~Rosenhaus,
``Multipoint Conformal Blocks in the Comb Channel,''
JHEP \textbf{02} (2019), 142
doi:10.1007/JHEP02(2019)142
[arXiv:1810.03244 [hep-th]].



\bibitem{Alkalaev:2023evp}
K.~Alkalaev and S.~Mandrygin,
``Torus shadow formalism and exact global conformal blocks,''
JHEP \textbf{11} (2023), 157
doi:10.1007/JHEP11(2023)157
[arXiv:2307.12061 [hep-th]].


\bibitem{Fei:2014xta}
L.~Fei, S.~Giombi, I.~R.~Klebanov and G.~Tarnopolsky,
``Three loop analysis of the critical O(N) models in 6-{\ensuremath{\varepsilon}} dimensions,''
Phys. Rev. D \textbf{91} (2015) no.4, 045011
doi:10.1103/PhysRevD.91.045011
[arXiv:1411.1099 [hep-th]].
    


\bibitem{Kakkar:2022hub}
A.~Kakkar and S.~Sarkar,
``On partition functions and phases of scalars in AdS,''
JHEP \textbf{07} (2022), 089
doi:10.1007/JHEP07(2022)089
[arXiv:2201.09043 [hep-th]].


\bibitem{Bishwakarma:2026xhq}
B.~Bishwakarma, A.~Kakkar and S.~Sarkar,
``Neumann scalars in AdS: partition functions and phases,''
[arXiv:2607.04417 [hep-th]].

\bibitem{harry}
``HIGHER TRANSCENDENTAL FUNCTIONS, Volume I Based, in part, on notes left by Harry Bateman''

\bibitem{wolfram}
``https://functions.wolfram.com/07.31.16.0001.01 ''


\end{thebibliography}
\end{document}